\documentclass[sn-mathphys]{sn-jnl}

\usepackage{multirow}
\jyear{2021}%
\theoremstyle{thmstyleone}%
\theoremstyle{thmstyletwo}%

\theoremstyle{thmstylethree}%

\usepackage{fancyhdr}

\begin{document}

\title[Mohamed et al.]{Intrinsic Voltage Offsets in Memcapacitive Bio-Membranes Enable High-Performance Physical Reservoir Computing}

\author[1]{\fnm{Ahmed S.} \sur{Mohamed}}\email{asm6015@psu.edu}

\author[2]{\fnm{Anurag} \sur{Dhungel}}\email{adhungel@go.olemiss.edu}

\author*[2]{\fnm{Md Sakib} \sur{Hasan}}\email{mhasan5@olemiss.edu}

\author*[1]{\fnm{Joseph S.} \sur{Najem}}\email{jsn5211@psu.edu}

\affil[1]{\orgdiv{Department of Mechanical Engineering}, \orgname{The Pennsylvania State University}, \orgaddress{\city{University Park}, \postcode{16802}, \state{Pennsylvania}, \country{USA}}}

\affil[2]{\orgdiv{Department of Electrical and Computer Engineering}, \orgname{University of Mississippi}, \orgaddress{\city{Oxford}, \postcode{38677}, \state{Mississippi}, \country{USA}}}
\abstract{

Reservoir computing is a brain-inspired machine learning framework for processing temporal data by mapping inputs into high-dimensional spaces. Physical reservoir computers (PRCs) leverage native fading memory and nonlinearity in physical substrates, including atomic switches, photonics, volatile memristors, and, recently, memcapacitors, to achieve efficient high-dimensional mapping. Traditional PRCs often consist of homogeneous device arrays, which rely on input encoding methods and large stochastic device-to-device variations for increased nonlinearity and high-dimensional mapping. These approaches incur high pre-processing costs and restrict real-time deployment. Here, we introduce a novel heterogeneous memcapacitor-based PRC that exploits internal voltage offsets to enable both monotonic and non-monotonic input-state correlations crucial for efficient high-dimensional transformations. We demonstrate our approach's efficacy by predicting a second-order nonlinear dynamical system with an extremely low prediction error ($1.80\times10^{-4}$). Additionally, we predict a chaotic Hénon map, achieving a low normalized root mean square error ($0.080$). Unlike previous PRCs, such errors are achieved without input encoding methods, underscoring the power of distinct input-state correlations. Most importantly, we generalize our approach to other neuromorphic devices that lack inherent voltage offsets using externally applied offsets to realize various input-state correlations. Our approach and unprecedented performance are a major milestone towards high-performance full in-materia PRCs.}




\maketitle

\section{Introduction}\label{sec1}

Reservoir Computing (RC) is a developing brain-inspired machine learning architecture that maps input features into a higher-dimensional space to classify and process temporal data. RC has achieved state-of-the-art performances for nontrivial time series processing tasks, particularly ones with short-term history dependence, such as speech recognition \cite{Verstraeten2006Reservoir-basedRecognition}, chaotic time series prediction \cite{Gauthier2021NextComputing, Zhong2021DynamicProcessing}, and many more \cite{Cucchi2022Hands-onImplementation, Nakajima2020PhysicalPerspective, Tanaka2019RecentReview, Schrauwen2007AnImplementations}.
At its outset, RC algorithms were implemented and extensively studied in silico using RNN-based reservoirs \cite{Lukosevicius2009ReservoirTraining, Appeltant2011InformationSystem} (\textbf{Supplementary Note 1}), demonstrating improved accuracies in temporal signal prediction. Unlike the hardware implementations of RNNs (i.e., Recurrent Neural Networks), typically requiring advanced technologies of neural network \cite{Misra2010ArtificialProgress} or neuromorphic hardware; theoretically, any dynamical system with sufficient short-term memory and nonlinearity can serve as a physical reservoir \cite{Cucchi2022Hands-onImplementation, Appeltant2011InformationSystem}. This concept introduced a new subdomain, Physical Reservoir Computing (PRC), promising faster information processing and lower learning costs \cite{Tanaka2019RecentReview}. The growing interest in PRCs led to the development of various implementations, including spintronic oscillators \cite{Torrejon2017NeuromorphicOscillators}, atomic switch networks \cite{Sillin2013AComputing}, silicon photonic modules \cite{Vandoorne2014ExperimentalChip, VanDerSande2017AdvancesComputing}, ferroelectric transistors \cite{Toprasertpong2022ReservoirTransistor, Duong2023DynamicProcessing}, and, more particularly, dynamic mem-elements (short for memory elements) \cite{Du2017ReservoirProcessing, Zhu2020MemristorAnalysis, Midya2019ReservoirMemristors, Zhong2021DynamicProcessing, Hossain2023Biomembrane-BasedProcessing} due to their increased functional density and energy efficiency. Mem-elements \cite{DiVentra2009CircuitMeminductors, Liu2021TheEmulator} refer to an array of nonlinear passive circuit elements with stimulus- and history-dependent states, such as memristors \cite{Chua1971MemristorTheElement, Strukov2008TheFound}, memcapacitors \cite{Najem2019DynamicalMembranes}, meminductors \cite{Dinavahi2023PhysicalElement}, and other higher-order mem-elements \cite{Biolek2016EveryTheorem, Abdelouahab2014Memfractance:Memory}. Unlike conventional reservoirs that achieve higher-dimensional feature spaces via RNN-based spatial nodes, dynamic memristors and memcapacitors leverage their ability to co-locate nonlinearity and short-term memory in their states to realize higher-dimensional mapping via temporal \lq\lq virtual nodes\rq\rq \cite{Appeltant2011InformationSystem, Du2017ReservoirProcessing}. In both architectures, effective mapping is accomplished when reservoir nodes provide linearly independent projections on the input since each node represents a dimension in the transformation process \cite{Appeltant2011InformationSystem, Dambre2012InformationSystems}. In conventional reservoirs, linear independence is attained by selecting random sparse connections between the nodes \cite{Appeltant2011InformationSystem, Du2017ReservoirProcessing}. Conversely, due to their parallel-nodes architecture, physical mem-element-based RCs achieve inter-node linear independence by either 1) using multiple input encoding schemes on homogeneous reservoirs with nominally similar devices that respond similarly to the same input \cite{Du2017ReservoirProcessing, Zhong2021DynamicProcessing}, or 2) using heterogeneous reservoirs with distinct, tunable devices \cite{Nishioka2022Edge-of-chaosReservoir, Ghenzi2024HeterogeneousMemristors, Armendarez2024Brain-InspiredPlasticity} that respond distinctively to the same input. In the former approach, different input encodings of the same input, such as masking \cite{Zhong2021DynamicProcessing} and pulse-width encoding \cite{Du2017ReservoirProcessing}, are employed to provide distinguishable, therefore, linearly independent responses from the same input. While this approach proved viable for many RC applications, relying on external input encoding methods incurs high overhead costs and restricts the reservoir's implementation for real-time tasks. In addition, such methods force the use of a large number of reservoir nodes with large arbitrary device-to-device  variation\cite{Du2017ReservoirProcessing, Zhu2020MemristorAnalysis, Midya2019ReservoirMemristors, Moon2019TemporalSystem}, which can cause over-fitting and, more generally, contradicts the efficiency and compactness requirements driving the realization of physical reservoirs. While heterogeneous reservoirs address these issues via unique memristors with variable responses for the same input \cite{Armendarez2024Brain-InspiredPlasticity, Ghenzi2024HeterogeneousMemristors}, they remain limited in regards to the extent of the linear independent nodes achieved, especially with tasks that require non-monotonic input-state correlations (i.e., a decrease followed by an increase in state in response to an increasing input sweep) \cite{Zhong2021DynamicProcessing}. Such non-monotonic tasks are usually addressed using pre-processing techniques like masking \cite{Appeltant2011InformationSystem, Zhong2021DynamicProcessing}, which forces non-monotony by applying positive and negative signals for the same input. Thus, the enlisted caveats in both approaches call for high-performance heterogeneous physical reservoirs comprising devices with non-monotonic input-state relationships. Such PRCs are expected to omit unnecessary pre-processing expenses, enable time-synchronous applications, and, most importantly, solve problems that require non-monotonic input-state relationships.

Here, we introduce a novel heterogeneous memcapacitor-based reservoir in which positive, negative, and negative-to-positive non-monotone input-state correlations, with varying degrees, all coexist across the reservoir nodes for the same input range. We achieve this via modular asymmetric biomembrane-based memcapacitors,\cite{Najem2019DynamicalMembranes} where an internal voltage offset arises due to a mismatch in dipole potentials of each side of the lipid membrane. By designing the degree of asymmetry between both sides of the membrane, we enable control over the internal voltage offset, which, as we show throughout, directly influences the input-state correlation of the device of interest. Such versatile plasticity, or state change, is commonly observed in biological computing systems, particularly synapses that exhibit both paired-pulse facilitation (PPF) and paired-pulse depression (PPD), with the ability to switch between both modes for efficient signal processing of neural inputs \cite{Rueda-OrozcoBriefNeurons, Debanne1996Paired-pulseRelease}. We fully characterize the devices' rich nonlinear dynamics and short-term memory features to fit the context of PRC. Then, to showcase its optimal high-dimensional mapping, we apply our PRC to predict a benchmark second-order nonlinear dynamical system \cite{Du2017ReservoirProcessing}. We leverage the varying degrees of non-monotonic input-state relations to solve the chaotic Hénon map, which, unlike this work, has only been solved using input masking techniques in the past \cite{Zhong2021DynamicProcessing, Wu2024ACuInP2S6, Pei2023Power-EfficientArrays, Chen2023All-ferroelectricComputing, Fang2024Oxide-BasedComputing, Feng2023FullyStates}. In solving all these nontrivial tasks, we limited our reservoir size to only twelve devices or less, confirming the nodes' high degree of linear independence while consuming ultra-low power and energy due to its memcapacitive nature as per \cite{Hossain2023Biomembrane-BasedProcessing}. Lastly, as this reservoir's uniqueness stems from the ability to carefully tune an internal voltage offset to achieve a desirable input-state relationship, we discuss the equivalence of applying external voltage offsets to devices that lack inherent ones. Since adding an external offset may not pose a significant pre-processing complication on the overall RC, we believe that paralleling the considerable performance improvements observed with internal voltage offsets to those resulting from external voltage offsets may have a substantial benefit-to-cost ratio. 

\section{Results}\label{sec2}

\subsection{Asymmetric biomembrane-based memcapacitors}
\begin{figure}
    \centering
    \includegraphics[width=4.5 in]{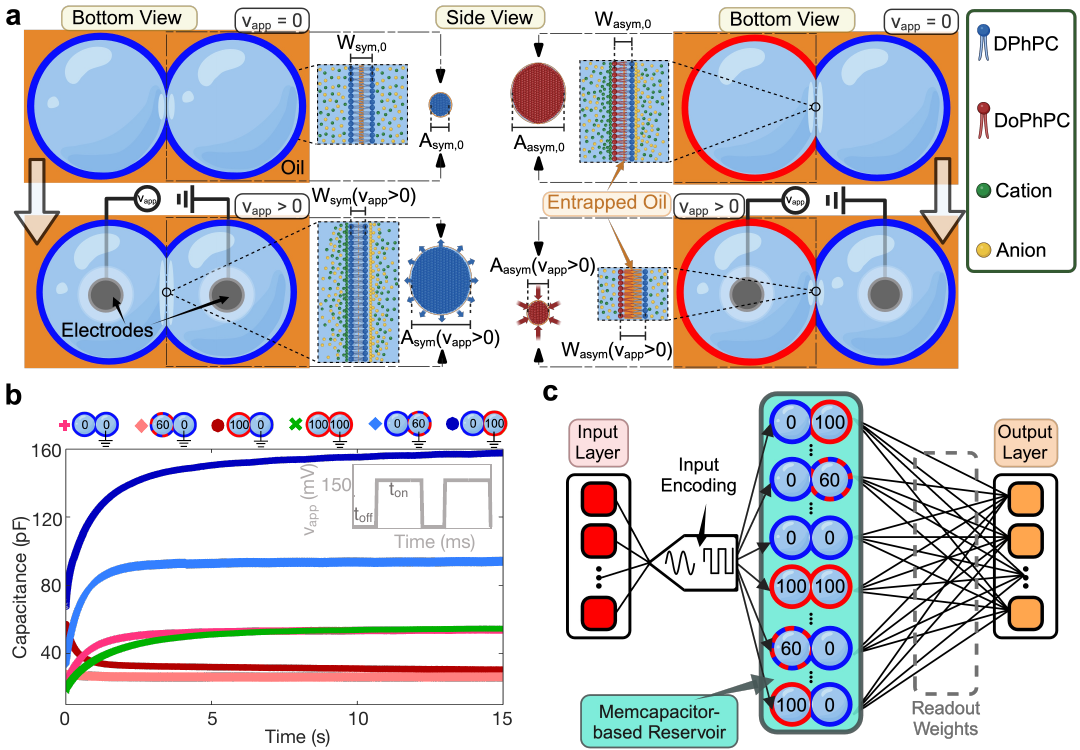}
    \caption{An illustration of symmetric and asymmetric memcapacitors, their response to voltage stimuli, and the resulting heterogeneous memcapacitor-based reservoir. We show in \textbf{Fig. S4} real device photos we took using an inverted microscope. \textbf{a} A symmetric memcapacitor consists of two leaflets of the same composition (e.g., DPhPC-DPhPC). In the case of a symmetric membrane, as shown on the left side, experiencing a voltage stimulus forces a faster increase in the interfacial area (electrowetting) and a slower decrease in hydrophobic thickness (electrocompression) due to oil expulsion \cite{Taylor2015DirectBilayer,Najem2019DynamicalMembranes}. For an asymmetrical bilayer, an intrinsic voltage offset (see \textbf{Supplementary Note 3} and \textbf{Supplementary Fig. S3}) arises from a mismatch between the two phospholipids' head dipoles \cite{Taylor2019ElectrophysiologicalFlip-flop}. At rest, an asymmetric membrane possesses higher capacitance than a symmetric membrane due to the internal offset, represented by the bilayer areas and thicknesses for the cases of $v_{app} = 0$. For a negative internal voltage offset shown on the right side, a positive applied stimulus acts against the internal offset, leading to a decrease in the membrane interfacial area and an increase in the hydrophobic thickness since the overall membrane potential has decreased. \textbf{b} Representative capacitance responses for six different devices, namely 0-0, 60-0, 100-0, 100-100, 0-60, and 0-100, to a pulse train of 5000 pulses (\textbf{Supplementary Fig. S6} shows the response of all twelve devices). Each input pulse is 3 ms long with 2 ms of ON time and 1 ms of OFF time, as shown by the top-right inset. As observed, the starting capacitance is larger for asymmetrical devices compared to symmetrical cases. While the steady-state capacitances for the symmetrical cases are comparable, their transient responses are different as they exhibit dissimilar temporal dynamics, as discussed later. \textbf{c} A visual representation of the heterogeneous memcapacitor-based reservoir. \textbf{Supplementary Fig. S5} displays a full guide on the twelve devices, and their representing visuals}
    \label{Figure1}
\end{figure}

Biomembrane-based memcapacitors, short for memory capacitors, are nonlinear, two-terminal electrical energy storage elements that exhibit short-term memory dependence on present and past stimuli \cite{Najem2019DynamicalMembranes}. They usually consist of insulating $(>100M\Omega \cdot cm^2)$ and capacitive phospholipid membranes formed between two aqueous lipid-monolayer-encased droplets ($\sim 200nL$ each) in oil \cite{Taylor2015DirectBilayer, Najem2019DynamicalMembranes}. 
These devices can be symmetric, where both membrane leaflets consist of the same lipid type, or asymmetric, where each leaflet consists of a different type of lipid, as previously described by Najem et al. \cite{Najem2019DynamicalMembranes}. A comparison between both device types is illustrated in \textbf{Fig. \ref{Figure1}a}. 
Upon stimulation of symmetric devices (left half of \textbf{Fig. \ref{Figure1}a}) with an externally applied potential $v_{app}$, irrespective of the voltage sign, the bilayer area, $A(v_{app}(t))$, increases (i.e., electrowetting \cite{Taylor2015DirectBilayer, Requena1975TheFilms}) and the thickness, $W(v_{app}(t))$, decreases (i.e., electrocompression \cite{Evans1975MechanicsMembranes, Taylor2015DirectBilayer}), resulting in a substantial increase in capacitance (see \textbf{Supplementary Note 2}). In contrast, for asymmetric memcapacitors, an internal voltage offset, $v_\Phi$, arises due to the difference in dipole potential magnitudes between DPhPC and DoPhPC lipids \cite{Najem2019DynamicalMembranes} (see \textbf{Supplementary Note 3} for more details). In both symmetric and asymmetric cases, the capacitance state solely depends on the absolute value of the membrane potential, $v_m$, which is the sum of the external and internal potentials ($v_m(t) =\lvert v_{app}(t) + v_{\Phi} \rvert$). Therefore, for an asymmetric memcapacitor with a negative internal offset ($v_{app}(t=0) = 0$, hence $v_m(t=0) = \lvert v_{\Phi} \rvert = \lvert -138 \rvert$ $mV$, for leaflet composition shown at the right side of \textbf{Fig. \ref{Figure1}a}), the capacitance at rest is more significant than that of a symmetrical memcapacitor with no internal offset ($v_{app}(t=0) = 0$, hence $v_m(t=0) = 0$). Accordingly, applying a positive external potential equal, yet opposite in sign, to the internal offset ($v_{app}(t>0) = 138$ $mV$) minimizes the membrane potential ($v_m(t>0) \rightarrow 0$), thus minimizing the membrane capacitance, $C_m$, to the lowest possible magnitude ($\mathrm{A_{asym}}$ and thicknesses $\mathrm{W_{asym}}$ on the right side of \textbf{Fig.} \ref{Figure1}\textbf{a}, $C_m \propto \mathrm{A_{asym}}$ and $C_m \propto 1/\mathrm{W_{asym}}$). In contrast, applying any negative potential ($v_{app}(t>0) < 0$) will result in a further increase in $C_m$ due to the resultant increase in membrane potential as both the applied and the internal potentials are acting in the same direction (see \textbf{Supplementary Fig. S2-4}).
\par Throughout this study, we used two types of lipids: DPhPC (1,2-diphytanoyl-sn-glycero-3-phosphocholine) and DoPhPC (1,2-di-O-phytanoyl-sn-glycero-phosphocholine). For the facilitated distinction between the two long acronyms, we will refer to DPhPC and DoPhPC as \lq\lq D\rq\rq and \lq\lq Do\rq\rq lipids, respectively. By combining these two lipid types, we have created six types of lipid leaflets: 1) pure 100\%:0\% Do:D, 2) 20\%:80\% Do:D, 3) 40\%:60\% Do:D, 4) 60\%:40\% Do:D, 5) 80\%:20\% Do:D, and 6) 0\%:100\% Do:D. Using these six leaflet types, we fabricated twelve memcapacitors (two symmetric and ten asymmetric) with varying degrees of asymmetry (see \textbf{Supplementary Fig. S5}). Each degree of asymmetry corresponds to a different internal potential, which, under the same input, will induce a different overall membrane potential. In \textbf{Fig. \ref{Figure1}b}, we display the unique capacitance responses of six out of the twelve devices (see \textbf{Supplementary Fig. S6} for all twelve devices' responses) to 5000, +150-$mV$ pulses as depicted by the inset at the left-top corner. It is important to note that in the figure legends throughout the text, the numbers in the cartoon droplets refer to the percentage of Do lipids in the droplet. Additionally, we use the Do percentage composition as the only reference for the device's naming convention throughout the text for shorter notation, as the difference between these two numbers in this naming convention effectively represents the degree of asymmetry in a device. For example, the legend corresponding to the red circular markers pertains to an utterly asymmetric memcapacitor comprised of a pure 100\%:0\% Do:D hot droplet interfaced with a pure 0\%:100\% Do:D grounded droplet. We refer to this configuration as the \lq\lq 100-0\rq\rq memcapacitor. Since each of the responses in \textbf{Fig. \ref{Figure1}b} corresponds to a different intrinsic offset under the same input (left-top corner inset); some memcapacitors exhibit PPF while others exhibit PPD. The main criteria for whether a device will exhibit PPF ($\Delta C_m>0$) or PPD ($\Delta C_m<0$) is only dependent on $v_m$ and is given by the following:
\begin{equation}
\Delta C_m = C_m(t_2) - C_m(t_1)\begin{cases}
>0,& \text{if } v_m(t_2) > v_m(t_1) \text{ for }t_2 > t_1\\
    <0,             & \text{otherwise}
\end{cases}
\label{PPF_PPD_Criteria_equation}  
\end{equation}
In \textbf{Fig. \ref{Figure1}b}, the 100-100 and 0-0 memcapacitors, due to their symmetric composition ($v_{\Phi,100-100} = v_{\Phi,0-0} = 0$), exhibit PPF and yield similar steady-state capacitance values. However, their response paths are unique due to the inherent difference in temporal dynamics between D and Do lipid types. Similarly, the 0-60 and 0-100 devices ($v_{\Phi,0-60} = 82.8$ $mV$ $v_{\Phi,0-100} = 138$ $mV$) also exhibit PPF, but at highly distinct degrees due to the dissimilar additive effects of the different internal potentials. Finally, the 100-0 and 60-0 cases, possessing unique negative internal offsets ($v_{\Phi,60-0} = -82.8$ $mV$ $v_{\Phi,100-0} = -138$ $mV$), demonstrate different degrees of PPD. As we later discuss, this resultant spectrum of responses from the same input is a highly desirable feature for developing physical reservoirs with efficient high-dimensional mapping. We use twelve unique devices (two symmetric and ten asymmetric) to construct a high-performance memcapacitor-based PRC (\textbf{Fig. \ref{Figure1}c}). 

\begin{figure}
    \centering
    \includegraphics[width=4.5 in]{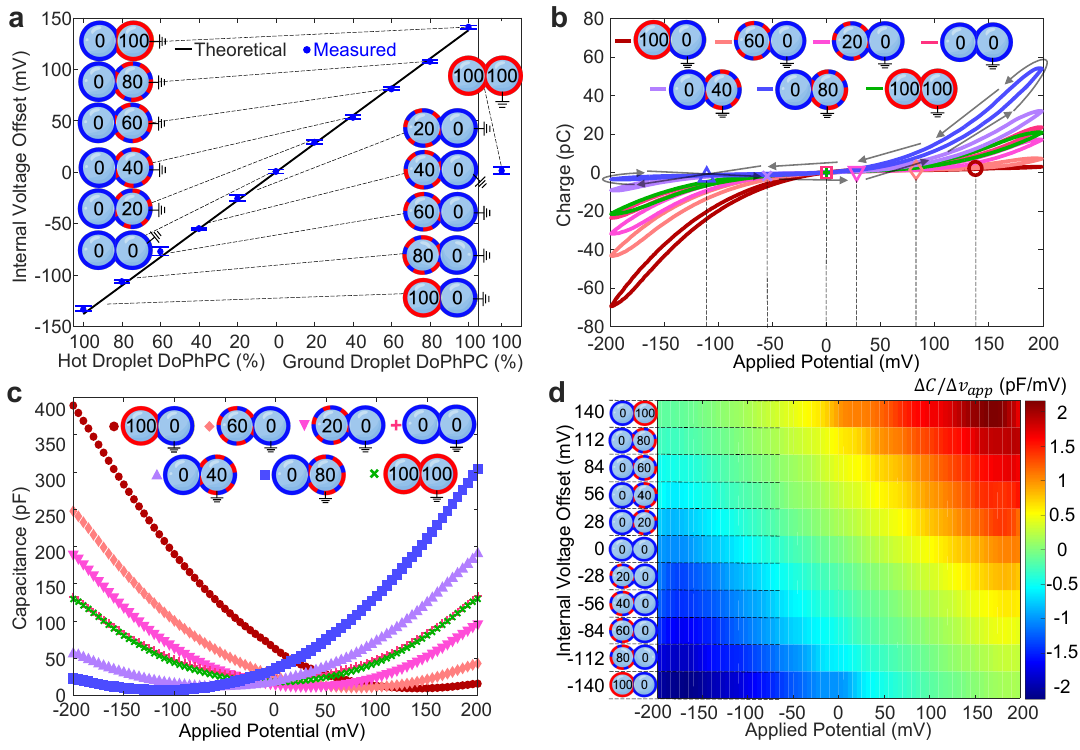}
    \caption{Characterizing the inherent voltage offsets and their influence on input-state correlations. \textbf{a} the internal voltage offset corresponding to different membrane compositions. Mirroring the membrane composition switches the sign of the obtained internal voltage offset. Regardless of the composition, symmetric membrane leaflets have no internal voltage offsets. \textbf{b} Charge versus applied potential curves for seven distinct devices, namely, 100-0, 60-0, 20-0, 0-40, 0-80, and 100-100. The applied potential sweeps are of 50-mHz frequency. \textbf{Supplementary Fig. S7} shows the curves for all twelve devices. The larger markers refer to the pinching point for curves of the same color. \textbf{c} Steady-state capacitance as a function of applied potential for same devices listed in panel \textbf{b}. As observed in this panel, the input-state correlation across different memcapacitors has varying degrees of positive, negative, and negative, followed by positive correlations. \textbf{Supplementary Fig. S8} shows the steady-state capacitance for all twelve devices. \textbf{d} The capacitance vs applied potential slopes ($\Delta C_m/\Delta v_{app}$) are shown to represent the diversity of input-state correlations qualitatively. The red and blue regions correspond to positive and negative input-state correlations.}
    
    \label{Figure2}
\end{figure}

\subsection{Controllable nonlinearity, memory, and input-state correlations}

As discussed in the previous section, the uniqueness of each device's response stems from its distinctive internal offset, which is a function of compositional asymmetry. By measuring the internal offset associated with each of the twelve device types, we observed that the internal offsets exhibit a linear relationship with the degree of membrane asymmetry. In \textbf{Fig.} \ref{Figure2}\textbf{a}, we plot the error bars representing the measured internal voltage offsets of ten versions of each device versus the percentage of Do composition in both the hot and grounded droplets. For clarity, we highlight that throughout this work, unless specified, the left droplet is always connected to the hot electrode while the right droplet is grounded. Interchanging the droplet terminals is equivalent to switching the internal voltage offset's sign (see \textbf{Supplementary Fig. S5}). According to Leon Chua's definition \cite{Romero2021MemcapacitorReview}, memcapacitors must exhibit a pinched hysteresis in the charge-voltage plane \cite{Najem2019DynamicalMembranes,DiVentra2009CircuitMeminductors}, as displayed in \textbf{Fig. \ref{Figure2}b} (see \textbf{Supplementary Fig. S9-12}). While the charge-voltage curve must pinch at 0, for the memcapacitors, the pinching point (outlined by the black markers in \ref{Figure2}b) occurs at $-v_{\Phi}$, which is 0 for the symmetrical cases. Further, the amount of charge stored in the memcapacitor as a function of the applied potential reflects this asymmetry. As seen in \textbf{Fig. \ref{Figure2}b}, membranes with positive $v_{\Phi}$ store more positive charge for the same positive input, while membranes with negative $v_{\Phi}$ hold more negative charge for the same positive input (see \textbf{Supplementary Fig. S7}). 

\par One important design requirement for efficient PRCs is input nonlinearity, which requires a nonlinear relationship between the state (capacitance) and the input (voltage). In this regard, we observed that the capacitance is a quartic (to the 4-th power) function of the applied potential in steady-state, consistent with established literature \cite{Najem2019DynamicalMembranes, Taylor2015DirectBilayer} (see \textbf{Supplementary Methods}). In \textbf{Fig. \ref{Figure2}c}, we display the average steady-state capacitance responses for seven different devices as a function of the applied potential (see \textbf{Supplementary Fig. S8} for all twelve devices). We also observed that the capacitance response shifts across devices, with the minimum occurring approximately at $-v_\Phi$. For the same input range, the capacitance may monotonically increase, monotonically decrease, or non-monotonically decrease, then increase (\textbf{Fig. \ref{Figure2}c}). This phenomenon implies that the input-state relationship can be positively, mostly negatively, or positively and then negatively correlated. \textbf{Fig. \ref{Figure2}d} illustrates the reservoir's ability to realize not only the listed different input-state correlations ($\Delta C/\Delta v_{applied}$) but also the different degrees of each of the listed. As observed in \textbf{Fig. \ref{Figure2}d}, the input-state slopes ($\Delta C/\Delta v_{applied}$) gradually vary from negative to positive across eleven devices with varying D:Do composition when subjected to a voltage sweep varying from -200 mV to 200 mV. These highly distinct input-state correlations, which are unachievable by all previous heterogeneous reservoirs \cite{Armendarez2024Brain-InspiredPlasticity, Nishioka2022Edge-of-chaosReservoir, Ghenzi2024HeterogeneousMemristors}, are crucial for ensuring efficient high-dimensional mapping \cite{Dambre2012InformationSystems, Du2017ReservoirProcessing}. While we only picked twelve devices for this work (see \textbf{Fig. \ref{Figure2}a}), there exists a whole continuum of input-state relationships spanning any device with internal voltage offset values between -138 mV and 138 mV. Given the relatively \cite{Du2017ReservoirProcessing, Midya2019ReservoirMemristors, Moon2019TemporalSystem, Zhu2020MemristorAnalysis} minimal variation ($\leq \pm 10\%$ from \textbf{Fig. \ref{Figure2}a}) associated with the internal voltage offset as a function of composition, we believe that biomembrane-based memcapacitors enable high design fidelity when engineering a heterogeneous RC for an application of interest.
\par Besides input nonlinearity and versatile input-state correlations, a crucial requirement for PRCs is fading memory \cite{HerbertJaeger2002Short-termNetworks}. To characterize the memory capabilities of these devices, we examined PPF and PPD indices \cite{Armendarez2024Brain-InspiredPlasticity} (PPF\% = PPD\% = $(C_{m,A} - C_{m,B})/C_{m,A} \%$ as a function of pulse duration, PD, (i.e., pulse on-time) and inter-pulse interval, IPI, (i.e., pulse off-time) for all twelve devices (see \textbf{Supplementary Fig. S13}). The PPF index measures the device's tendency to retain its state as a function of pulse parameters. A positive PPF refers to increased capacitance, while PPD refers to decreased capacitance due to an applied pulse train. The applied pulse train has a fixed amplitude of 150 mV but varied frequency as represented by its PD and IPI. This metric informs on the frequencies and duty cycles at which the memcapacitors' memory properties are most effective, as observed from the maps and their 1-D projections in \textbf{Fig.\ref{Figure3}a} and \textbf{Fig.\ref{Figure3}b} (see \textbf{Supplementary Fig. S13-14}). The applied pulse is too short for small PD values for the memcapacitors to respond. However, as the PD increases, the PPF or PPD profiles grow in magnitude until their respective magnitudes reach their peak values. Any further increase in the PD elicits a reduced PPF or PPD until the PD is long enough for the memcapacitors to reach a steady state, where any successive stimulation results in no increase or decrease in the measured capacitance. Due to the long-term dynamics stemming from electrocompression \cite{Najem2019DynamicalMembranes} (\textbf{Supplementary Methods}), the PD is required to reach a complete steady state. Consequently, no visible PPF or PPD is beyond the 1000-ms window displayed in \textbf{Fig. \ref{Figure3}a-b}. Conversely, the pulse train is closer for a small IPI, corresponding to more elongated stimulation and larger PPF or PPD. Any further increase in the IPI yields a decay in the PPF or PPD until the memcapacitor dissipates all the memory from the previous pulse, corresponding to 0\% PPF or PPD.  Accordingly, the memcapacitors appear to best exploit their memory features at PDs and IPI values between approximately 100 ms and 500 ms. Therefore, we determined that the frequency of the input signal must be limited to 2-10 Hz with no less than 50\% duty cycle for square wave pulses. In addition to the PPF and PPD indices, we assessed the effect that the internal offsets have on the hysteresis lobe area of the capacitance vs applied voltage curves (see \textbf{Supplementary Fig. S15}), which is another memory-related metric \cite{Chua1971MemristorTheElement}. We note that, when sweeping the input from  -250 mV to +250 mV, the hysteresis lobe area reverses its rotation from counterclockwise, for devices exhibiting PPF, to clockwise, for devices exhibiting PPD \textbf{Supplementary Fig. S15}).

Furthermore, the devices' temporal dynamics, represented by the parameter $\zeta_{ew}$, exhibit cubic sensitivity to a change in the membrane potential $\Delta v_m$. We remark that the data shown is per membrane type; i.e. this data also pertains to mirror-image asymmetrical devices (i.e., 0-100, 0-80, 0-60, 0-20). The parameter $\zeta_{ew}$ is indirectly related to the electrowetting time constant as per the devices' model (see \textbf{Supplementary Methods}). \textbf{Fig.\ref{Figure3}d} shows capacitance decays for various voltage drops within the same 0-0 memcapacitor. This $\zeta_{ew}$ versus $\Delta v_m$ correlation tends to vary across various membrane compositions as observed in \textbf{Fig.\ref{Figure3}c}. Temporal dynamics' voltage sensitivity within a device and its variation across devices for the same input range have proven advantageous for the reservoir's performance \cite{Armendarez2024Brain-InspiredPlasticity}. 

\begin{figure}
    \centering
    \includegraphics[width=4.5 in]{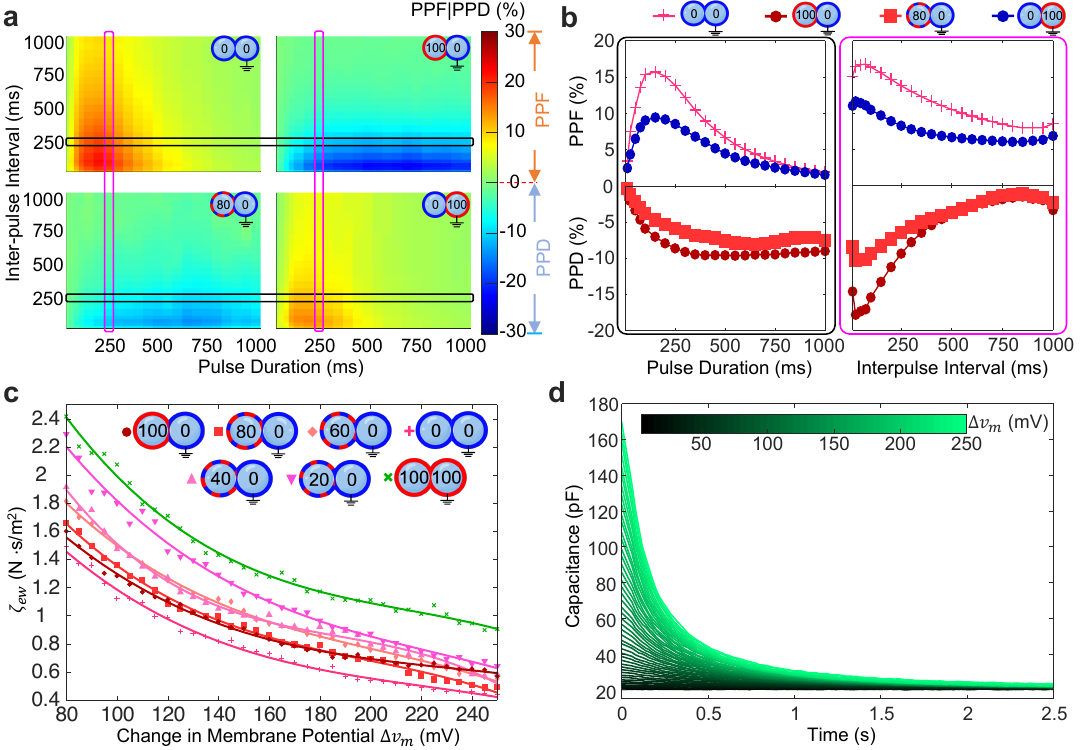}
    \caption{PPF and PPD indices as functions of PD and IPI and memory-related characteristics as functions of change in transmembrane potential across varying memcapacitor compositions. We averaged all displayed experimental data across five samples, representing the memcapacitors used in the RC tasks. \textbf{Supplementary Fig. S16-S18} display the information and statistics on the cycle-to-cycle and device-to-device variability of these memcapacitors. \textbf{a} 2D maps of the PPF and PPD indices as functions of PD and IPI for four devices, namely 100-100, 100-0, 80-0, and 0-100. \textbf{Supplementary Fig. S13} showcases the 2D maps for all twelve devices. \textbf{b} Horizontal and vertical projections of the four maps. In the horizontal case, we fixed the IPI to 250 ms, and the PPF and PPD are plotted against varying PD in the top and bottom panels, respectively. We fixed The PD to 250 ms for the vertical projection, and the PPF and PPD are plotted against varying IPI in the top and bottom panels, respectively. The PPF and PPD projections for the remaining eight devices can be found in \textbf{Supplementary Fig. S14}. \textbf{c} A plot of the temporal parameter $\zeta_{ew}$ as a function of membrane potential change $\Delta_{v_m}$ for seven devices, namely 100-0, 80-0, 60-0, 40-0, 20-0, 0-0, and 100-100. Since $\zeta_{ew}$ only varies with the absolute membrane composition, mirroring the listed asymmetric membranes to yield devices 0-100, 0-80, 0-60, 0-40, and 0-20 generates similar $\zeta_{ew}$ data, in the same listing order. \textbf{d} A qualitative representation of the capacitance drop rate, which is indirectly influenced by $\zeta_{ew}$ (see \textbf{Supplementary Methods}), for different $\Delta_{v_m}$.}
    \label{Figure3}
\end{figure}

\subsection{Second-Order Nonlinear Dynamical System Prediction}

Nonlinear dynamical systems are mathematical models that capture complex dynamics of natural and engineered systems with time-dependent interactions, such as fluid dynamics \cite{Richtmyer1978NonlinearDynamics}, control systems \cite{Isidori1995NonlinearSystems}, population biology \cite{Adachi2022UniversalDynamics}, among others \cite{Gross2006GeneralizedSystems}. Second-order nonlinear dynamical systems are exciting due to their prevalence in electrical and mechanical systems, i.e., RLC circuits and spring-damper mechanics. 
In addition, these systems are inherently temporal for computational purposes and require memory and nonlinearity to predict their response accurately. Therefore, predicting their response is a suitable task for evaluating the performance of an RC system. 
\par In this section, we employ our heterogeneous memcapacitor-based RC to solve a benchmark formulation of a discrete-time second-order nonlinear dynamical system \cite{Du2017ReservoirProcessing, Nishioka2022Edge-of-chaosReservoir, Armendarez2024Brain-InspiredPlasticity}, commonly referred to as SONDS. This evaluation aims to assess the effectiveness of our RC in nonlinearly transforming temporal data. This function gives the formulation of SONDS:

\begin{equation}
y(k)=0.4y(k-1)+0.4y(k-1)y(k-2)+0.6u^3(k)+0.1
\label{SONDS_equation}  
\end{equation}
\newline
\noindent In this formulation, the output signal, $y(k)$, depends on the present input, $u(k)$, as well as the product of the previous two inputs, $y(k-1)$ and $y(k-2)$ (i.e., a time lag of two-time steps), where $k$ is the time step indexing variable. This function is considered a second-order nonlinear system due to the second term on the right-hand side of \textbf{Eq. \ref{SONDS_equation}}. 
The objective of this task is to use the random input $u$ to train the memcapacitor-based RC system to predict the hidden function $y$ with no knowledge of the underlying relationship between the input $u$ and output $y$. This is achieved by the reservoir layer by effectively nonlinearly mapping the one-dimensional input $u$ into a higher dimensional state space. This process significantly enhances the linear separability of the input, making it easier for the linear regression layer \cite{Appeltant2011InformationSystem, Cover1965GeometricalRecognition}.
\par
\begin{figure}
    \centering
    \includegraphics[width=4.5 in]{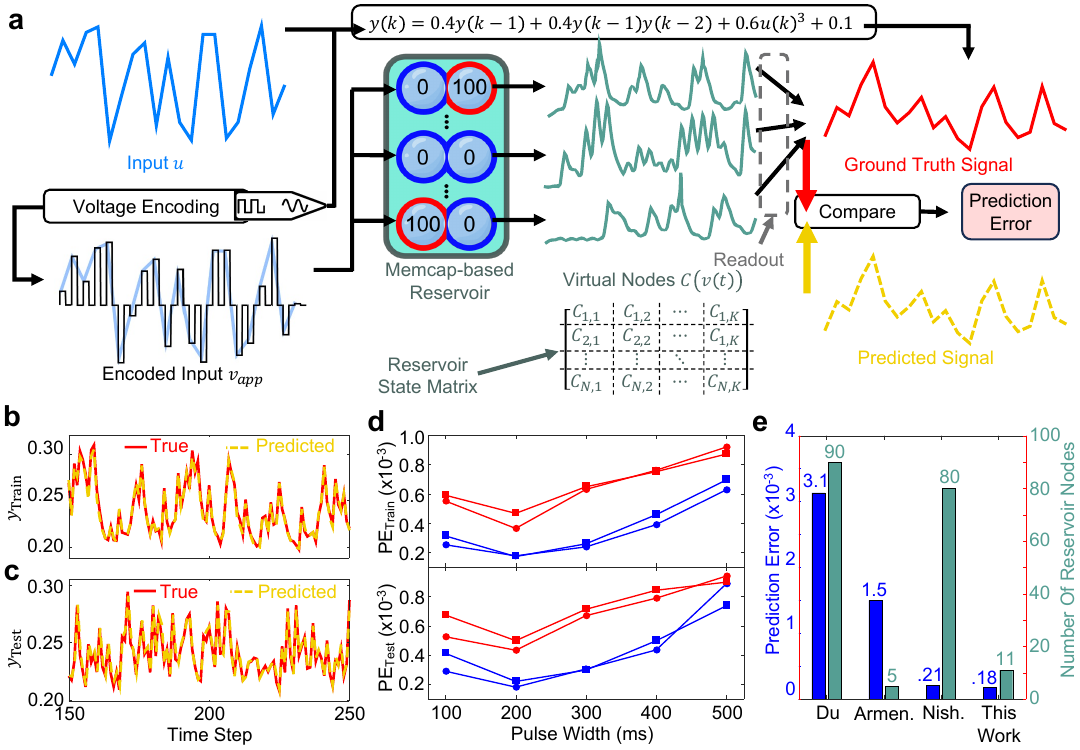}
    \caption{A visual representation of the process flow of solving SONDS and a summary of the prediction results obtained by the heterogeneous memcapacitor-based RC system. \textbf{a} The process flow of solving SONDS. The training random input stream $u$\(\mathrm{_{Train}}\) is linearly interpolated and sampled into pulses with a specific voltage range, pulse width, and duty cycle using a voltage encoder. We channeled the resulting pulses into the reservoir layer comprising twelve distinct memcapacitors, and the capacitance is recorded at the end of each pulse, yielding an N-by-K state matrix (N = 12 and K = 1000 in this case). The state matrix trains a linear readout layer using linear regression. The trained readout layer is then multiplied by the training state matrix to yield a prediction for the $y$\(\mathrm{_{Train}}\). The resulting $y$\(\mathrm{_{Train}}\) is then compared to the true $y$\(\mathrm{_{Train}}\) using a defined prediction error metric (see \textbf{Supplementary Methods} section). \textbf{b} The predicted training data set $y$\(\mathrm{_{Train}}\) over the true values of $y$\(\mathrm{_{Train}}\) for time steps 150-250. These results correspond to a $1.73\times10^{-4}$ prediction error. \textbf{c} The predicted testing data set $y$\(\mathrm{_{Test}}\) over the true values of $y$\(\mathrm{_{Test}}\) for time steps 150-250. These results correspond to a $1.80\times10^{-4}$ prediction error.
    \textbf{d} The training (top) and (bottom) testing prediction errors for the five different pulse widths for voltage ranges of 1) -150 mV to +150 mV (circles) and 2) -200 mV to +200 mV (squares) for a heterogeneous reservoir with internal voltage offsets (blue) and a homogeneous reservoir with externally applied offsets (red). \textbf{e} A visual comparison of testing prediction errors achieved previously by \cite{Du2017ReservoirProcessing, Nishioka2022Edge-of-chaosReservoir, Armendarez2024Brain-InspiredPlasticity} and this work as well as the number of reservoir nodes needed to achieve these testing prediction errors.  
    }
    \label{Figure4}
\end{figure}

 In \textbf{Fig. \ref{Figure4}a}, the process flow of SONDS is displayed. A uniformly random sequence $u$, comprising 2000 time steps (the first 1000 for training and the remaining 1000 for testing) in which each time step ranges from 0–0.5, is generated and channeled into a voltage encoder. The voltage encoder linearly interpolates each point in $u$ to a voltage pulse with an amplitude range of -150–150 mV for one set of experiments and -200–200 mV for another. For each of the two sets, we formulated the five subsets where we held the pulses for five different pulse widths: from 100–500 ms in increments of 100 ms as guided by the PPF earlier analysis (\textbf{Fig. \ref{Figure3}a-b}), hence resulting in a total of ten subsets (five pulse widths for two different encoding ranges). In all subsets, the pulses exhibit a duty cycle of 50\%. For clarity, throughout this manuscript, we use the term \lq\lq pulse width\rq\rq to refer to the entirety of pulse time, encompassing both the pulse's active (on-time) and inactive (off-time) times. The resulting pulses for each listed ten subsets are fed independently to the eleven unique memcapacitors (all devices mentioned except for the 100-100 device), one subset at a time. Due to the voltage stimulation, the memcapacitors' states, i.e., capacitance, either amplify or diminish in magnitude as governed by the overall magnitude and sign of membrane potential. The capacitance for each input time step is measured at the end of the on-time of each pulse, yielding a total of two 11,000 states spanning an 11-by-1000 state matrix for 1000 training inputs. This matrix and the true training output are then calculated from \textbf{Eq.} \ref{SONDS_equation} to digitally train a 12-by-1 linear regression (one for each device in addition to a bias weight) layer to minimize the prediction error (PE, defined in the \textbf{Supplementary Methods} section) and consequently obtain a prediction of the training output $y$\(\mathrm{_{Train}}\). A 100-point sample of true training output and prediction are plotted against each other in \textbf{Fig. \ref{Figure4}b} for the lowest achieved training PE of $1.73\times10^{-4}$ pertinent to the subset with a pulse width of 200 ms and encoding amplitudes of $\pm$200 mV. At this stage, the training of the readout layer is deemed complete and ready to receive the test data set. Akin to the training process, the test data set is initially encoded and then inputted into the reservoir layer, and the resulting capacitance state matrix, also sized at 1000-by-12 for 1000 test inputs, is directly multiplied by the trained readout layer to obtain a prediction for $y$\(\mathrm{_{Test}}\). The predicted output is then quantitatively compared to the true test output obtained from \textbf{Eq. \ref{SONDS_equation}} using the PE. Similarly, a 100-point sample of the true and predicted testing outputs is shown in \textbf{Fig. \ref{Figure4}c} for the lowest achieved testing PE of $1.80\times10^{-4}$ pertinent to the subset with a pulse width of 200 ms and encoding amplitudes of $\pm$150 mV. We attribute such low PEs to the devices' ability to realize different types of input-state correlations (\textbf{Fig. \ref{Figure2}d}) stemming from the distinct internal voltage offsets. We also expect that temporal dynamics variations across the devices (\textbf{Fig. \ref{Figure3}c-d}) to enhance the reservoir's quality slightly \cite{Armendarez2024Brain-InspiredPlasticity}, yet not as predominantly as the internal voltage offsets' effect. To test the impact of the different input-state correlations, we re-solve SONDS with the same ten subsets above, using a homogeneous reservoir consisting of eleven 0-0 symmetrical devices yet with eleven externally applied offsets (from -138 mV to 138 mV in increments of +27.6 mV). With such an approach, we can ensure that temporal dynamics variations across the devices are minimized and that the overall performance is attributed to the offsets. With the homogeneous reservoir, we achieved testing PEs as low as $3.65\times10^{-4}$ pertinent to the subset with a pulse width of 200 ms and encoding amplitudes of $\pm$150 mV. A summary of training and testing accuracies for all ten subsets using a heterogeneous and an externally biased homogeneous reservoir is presented in \textbf{Fig. \ref{Figure4}d}. Finally, to prove that the effects of variable temporal dynamics are minimal, we solve SONDS using the twelve different devices but with externally applied offsets that counteract their internal voltage offsets. With such a setup, any internal voltage offset effect is eliminated, and the reservoir quality is attributed to the versatile temporal dynamics of the twelve devices. As expected, the PEs are approximately two orders larger than those obtained using internally and externally applied offsets (\textbf{Supplementary Fig. S19}).
\par It is important to emphasize that this approach of employing a heterogeneous RC consisting of eleven unique devices for effective mapping purely relies on the highly variable responses generated by the reservoir layer, thus eliminating any overhead and energy costs associated with extensive input encoding pre-processing techniques \cite{Du2017ReservoirProcessing}. Such a method is quite dissimilar to the majority of antecedent studies which rely on extensive input encoding methods, such as feeding in the same input with different pulse widths to achieve distinct responses from nominally homogeneous RCs. In addition, these RCs depend on either large arbitrary device-to-device variations \cite{Du2017ReservoirProcessing} or cycle-to-cycle variations \cite{Guo2023GenerativeVariability} to increase the reservoir's dimensionality rather than leveraging a design feature in the reservoir layer. Further, we are aware of previous works involving heterogeneous memristor-based reservoir layers with controllable tuning \cite{Nishioka2022Edge-of-chaosReservoir, Armendarez2024Brain-InspiredPlasticity}. However, these works can tune their reservoir devices to generate responses that are distinct yet limited to the same input-state correlation, i.e., for an increasing input, the distinct states are either all increasing or all decreasing depending on whether the device's responses are positively or negatively correlated to the input stimulus, respectively. On the contrary, our RC system can realize input-state correlations that only positive (monotonically increasing), only negative (monotonically increasing), or both positive and negative (non-monotone) for the same input (\textbf{Fig. \ref{Figure2}d}). This wide spectrum of input-state relationships significantly reduces the chance of linear dependencies across the distinct reservoir states, yielding such low PEs with a low number of reservoir nodes as observed in both \textbf{Fig. \ref{Figure4}b} and  \textbf{Fig. \ref{Figure4}c}, insensitive to small variations in input encoding parameters (\textbf{Fig. \ref{Figure4}d}). We compare our testing prediction error with previously reported results and the number of reservoir nodes needed to achieve these testing prediction errors visually in \textbf{Fig. \ref{Figure4}e}. As shown, the testing PE achieved in this work is among the lowest in the literature of PRCs using only eleven nodes, underscoring the power of versatile input-state correlations.

\subsection{Chaotic Time Series Prediction}
In addition to SONDS, we opted to further assess the competence of our RC's computational capabilities by solving a chaotic time series. Chaotic systems are also commonly observed in a multitude of fields such as biological systems \cite{Kaplan1996SubthresholdAxons}, celestial mechanics \cite{Linde1986EternallyUniverse}, chemical reactions \cite{Hudson1981ChaosReaction}, and more \cite{Toker2020ANature}. Computationally, chaotic systems are known to be sensitive to minor errors as these errors innately accumulate to cause a significant divergence between the prediction and the actual output, rendering them inherently complex systems to predict. This section uses our memcapacitor-based RC to predict the Hénon Map \cite{Wen2014AInterpretations}, a benchmark \cite{Zhong2021DynamicProcessing} chaotic 2D map. Hénon Map is a nonlinear discrete 2D map that takes a point $(x(n),z(n))$ on the plane and transforms it into a new forward point $(x(n+1),z(n+1))$, where $n$ is the indexing variable. The following equations mathematically describe the map: 

\begin{equation}
x(n+1) = z(n) - 1.4x(n)^2
\label{eq:HénonMapEquation1}  
\end{equation}

\begin{equation}
z(n+1) = 0.3x(n) + w(n)
\label{eq:HénonMapEquation2}  
\end{equation}
\newline
\noindent where represents $w(n)$ is a Gaussian noise term with a mean and standard deviation of 0 and 0.05, respectively. The main objective is to forecast the system's position at the next time step $(n+1)$, using the input of the current time step at $(n)$. The system can be represented by an equation involving $x$ by merging \textbf{Eq.} \ref{eq:HénonMapEquation1} and \ref{eq:HénonMapEquation2} into one equation in terms of $x$ only. As a result of this simplification, the input for this task is $x(n)$, and the desired output is $x(n + 1)$. 
\par
\begin{figure}
    \centering
    \includegraphics[width=4.5 in]{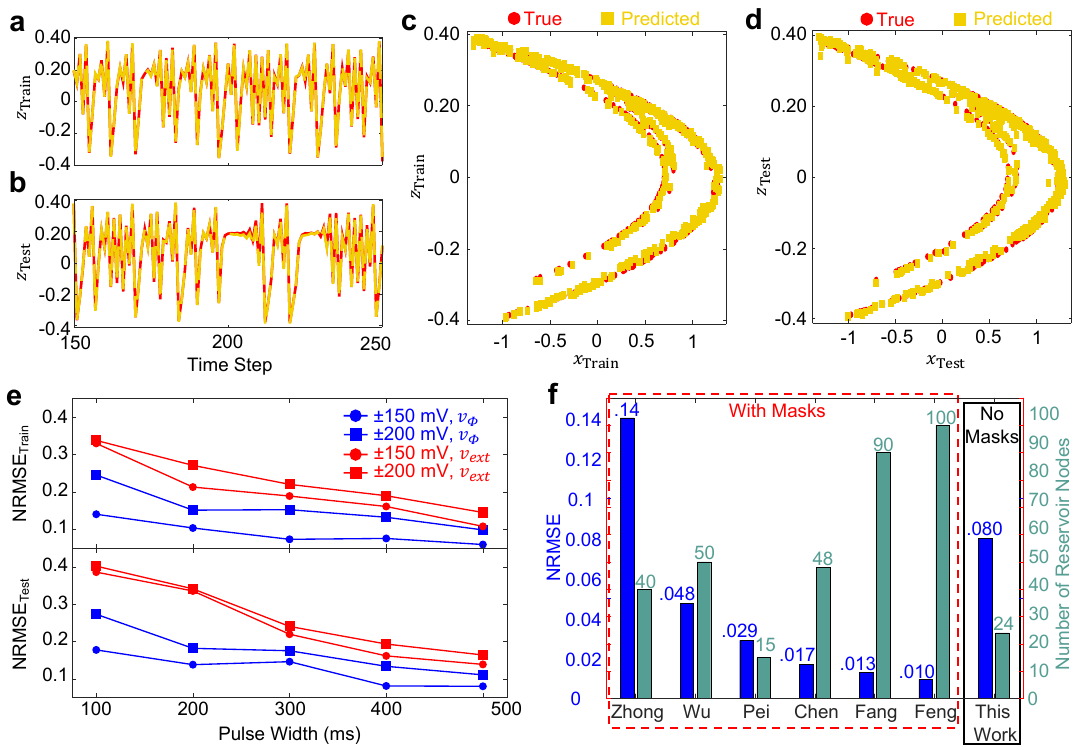}
    \caption{A summary of the results obtained by the heterogeneous memcapacitor-based RC system for the Hénon Map. \textbf{a} The predicted training data set $z$\(\mathrm{_{Train}}\) over the true values of $z$\(\mathrm{_{Train}}\) for time steps 150-250. These results correspond to an NRMSE of 0.059. \textbf{b} The predicted testing data set $z$\(\mathrm{_{Test}}\) over the true values of $z$\(\mathrm{_{Test}}\) for time steps 150-250. These results correspond to an NRMSE of 0.080. \textbf{c} The predicted training data 2-D map set $z$\(\mathrm{_{Train}}\) against $x$\(\mathrm{_{Train}}\) over the true 2-D map for all 1000 time steps. \textbf{d} The predicted testing data 2-D map set $z$\(\mathrm{_{Test}}\) against $x$\(\mathrm{_{Test}}\) over the true 2-D map for all 1000 time steps. \textbf{e} The training (top) and (bottom) testing NRMSEs for the five different pulse widths for voltage ranges of 1) -150 mV to +150 mV (circles) and 2) -200 mV to +200 mV (squares) for a heterogeneous reservoir with internal voltage offsets (blue) and a homogeneous reservoir with externally applied offsets (red). \textbf{f} A visual comparison of NRMSEs achieved previously by \cite{Zhong2021DynamicProcessing, Wu2024ACuInP2S6, Pei2023Power-EfficientArrays, Chen2023All-ferroelectricComputing, Fang2024Oxide-BasedComputing, Feng2023FullyStates} and this work as well as the number of reservoir nodes needed to achieve these NRMSEs.  
    }
    \label{Figure5}
\end{figure}

 The process flow of solving the Hénon map is very similar to that of SONDS shown in \textbf{Fig. \ref{Figure4}a}. First, 2000 time steps  (the initial 1000 are used for training while the other 1000 for testing) of the input time series $x$ are generated from the combined equation, where the first two entries, $x(1)$ and $x(2)$, are equal to 0. Analogous to SONDS, we encode the data in two voltage ranges: -150 mV to +150 mV and -200 mV to +200 mV. For each of the two sets, five subsets are formulated in which the pulses are held for five different pulse widths: from 100–500 ms in increments of 100 ms, yielding a total of ten subsets. The pulses alternate a duty cycle of 50\% in all subsets. Unlike in SONDS, we sample two for effective prediction instead of electing only one virtual node per input (one at the start and one at the end of the pulse ON time). The training pulses are then loaded to all twelve memcapacitors to extract a 25-by-1000 (two virtual nodes per device in addition to a bias node) state matrix and train the readout layer through the minimization of a normalized root mean squared error (NRMSE, defined in the \textbf{Supplementary Methods} section). The trained readout produces a prediction for, denoted $x$\(\mathrm{_{Train}}\), which we then used to calculate the vertical axis prediction $z$\(\mathrm{_{Train}}\). In \textbf{Fig. \ref{Figure5}a}, 100 samples of $z$\(\mathrm{_{Train}}\) are depicted, and the full 2-D map, including all 1000 training inputs, is shown in \textbf{Fig. \ref{Figure5}c}, where the horizontal axis represents $x$\(\mathrm{_{Train}}\). Finally, the trained readout is used to predict the testing outputs $x$\(\mathrm{_{Test}}\) and, subsequently, $z$\(\mathrm{_{Test}}\) using the test state matrix formed from the RC responses to the test input set. In \textbf{Fig. \ref{Figure5}b}, 100 samples of $z$\(\mathrm{_{Test}}\) are displayed as well as a 2-D map comprising all 1000 testing inputs is shown in \textbf{Fig. \ref{Figure5}d}, where the horizontal axis represents $x$\(\mathrm{_{Test}}\). The lowest achieved training and testing NRMSEs from all ten subsets are 0.059 and 0.080, respectively, using a pulse width of 500 ms. We attribute such low NRMSEs to the reservoir's ability to implement distinct input-state correlations. To test the impact of distinct input-state correlations, we re-predict the Hénon map using a homogeneous reservoir with externally applied offsets, similar to SONDs. A summary of the training and testing NRMSEs corresponding to all ten runs is demonstrated in \textbf{Fig. \ref{Figure5}e}. We observe that the reduction in error is mainly dominated by the input-state correlations stemming from the offsets (\textbf{Supplementary Fig. S20}). 
\par We note that, to the best of our knowledge, this is the first PRC that accurately predicts the Hénon map without employing any masking to the input signal, unlike prior PRC studies \cite{Zhong2021DynamicProcessing, Wu2024ACuInP2S6, Pei2023Power-EfficientArrays, Chen2023All-ferroelectricComputing, Fang2024Oxide-BasedComputing, Feng2023FullyStates}, which predominantly employ masking techniques to achieve competitive NRMSEs when predicting the Hénon map. Masking is an input encoding technique in which each point in the input sequence is multiplied by a random \cite{Zhong2021DynamicProcessing}, or binary \cite{Appeltant2011InformationSystem} matrix to increase the number of virtual nodes and, thus, the number of dimensions per input. Notably, implementing binary masks forces positive and negative reservoir states per the same input, which facilitates predicting non-monotone temporal sequences such as in the Hénon map (\textbf{Fig. \ref{Figure5}b}). While successful in achieving efficient high-dimensional projection of the input, masking and similar pre-processing approaches impose significant pre- and post-processing costs in hardware implementations \cite{Zhong2022AProcessing} and further complicate the use of the reservoir for time-synchronous applications.

\section{Discussion}\label{sec3}
This study presented a novel heterogeneous memcapacitor-based reservoir that enables monotonic and non-monotonic input-state correlations from the same input range via intrinsic voltage offsets. In addition to the pre-processing cost riddance, we have shown that the emergence of both monotonic and non-monotonic input-state correlations from the same input range leads to effective high-dimensional mapping. This is evident in the unprecedented prediction error achieved in SONDS using only eleven nodes. Additionally, we have shown that the system's versatility in input-state relationships can achieve an effect equivalent to that of a masking technique \cite{Appeltant2011InformationSystem} to diversify the reservoir's temporal nodes. This was proved in predicting the Hénon map, achieving NRMSEs comparable to those achieved in previous works that employ masking \cite{Zhong2021DynamicProcessing, Wu2024ACuInP2S6, Pei2023Power-EfficientArrays, Chen2023All-ferroelectricComputing, Fang2024Oxide-BasedComputing, Feng2023FullyStates}.

\par Since these results majorly stem from the devices' inherent voltage offsets, we assessed the effect of applying external offsets to the input signal as shown in the SONDS and Hénon Map prediction tasks(\textbf{Fig. \ref{Figure4}d} and \textbf{Fig. \ref{Figure5}e}). The results conclude that the marked performance is due to the emergence of diverse input-state correlations in both the homogeneous and heterogeneous reservoir cases, with the heterogeneous RC performing slightly better due (\textbf{Fig. \ref{Figure4}d} and \textbf{Fig. \ref{Figure5}e}) to the diverse temporal dynamics stemming from different lipid types (\textbf{Fig. \ref{Figure3}c}). Given the success of the external-offsets approach on a symmetric memcapacitor and its ease of implementation, we sought to expand the applicability of this approach to other dynamic devices used in PRC. We examined the models of two memristors that were previously used in PRC, namely those by Du \textit{et al.} \cite{Du2017ReservoirProcessing} and Armendarez \textit{et al.} \cite{Armendarez2024Brain-InspiredPlasticity}, to solve both SONDS and Hénon Map using external offsets. We chose these two models to show that our approach is applicable even for devices with varying temporal dynamics since the former memristor model assumes an invariable time-constant \cite{Du2017ReservoirProcessing}, while the latter possesses an input-sensitive time constant \cite{Armendarez2024Brain-InspiredPlasticity}.  The results (see \textbf{Supplementary Discussion}) show that, regardless of the choice of input encoding parameters, both devices achieve prediction errors and NRMSEs that are either comparable or better than those achieved using pulse-width encoding as implemented by \cite{Du2017ReservoirProcessing}. Predicting the Hénon map with external offsets produces NRMSEs that are starkly lower than those obtained from pulse-width encoding. It is essential to note that testing these two memristors is inadequate to claim that adding voltage offsets to inputs will yield improved reservoir performance for every device in the literature. Nonetheless, we claim that these results confirm that this approach is not limited to the memcapacitor presented in this work and, given its trivial implementation requirements and low overhead cost, can replace other commonly used input-encoding methods like pulse-width encoding and masking, which are practically more expensive to implement. And through the analysis in the \textbf{Supplementary Discussion}, we suggest, with no strict rules, that this approach is most effective with devices that 1) exhibit non-monotonic input-state relationships (similar to that of symmetric memcapacitors C-V relationships) and 2) possess non-switching, continuous input-state relationships (similar to that of symmetric memcapacitors C-V relationships) in addition to the nonlinearity and fading memory requirements imposed for a successful PRC implementation.

Our approach, which relies on native reservoir features rather than pre-processing methods, not only reduces overhead costs but also offers the advantage of enabling real-time deployment of dynamic reservoirs. It is important to note that in such times of rising demands for efficient computing systems, reduction in pre-processing circuitry and reservoir size is imperative, highlighting the significance of this work. In addition to the resultant overhead and reservoir size reduction, we underscore that this device's memcapacitive nature, which not only the energy consumption is independent of the pulse duration \cite{Hossain2023Biomembrane-BasedProcessing}, but also the overall power consumption is in the $\sim10^2$ fW range \cite{Hossain2023Biomembrane-BasedProcessing}, the lowest among devices used for RC \cite{Hossain2023Biomembrane-BasedProcessing}. The ultra-low power and energy expenditure arises from the low operating voltage ranges ($\sim \pm250$ mV) and currents ($\sim10^2$ nA). Besides the biologically relevant time scales ($\sim 10^2$ ms), these input and output ranges are well-suited for integrating this device into biological applications. In such applications, the structural rigidity, electronic nature, high operating voltage ranges, very fast switching dynamics, and bio-incompatible material substrates of solid-state devices are a mismatch for interfacing with biological environments that operate ionically, within low voltage ranges, and much slower switching dynamics.
On the other hand, soft, ionic, and biocompatible devices, such as our memcapacitor, are a great match for interfacing with biological environments. We emphasize that energy storage elements, such as memcapacitors, are much more energy-efficient and inherently well-suited for information processing applications than energy-dissipative elements, such as memristors. This is because we can only decide the information from an input signal at points that vary with time, as static points reveal no information. Energy storage elements, such as a memcapacitor, consume energy only at points at which the signal varies (more precisely, only at points at which the element is storing energy \cite{Hossain2023Biomembrane-BasedProcessing}), implying extreme energy-efficiency as the system only consumes energy when performing computation. In contrast, memristors and other energy-dissipative elements consume energy at any point throughout the input signal, static or dynamic. Similar reduction in static power consumption is a primary reason complementary metal oxide semiconductor (CMOS) logic overtook N-type metal–oxide–semiconductor (NMOS) logic as the dominant metal–oxide–semiconductor field-effect transistors (MOSFET) fabrication process for very large-scale integration (VLSI) chips (99\% of integrated circuit (IC) chips were fabricated using CMOS technology as of 2011 \cite{Voinigescu2013High-FrequencyCircuits}).
Furthermore, biomembrane-based memcapacitors are not as scalable as many solid-state memristive devices. Nevertheless, given the exciting prospects of this emerging interdisciplinary field of bio-inspired nanofluidics \cite{Hou2023LearningNanofluidics}, there have been numerous efforts in scaling up bio-membrane-based devices. One approach with high potential is microfluidic droplet-based network assemblies \cite{Schimel2021Pressure-drivenNetworks, Nguyen2016MicrofluidicBilayers, Nguyen2016HydrodynamicArrays, Nguyen2017AMembranes, Alcinesio2022FunctionalNetworks, Sarles2010PhysicalNetworks, Challita2018EncapsulatingOrganogel}. In these works, pressure-driven microfluidic chips \cite{Schimel2021Pressure-drivenNetworks} incorporate multiple droplet capture registers that are designed to catch two droplets and bring them in contact to form a lipid bilayer while leaving a channel for more droplets to pass through to be caught by other succeeding registers. In addition to microfluidic-based approaches, a droplet-based 3-D bioprinter \cite{Villar2013AMaterial} was invented and deployed to build 3-D multivesicular structures \cite{Alcinesio2022FunctionalNetworks}, synthetic tissues \cite{Alcinesio2020ControlledTissues}, and many more \cite{Graham2017High-ResolutionPrinting}. Moreover, we note several efforts that target the stability of biomembrane-based devices and structures \cite{Sarles2010PhysicalNetworks, Challita2018EncapsulatingOrganogel}. Furthermore, we highlight that memcapacitors can also be constructed using Electrowetting-on-dielectrics (EWODs) \cite{Mugele2005Electrowetting:Applications}, unlike the droplet-interface bilayer-based method used to build the memcapacitors in this work. Memcapacitive EWODs can potentially be an alternative route that addresses the scalability limitations of droplet-interface bilayer-based methods \cite{Li2020CurrentMicrofluidics}. 

Lastly, this pioneering study can serve as a guide for heterogeneous PRC systems framework and significantly adds to the PRC literature body aiming to increase PRC systems' efficacy. Regardless of its practical implementation status, our work will pave the way for energy-efficient solutions to temporal and classification problems. It can also be considered a vital asset in an era of rising demands for efficient systems. 

\section{Methods}\label{sec4}

\subsection{Lipid Solutions Preparation and Membrane assembly}\label{method_1}

To obtain the twelve different memcapacitors comprising our RC system, six different stock solutions were prepared using a mixture of Diphytanoyl-sn-glycero-3-phosphocholine (DPhPC, Avanti) and 1,2-di-O-phytanoyl-sn-glycero-phosphocholine (DoPhPC, Avanti) lipids, namely, 1) 0:1 DoPhPC:DPhPC, 2) 1:4 DoPhPC:DPhPC, 3) 2:3 DoPhPC:DPhPC, 4) 3:2 DoPhPC:DPhPC, 5) 4:1 DoPhPC:DPhPC, and 6) 1:0 DoPhPC:DPhPC. To obtain these lipid solutions in the listed respective order, the following six vials of lipids in chloroform solution were prepared: 1) 160 $\mu L$ of 25-$mg/mL$ DPhPC, 2) a mixture of 80 $\mu L$ of 10-$mg/mL$ DoPhPC and 128 $\mu L$ of 25-$mg/mL$ DPhPC, 3) a mixture of 160 $\mu L$ of 10-$mg/mL$ DoPhPC and 96 $\mu L$ of 25-$mg/mL$ DPhPC, 4) a mixture of 240 $\mu L$ of 10-$mg/mL$ DoPhPC and 64 $\mu L$ of 25-$mg/mL$ DPhPC, 5) a mixture of 320 $\mu L$ of 10-$mg/mL$ DoPhPC and 32 $\mu L$ of 25-$mg/mL$ DPhPC, and 6) 400 $\mu L$ of 10-$mg/mL$ DoPhPC were all evaporated under clean dry air, resulting in 4 $mg$ of the respective mixture of residual lipid cake at the bottom of six 4-$mL$ vials. The vials were then placed under vacuum for at least two hours and hydrated with 2 $mL$ of stock aqueous solution, producing six vials with 2 mg/mL multilamellar lipid liposomes. 
\par We prepared the stock aqueous solution by dissolving 500 mM potassium chloride (KCl, Sigma), 10-$mM$ 3-morpholino propane-1-sulfonic acid (MOPS, Sigma) with a measured pH of 5.8, in deionized water (18.2 $M\Omega$$\cdot$cm) and then stored in -40 $^oC$. To convert the multilamellar liposomes in the vials to unilamellar liposomes, all the lipid solutions underwent six freeze/thaw cycles. They were individually extruded through a 100-$nm$-pore polycarbonate membrane (Whatman) in eleven successive passes using an Avanti Mini Extruder. Finally, the extruded solutions were sonicated for five minutes and vortex-mixed for 60 seconds \cite{Najem2019AssemblyMembranes}. The resulting solutions were either stored for weeks at 4$^oC$ or directly used for experimentation. 
\par For electrical interrogation, a micropipette was used to pipette two 200-$nL$ droplets from the prepared lipid stock solutions onto two 125 $\mu m$-diameter, ball-end silver/silver chloride (Ag/AgCl 99.99\%, Goodfellow) wires submerged in a decane ($\geq$99\%, Sigma-Aldrich) oil-filled, transparent acrylic reservoir. Before droplet deposition, the wires were coated with 1\% agarose gel to avoid droplets detaching from the wires due to the decreasing surface tension associated with the lipid monolayer formation. 
\par Initially, the droplets were suspended on the wires free of contact with each other, the acrylic substrate, and the oil/air interface for 5-7 minutes, allowing for a packed lipid monolayer to form at the droplets’ water/oil interface. The monolayer formation was monitored visually via a 4x objective lens on an Olympus IX73 inverted microscope. Once the monolayers were formed, which was detected by observing the droplets leave the lens’ focal plane as they vertically sagged from the wires, the droplets were brought in contact with each other by moving the wires using 3-axis micro-manipulators to form a bilayer at the contact interface spontaneously. 

\subsection{Electrical Measurements Setup}\label{method_2} 
Before applying any problem-specific transmembrane voltage signal, we verified the successful formation of the interfacial bilayer by using a triangular voltage signal with a frequency of 100 $Hz$ and an amplitude of 10 $mV$ to the electrodes through a Tektronix AFG31022 function generator. Due to the membrane’s large electrical insulation ($>$100 $M\Omega$$\cdot$cm) and capacitive interfacial area, a small 100-$Hz$ square ($\sim 50-300$ $pA$, depending on the lipid leaflets asymmetry and resulting internal voltage offset) current response is expected as an output from a non-leaking bilayer. To capture the dynamic and steady-state changes in capacitance as a function of voltage (Fig. \ref{Figure2}), a voltage waveform consisting of a 2-$mHz$, 250-$mV$ amplitude sinusoidal waveform superimposed on a 50-$Hz$, 10-$mV$ triangular waveform was supplied to the twelve types membranes of interest. The slow-frequency component of the waveform drove the geometric reconfiguration of the membrane (i.e., electrowetting and electrocompression), while the fast-frequency component was used to obtain the capacitance magnitude at every semi-period. For the capacitance step response in \textbf{Supplementary Fig. S2}, the low-frequency component was a 10-second, 250-$mV$ square wave. A custom MATLAB script (available upon request) was used to compute the capacitance at every semi-period by fitting the analytical solution of a parallel RC circuit current response to the measured current. Simultaneously, the bilayer area changes (\textbf{Supplementary Fig. S2}) were monitored at 60 fps using a camera attached to the inverted microscope. The corresponding videos, obtained via Olympus CellSens software, were then post-processed using a custom MATLAB script (available upon request) to extract the bilayer’s interfacial minor axis radius, which was used to compute the interfacial area changes (\textbf{Supplementary Fig. S2}) and subsequently, the bilayer hydrophobic thickness changes (\textbf{Supplementary Fig. S2}).
A custom MATLAB script (available upon request) was used to control an NI 9264 voltage output module to send an arbitrary transmembrane voltage signal. The capacitance magnitude at every pulse was computed from the current responses obtained using a custom MATLAB script (available upon request). 
 \par{All the current measurements were recorded and digitized at 50,000 samples/second (to avoid capacitive-spike aliasing) using a patch-clamp amplifier Axopatch 200B and Digidata 1440A data acquisition system (Molecular Devices), respectively. All current recordings are conducted on an active vibration isolation table and under appropriate shielding, using a lab-made Faraday cage to reduce the noise to less than 2 $pA$.}
\par{All model simulations and energy consumption calculations were implemented using a custom MATLAB script (available upon request).}

\newpage





\maketitle
\newpage
\section{Supplementary Information}
\section*{Supplementary Notes}\label{SectionS1}
\subsection*{Supplementary Note 1: Conventional versus Physical Reservoir Computing Architectures}\label{SubsectionS1.1}

\begin{figure}[h]
    \centering
    \includegraphics[width=4.5 in]{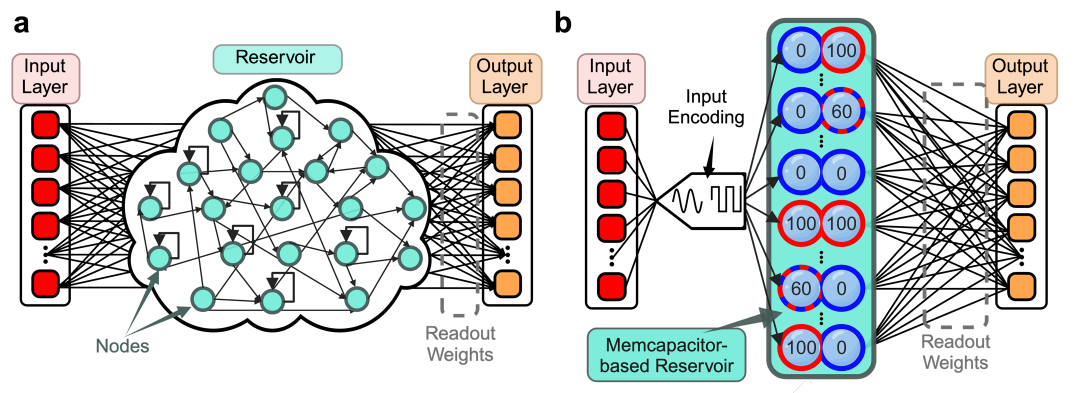}
    \caption{Reservoir computing architectures. \textbf{a} A conventional reservoir computing architecture, i.e., ESN, with the input directly fed into a sparse network of nodes (in most cases nonlinear activation nodes). The randomly connected nodes nonlinearly map the input into a higher dimensional state space. The high-dimensional output from the nodes is then used to train a regression readout layer. \textbf{b} A parallel-nodes reservoir architecture, which is more commonly observed with physical reservoirs due to fabrication convenience. The input is usually encoded first to fit the physical reservoir's operating conditions. The reservoir parallel nodes nonlinearly transform the input into a higher dimensional state space, and, similar to ESNs, the output from the nodes is then used to train a regression readout layer. }
    \label{FigureS1}
\end{figure}

A reservoir computing system, in principle, has three main layers: 1) the input layer, 2) the reservoir layer, and 3) the output layer (including the readout). For different problems and implementations, these three layers may vary in structure. For instance, the input layer can inject the raw input directly into the reservoir layer without implementing any encoding or interpolation on the input as in conventional echo state networks (ESNs) as depicted by \textbf{Fig. S1a}, or it may incorporate an encoding scheme as seen in \textbf{Fig. S1b}, where the input sequence gets interpolated and sampled \cite{Du2017ReservoirProcessing}, masked \cite{Appeltant2011InformationSystem, Zhong2021DynamicProcessing}, or even fully mapped to another space \cite{Du2017ReservoirProcessing, Hossain2023Biomembrane-BasedProcessing} before channeling into the reservoir. Similarly, the reservoir layer may be constructed using sparsely connected spatial nodes as in ESNs \cite{Jaeger2001TheNetworks} as represented in \textbf{Fig. \ref{FigureS1}S1a}, using temporal reservoirs as in a nonlinear node with delayed feedback \cite{Appeltant2011InformationSystem}, or using parallel connected physical nodes with inherent nonlinearity and fading memory \cite{Du2017ReservoirProcessing, Hossain2023Biomembrane-BasedProcessing} as in the memcapacitor-based PRC in \textbf{Fig. S1b}. Finally, the output layer, which includes a readout layer, can include traditional regression layers akin to these in feed-forward neural networks \cite{Jaeger2001TheNetworks} as in \textbf{Fig. S1a,b}, for implementing linear or logistic regression on the state space or other more complex networks like convolutional neural network layers \cite{Armendarez2024Brain-InspiredPlasticity, Zhu2020MemristorAnalysis}.
These listed variations of each layer are among many others \cite{Gauthier2021NextComputing, Liang2022RotatingComputing} that can be found in the vast literature on reservoir computing. Different permutations of the listed forms of input, reservoir, and output layers have been implemented depending on the tasks of interest as well as the nature of the substrate used to solve these tasks, which range from in silico implementations \cite{Jaeger2001TheNetworks} to a plethora of physical substrates \cite{Tanaka2019RecentReview, Cucchi2022Hands-onImplementation}.

\newpage
\subsection*{Supplementary Note 2: Dynamics of Lipid Membrane-based Memcapacitors}\label{SubsectionS1.2}

\begin{figure}[h]
    \centering
    \includegraphics[width=4.5 in]{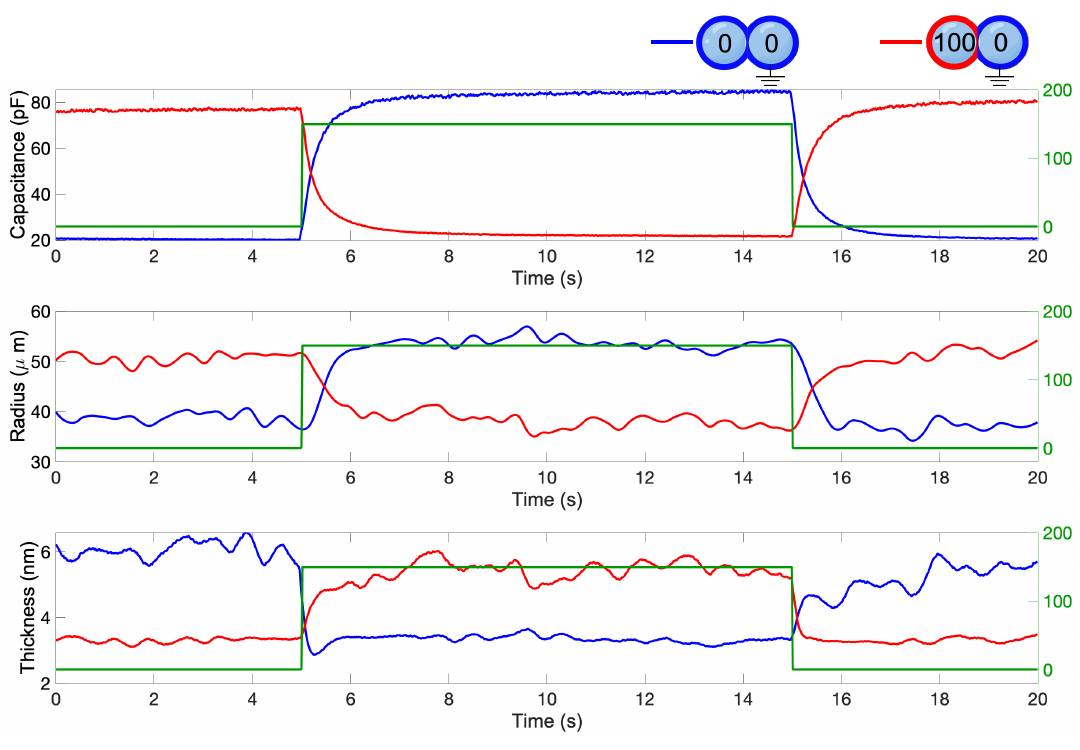}
    \caption{Dynamic responses of membrane capacitance, interfacial area, and hydrophobic thickness to a positive step voltage input. The capacitance and area values are measured as described in the Methods section of the main text, while the thickness is calculated from the obtained capacitance and area values. \textbf{a} For a symmetric memcapacitor (0-0), the membrane capacitance (blue) increases in response to a positive 150-mV transmembrane voltage step input. For an asymmetric memcapacitor (100-0), the membrane capacitance (red) decreases in response to a positive 150-mV transmembrane voltage step input. \textbf{b} For a symmetric memcapacitor (0-0), the interfacial area's major axis radius (blue) increases in response to a positive 150-mV transmembrane voltage step input due to electrowetting. For an asymmetric memcapacitor (100-0), the interfacial area's major axis radius (red) decreases in response to a positive 150-mV transmembrane voltage step input as the overall transmembrane potential drops in magnitude. \textbf{c} For a symmetric memcapacitor (0-0), the hydrophobic thickness (blue) decreases due to charge-induced electrocompression resulting from the positive 150-mV transmembrane voltage step input. For an asymmetric memcapacitor (100-0), the hydrophobic thickness (red) is relaxed and grows in response to a positive 150-mV transmembrane voltage step input as the overall transmembrane potential drops in magnitude.}
    \label{FigureS2}
\end{figure}

Memcapacitors, also known as memory capacitors, are energy storage components with two terminals. They exhibit memory properties, meaning their capacitance depends nonlinearly on internal states and can be adjusted based on past and present external stimulation. Similar to memristors, memcapacitors can be classified as either nonvolatile (maintaining their states upon removal of external stimulus) or volatile (dissipating their states upon removal of external stimulus) \cite{Yang2013MemristiveComputing}. In this study, we construct a memcapacitor from the parallel leaflets of a lipid bilayer, as conducted previously \cite{Najem2019DynamicalMembranes}. By interfacing two lipid monolayer-coated water droplets ($\sim 200$ $nL$ each) immersed in oil, a capacitive, elliptical lipid bilayer ($\sim 50$ $\mu m $ in radius) forms spontaneously  $(\sim 0.1-1)$ $\mu F \cdot cm^{-2}$ \cite{Sarles2010RegulatedSubstrates, Najem2018MemristiveMimics}. The bilayer is highly electrically insulating $(>100$ $M\Omega \cdot cm^2)$ and comprises a dielectric core of hydrophobic lipid tails and trapped oil. When a voltage is applied across the bilayer, it undergoes geometric changes due to electrowetting (EW) \cite{Requena1975TheFilms, Taylor2015DirectBilayer} and electrocompression (EC) \cite{Evans1975MechanicsMembranes, Najem2018MemristiveMimics}. Electrowetting, driven by ionic charge-induced electrostatic forces, increases the bilayer area, while electrocompression reduces its hydrophobic thickness by expelling trapped oil. Charge-induced electrostatic forces and the expulsion of entrapped oil drive these effects. The interfacial area radius ($R(t)$) and hydrophobic thickness ($W(t)$) dynamics are captured by the coupled state equations below \cite{Najem2019DynamicalMembranes}: 

\begin{equation}
\frac{dR(t)}{dt} =  \frac{1}{\zeta_{ew}}((\frac{a\epsilon\epsilon_{0}}{2W(t)}) (v_{app}(t) + v_{\Phi})^2- k_{ew}(R(t)-R_0))
\label{equationR}
\end{equation}	

\begin{equation}
\frac{dW(t)}{dt} =  \frac{1}{\zeta_{ec}}((\frac{-a\epsilon\epsilon_{0}\pi R(t)^2}{2W(t)^2}) (v_{app}(t) + v_{\Phi})^2+ k_{ec}(W_0-W(t)))
\label{equationW}
\end{equation}

Where $a$ is the eccentricity of an ellipse, $\epsilon$ is the relative dielectric constant for the hydrophobic tails and residual oil mixture, $\epsilon_{0}$ is the permittivity of free space, $R_0$ ($m$) is the zero-volt, interfacial area minor axis radius, $W_0$ ($m$) is the zero-voltage hydrophobic thickness, $v_{app}$ and $v_{\Phi}$ are the applied transmembrane potential and dipole offset potential (see \textbf{Supplementary Note 3}), respectively, $\zeta_{ew}$  and $k_{ew}$ are the EW effective damping ($Nsm^{-2}$) and stiffness ($Nm^{-2}$) coefficients in the tangential direction, respectively, and $\zeta_{ec}$ and $k_{ec}$ are the EC effective damping ($Nsm^{-1}$) and stiffness ($Nm^{-1}$) coefficients in the normal direction, respectively. Similar to a standard parallel-plate capacitor, where $R(t)$ $\gg$ $W(t)$, the dynamic membrane capacitance, $C_m(R(t), W(t))$, can be expressed as: 
				
\begin{equation}
C_m(R(t),W(t))= \frac{\varepsilon \varepsilon_0 A(t)}{W(t)} =  \frac{\varepsilon \varepsilon_0 (a \pi R(t)^2)}{W(t)}
\label{equationC}
\end{equation}

\par \textbf{Fig. S2} displays the capacitance, interfacial radius, and hydrophobic thickness responses to an applied transmembrane membrane potential 150-mV square wave for both a symmetric and an asymmetric membrane. As observed, the symmetric membrane (blue curve) increases $\sim$2 times in area ($\sim$1.4 times in radius due to EW). It decreases by $\sim$1/2 in thickness (EC), corresponding to an analog $\sim$4 times growth in capacitance. In contrast, a 100-0 asymmetric membrane  (red curve) decreases to $\sim$1/2 its resting area ($\sim$0.7 times in radius) and increases by $\sim$1.7 in thickness (EC), corresponding to an analog decrease in capacitance to $\sim$0.25 of its resting capacitance. For the symmetric case, starting capacitance ($C_m(t = 0)$) is at its lowest possible value as the overall membrane potential ($v_m(t) = v_{app}(t) + v_{\Phi} = 0$) is zero. Whereas, for the asymmetric case, the starting capacitance ($C_m(t = 0)$) is at an excited state due to the existence of asymmetry-induced dipole offset ($v_m(t) = v_{app}(t) + v_{\Phi} = -138 mV$). Applying a potential to the asymmetric membrane results in an overall decrease in the membrane potential ($v_m(t) = v_{app}(t) + v_{\Phi}$ $\rightarrow  \vert v_m(t)\vert < -138 mV$), thus a diminishing the membrane capacitance as manifested in \textbf{Fig. S2}. The capacitance responses can be approximated by a summation of two exponential functions (i.e. $C_m \simeq a_1e^{-b_1t} + a_2e^{-b_2t}$) where one term approximates the faster electrowetting effects and while the second term approximates the slower electro-compression effects \cite{Najem2019DynamicalMembranes}. At the same time, it takes $\sim2.5$ for both the symmetric and asymmetric cases shown in \textbf{Fig. S2} to return to their initial state (memory dissipation) upon stimulus removal, we note that the memcapacitors' electrowetting and electrocompression dynamics are voltage-dependent (see \textbf{Supplementary Methods}). The devices exhibit pinched hysteresis loops, with varying pinching points, for periodic voltage sweeps as observed in its $Q-v$ curves (Main Text Fig. 2b and Supplementary Fig. S7). It is also important to note that the device is slightly stochastic with native cycle-to-cycle variations (\textbf{Supplementary Fig. S16-S17)}. 


\newpage
\subsection*{Supplementary Note 3: Leaflet Asymmetry-induced Potential}\label{SubsectionS1.3}

\begin{figure}[h]
    \centering
    \includegraphics[width=4.5 in]{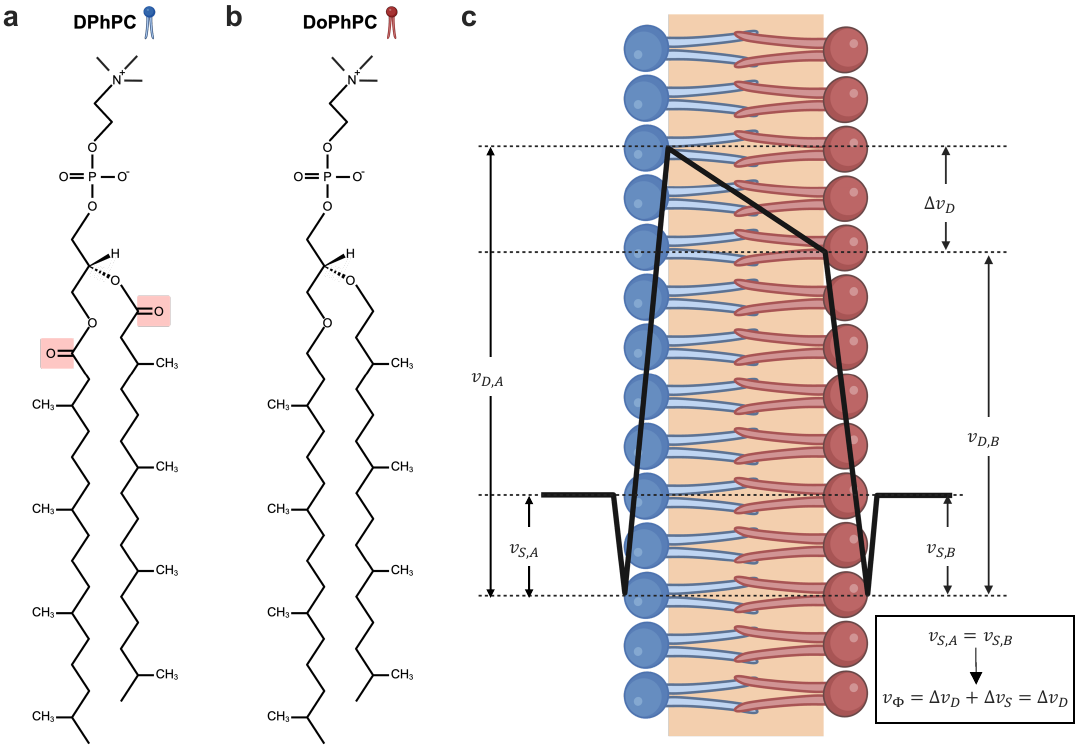}
    \caption{DPhPC and DoPhPC lipid structural formulae and membrane potential profiles explaining the emergence of lipid internal offset. \textbf{a} The structural formula of DPhPC. \textbf{b} The structural formula of DoPhPC. \textbf{c} Membrane potential profiles in leaflets of DPhPC and DOPhPC forming an asymmetric bilayer. The internal offset results from differences in surface potential (zero in this case) and dipole potential differences.}
    \label{FigureS3}
\end{figure}

Throughout this work, the lipid bilayers are comprised of varying combinations of DPhPC (1,2-diphytanoyl-sn-glycero-3-phosphocholine) and DoPhPC (1,2-di-O-phytanoyl-sn-glycero-phosphocholine) lipids, with structural formulae displayed in \textbf{Fig. S3a} and \textbf{Fig. S3b}, respectively. Due to the difference in lipid head structures (highlighted with translucent red in {Fig. S3b}), and thus electric dipole differences, an overall dipole potential difference $\Delta v_{D}$ arises across the membrane ({Fig. S3c}). This difference in lipid dipole potentials $\Delta v_{D}$ in addition to differences in surface potential $\Delta v_{S}$ yield the overall internal potential offsets $v_\Phi$ discussed throughout this study, as given by \cite{Taylor2019ElectrophysiologicalFlip-flop}:

\begin{equation}
v_\Phi = \Delta v_{D} + \Delta v_{S} 
\end{equation}	

For all devices used in this work, there are no surface potential differences ($\Delta v_{S} = 0$) between the two bilayer leaflets, as the salt concentrations used in any two interfacing droplets are identical. Accordingly, the dipole potential is the only contributor to the internal offsets ($v_\Phi = \Delta v_{D}$). Thus, we use the terms dipole offset and internal offset interchangeably. 

\newpage
\maketitle
\section*{Supplementary Figures}\label{SectionS2}

\begin{figure}[h]
    \centering
    \includegraphics[width=4.5 in]{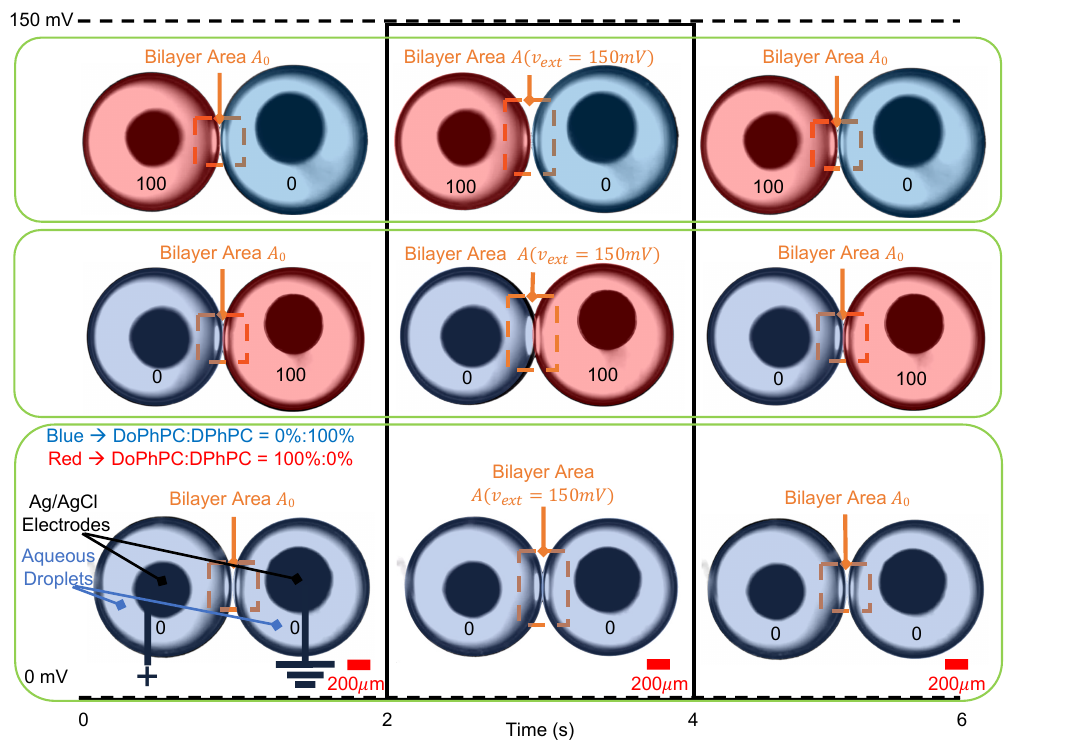}
    \caption{Bottom-view snapshots of experimental symmetric and asymmetric memcapacitors at rest and under voltage application obtained using an inverted microscope. The snapshots are colored blue and red to match the corresponding device colors in cartoon figures displayed throughout the paper. Scale bars are shown in red. For time $0<t<2$, the memcapacitors are at rest, where the symmetric 0-0 memcapacitor (bottom-left) has a small interfacial area relative to the two asymmetric memcapacitors (mid-left and top-left). The two asymmetric memcapacitors are initially stressed due to the internal offsets (138 mV for 0-100 and -138 mV for 100-0). After applying step transmembrane potential of 150 mV ($2<t<4$), the symmetric 0-0 memcapacitor exhibits a rise in capacitance and area (bottom-mid). Similarly, the asymmetric memcapacitor 0-100 (mid-mid) also increases capacitance and area, with a steady-state capacitance and area much larger than that of the symmetric 0-0 case. In contrast, the asymmetric memcapacitor 100-0 (top-mid) also decreases capacitance and area as the applied potential opposes the internal offset. Finally, the membranes are unstimulated for time $4<t<6$ and return to their resting states.} 
    \label{FigureS4}
\end{figure}

\newpage

\begin{figure}[h]
    \centering
    \includegraphics[width=4.5 in]{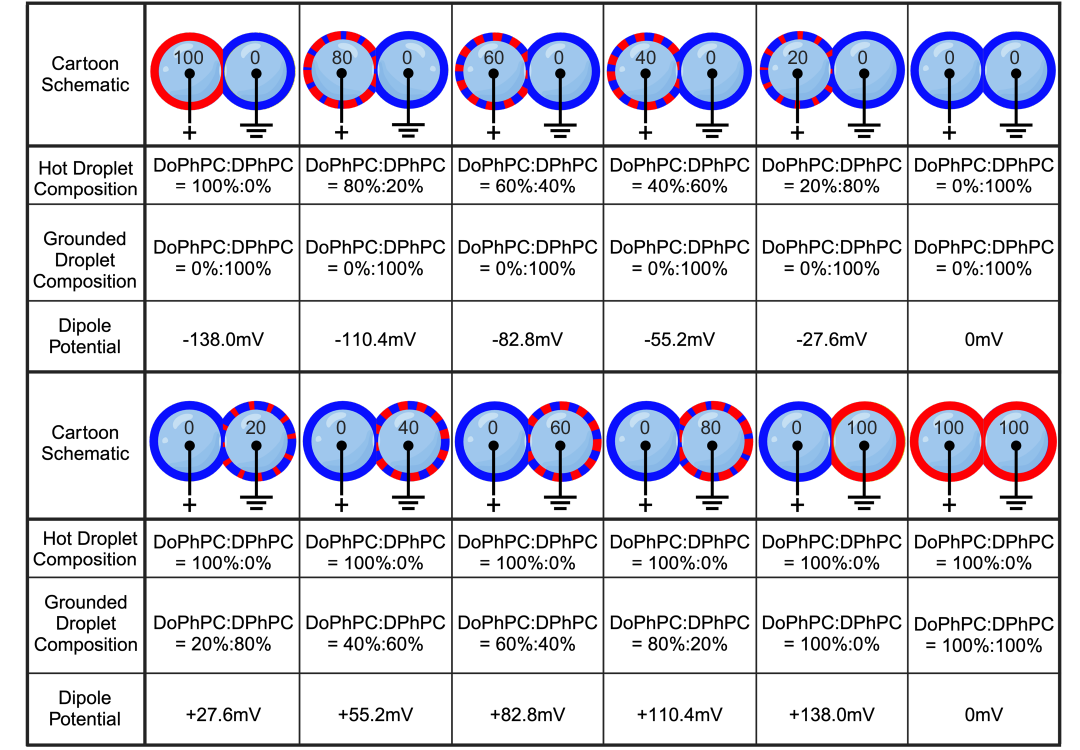}
    \caption{A guide to distinguishing between the 12 memcapacitors used in this work. For all 12 devices, the guide displays the representative cartoons of each device, the corresponding leaflet composition of each droplet, and the corresponding internal dipole potential.}
    \label{FigureS5}
\end{figure}

\newpage

\begin{figure}[h]
    \centering
    \includegraphics[width=4.5 in]{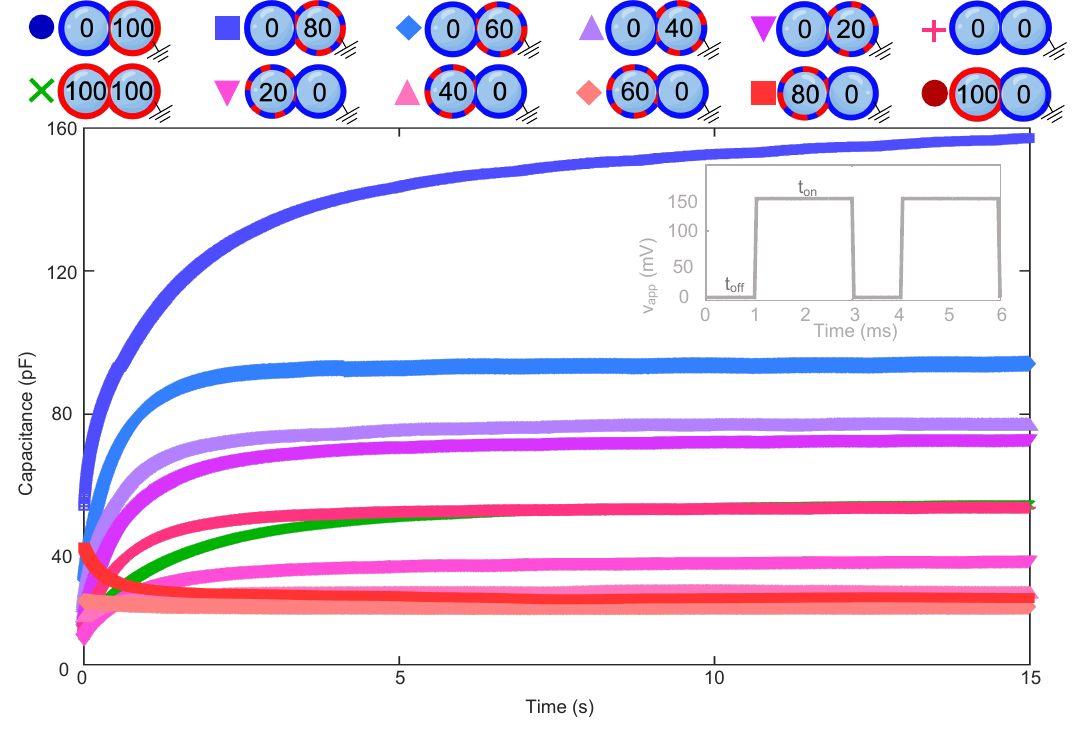}
    \caption{Capacitance responses for all 12 devices to a pulse train of 5000 pulses. Each input square pulse is 3-ms-long with 2 ms of on-time and 1 ms of off-time, as shown by the top-right inset. As observed, the starting capacitance is larger for more asymmetrical devices than symmetrical cases. While the steady-state capacitance values for the symmetrical cases are very similar, each symmetric device's path differs due to different temporal dynamics, as discussed later. Asymmetric devices' responses vary from PPF to PPD depending on the membrane potential before and after the pulse is applied, as explained by \textbf{Eq. 1} in the main text.}
    \label{FigureS6}
\end{figure}

\newpage

\begin{figure}[h]
    \centering
    \includegraphics[width=4.5 in]{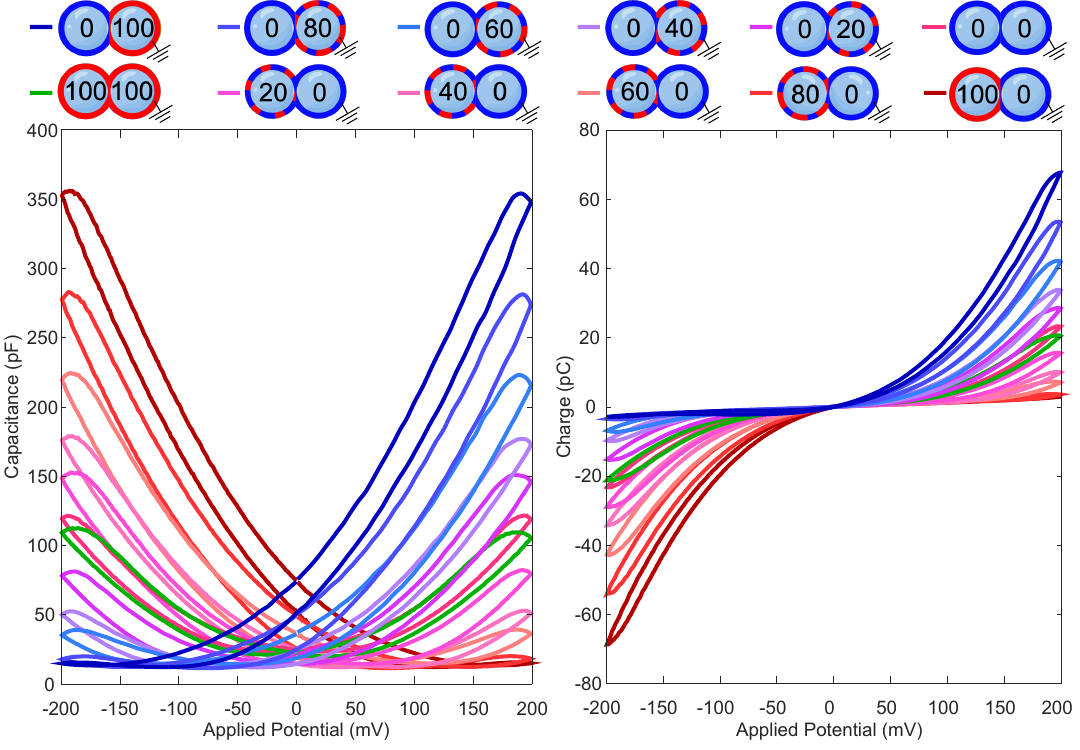}
    \caption{ Capacitance-voltage and charge-voltage hysteresis curves for all 12 devices. \textbf{a} Capacitance responses of all 12 memcapacitors as a function of a voltage sweep ranging from -200 mV to 200 mV at 50 mHz. As observed, as voltage increases from -200 mV to 200 mV, the capacitance only increases, decreases, or decreases then increases depending on the degree and orientation of asymmetry. The hysteresis observed is a fingerprint of memory in memcapacitors. \textbf{b} The corresponding charge responses of all 12 memcapacitors as a function of a voltage sweep ranging from -200 mV to 200 mV at 50 mHz. By definition, memcapacitors must exhibit hysteresis for their Q-V curves, which is the case for all devices, with the pinching point occurring at $-v_\Phi$.}
    \label{FigureS7}
\end{figure}

\newpage

\begin{figure}[h]
    \centering
    \includegraphics[width=4.5 in]{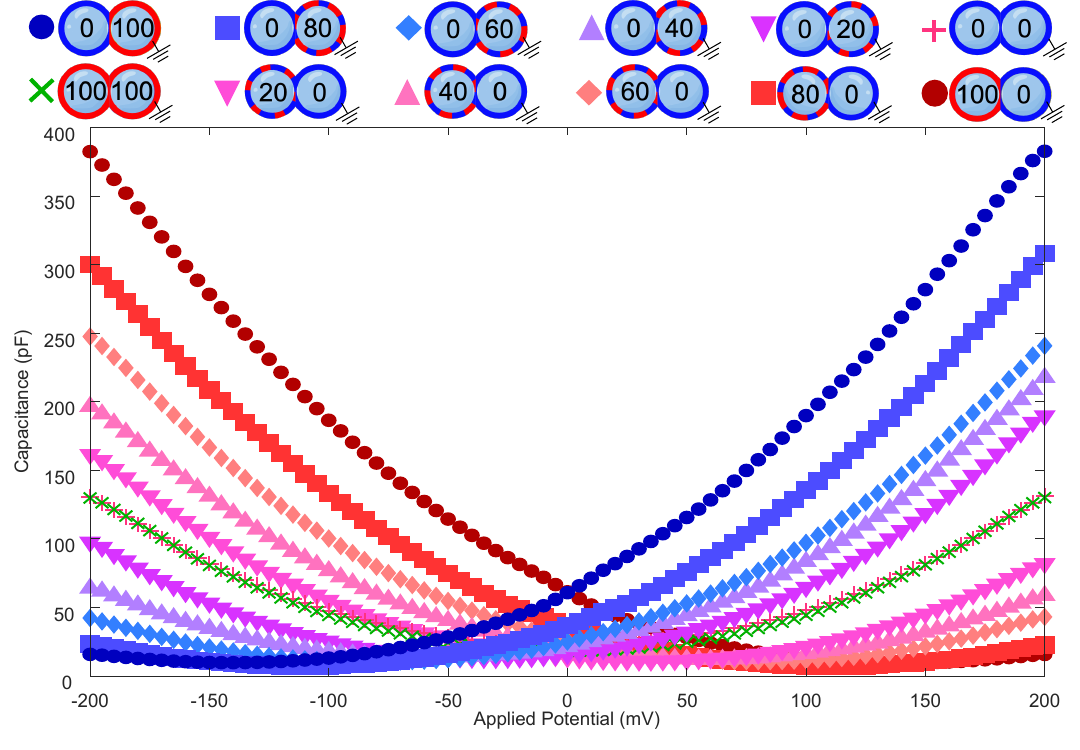}
    \caption{Steady-state capacitance responses of all 12 memcapacitors as a function of voltage step inputs ranging from -200 mV to 200 mV in increments of 5 mV. Each point in this plot was obtained by measuring the capacitance at the end of the on-time of the applied step (as in \textbf{Fig. S2} and as described in the Methods section of the main text). As observed with the dynamic responses in \textbf{Fig. S7a}, as voltage increases from -200 mV to 200 mV, the capacitance only increases, decreases, or decreases then increases depending on the degree and orientation of asymmetry.}
    \label{FigureS8}
\end{figure}

\newpage

\begin{figure}[h]
    \centering
    \includegraphics[width=4.5 in]{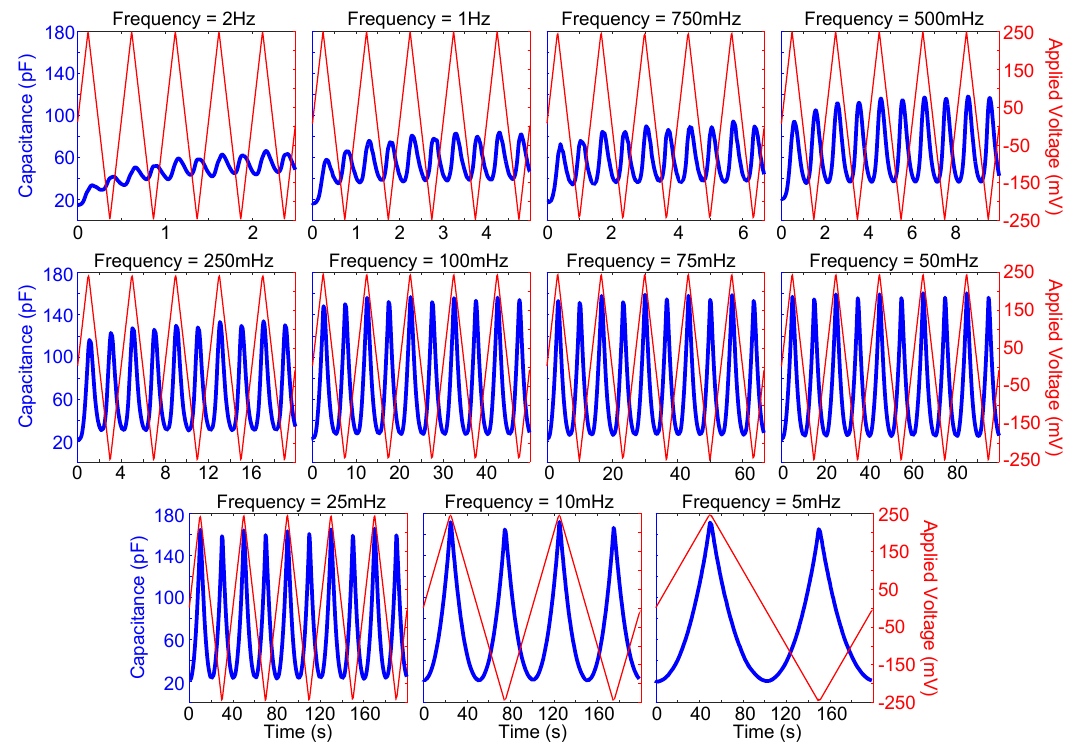}
    \caption{Dynamic capacitance temporal response for a symmetric 0-0 memcapacitor as a function of voltage sweeps for frequencies ranging from 2 Hz to as low as 5 mHz.}
    \label{FigureS9}
\end{figure}

\newpage

\begin{figure}[h]
    \centering
    \includegraphics[width=4.5 in]{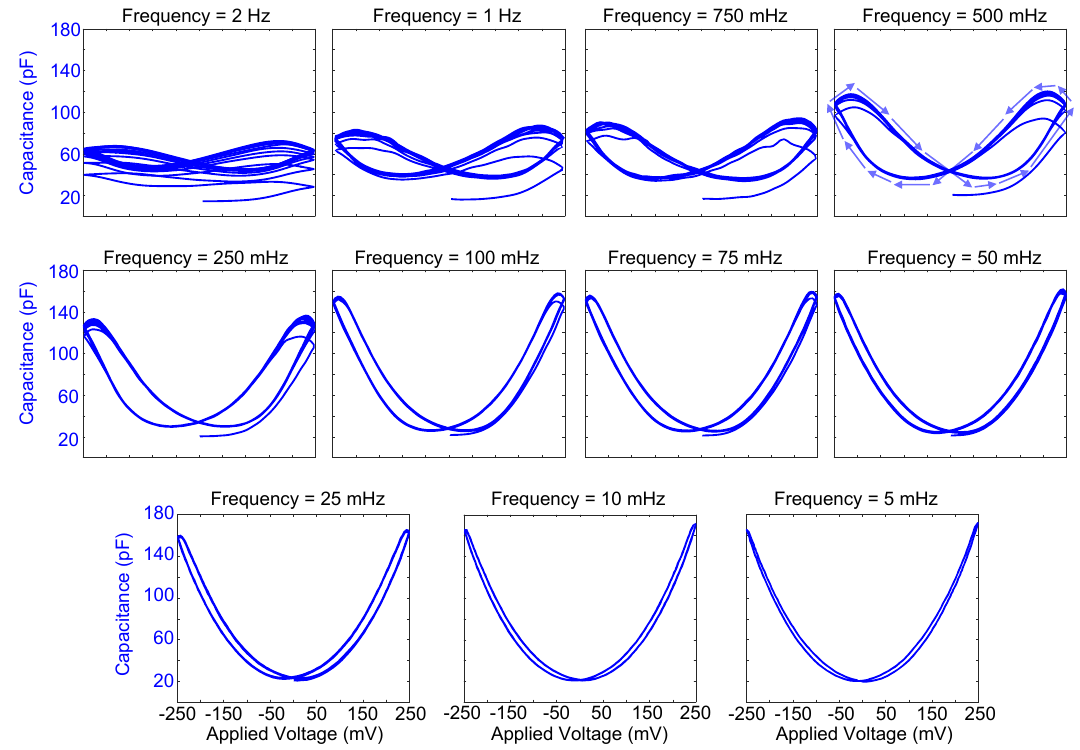}
    \caption{Capacitance-voltage response for a symmetric 0-0 memcapacitor to voltage sweeps with frequencies ranging from 2 Hz to as low as 5 mHz. As observed, the hysteresis lobe area varies for different frequencies, denoting the dependence of memory on the input frequency. As a fingerprint of mem-elements, the state variable's hysteresis lobe area must diminish for very large and very small input frequencies \cite{Chua1971MemristorTheElement} as observed for frequencies greater than 2 Hz and less than 5 mHz. Most pronounced memory effects are obtained at frequencies between 2 Hz and 5 mHz as qualitatively represented by the lobe area. }
    \label{FigureS10}
\end{figure}

\newpage

\begin{figure}[h]
    \centering
    \includegraphics[width=4.5 in]{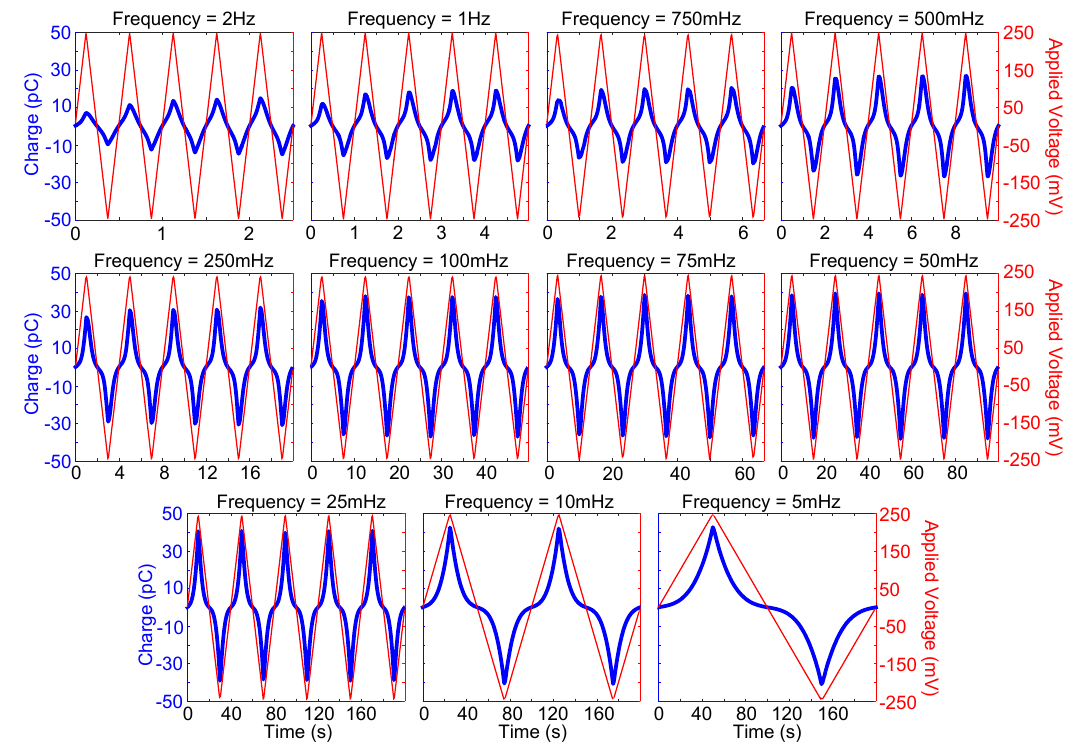}
    \caption{Dynamic charge temporal response for a symmetric 0-0 memcapacitor as a function of voltage sweeps for frequencies ranging from 2 Hz to as low as 5 mHz.}
    \label{FigureS11}
\end{figure}

\newpage

\begin{figure}[h]
    \centering
    \includegraphics[width=4.5 in]{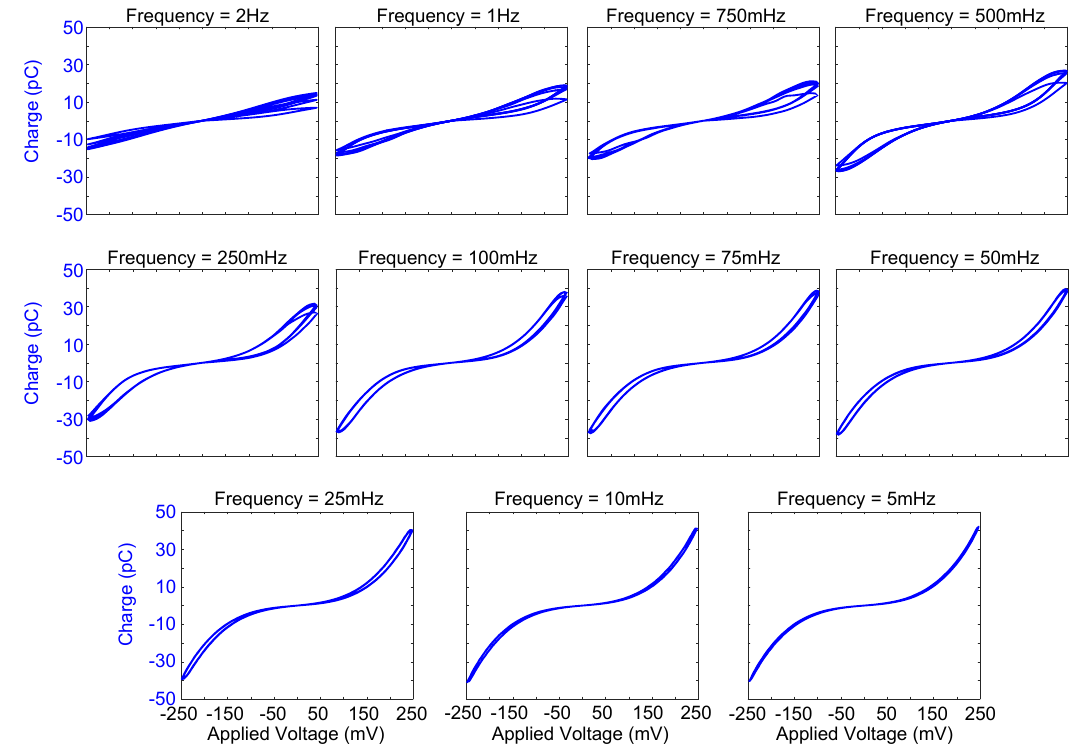}
    \caption{Charge-voltage response for a symmetric 0-0 memcapacitor to voltage sweeps with frequencies ranging from 2 Hz to as low as 5 mHz. As a fingerprint of memcapacitors, the charge-voltage hysteresis lobe area must diminish for very large and very small input frequencies \cite{Chua1971MemristorTheElement} as observed for frequencies greater than 2 Hz and less than 5 mHz. Most pronounced memory effects are obtained at frequencies between 2 Hz and 5 mHz as qualitatively represented by the lobe area.}
    \label{FigureS12}
\end{figure}

\newpage

\begin{figure}[h]
    \centering
    \includegraphics[width=4.5 in]{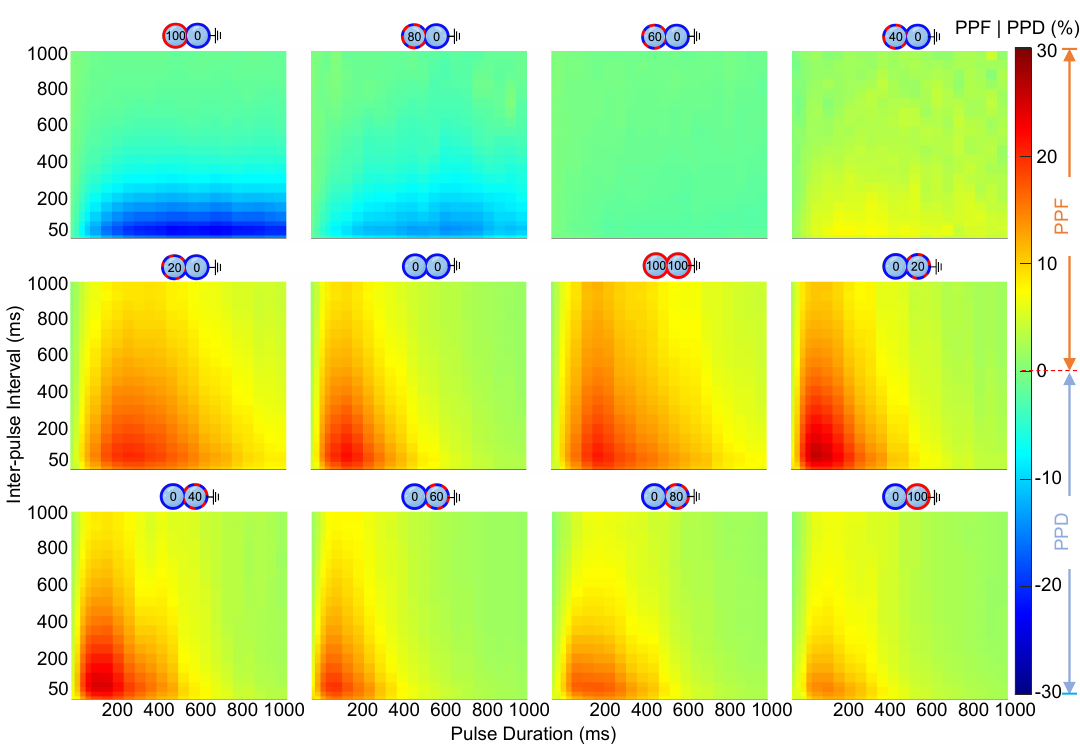}
    \caption{2-D maps of the PPF and PPD indices as functions of pulse duration and inter-pulse intervals for all 12 devices. For all pulse durations and inter-pulse intervals, the pulse was of an amplitude of 150 mV. Whether a device exhibits PPF or PPD depends on the membrane potential before and after the pulse is applied, as explained by \textbf{Eq. 1} in the main text.}
    \label{FigureS13}
\end{figure}

\newpage

\begin{figure}[h]
    \centering
    \includegraphics[width=4.5 in]{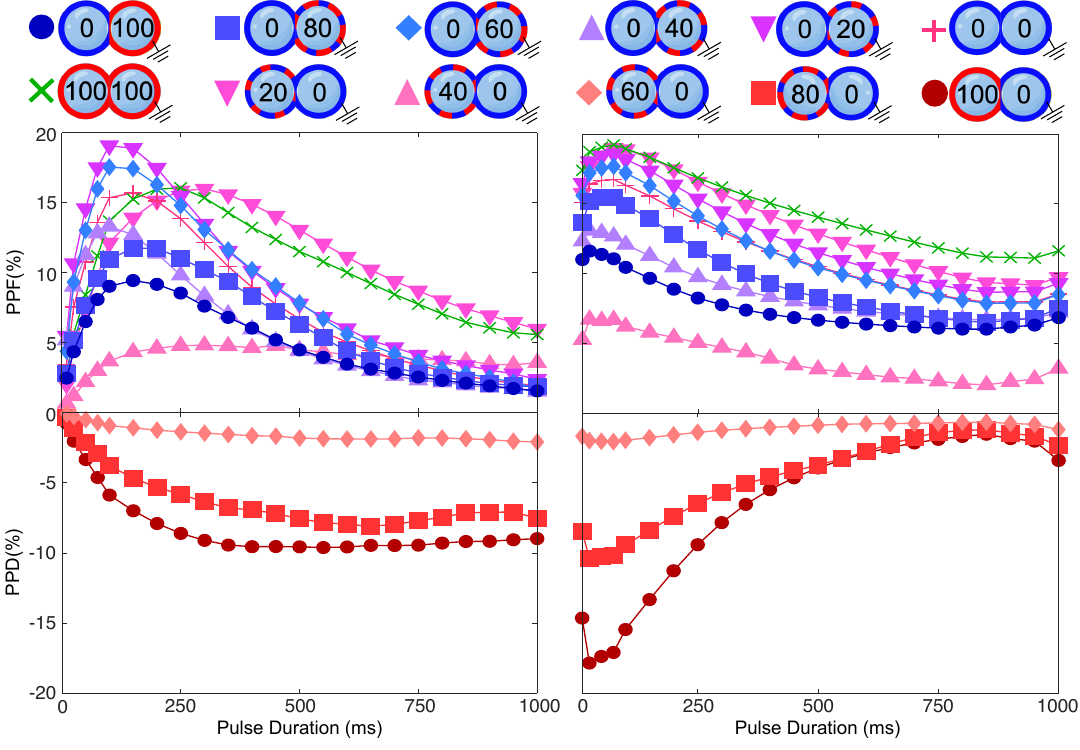}
    \caption{Horizontal and vertical projections of all 12 maps shown in \textbf{Fig. S13}. The inter-pulse interval is fixed to 250 ms, and the PPF and PPD are plotted against varying pulse durations in the top and bottom panels for all 12 devices. \textbf{b} The pulse duration is fixed to 250 ms, and the PPF and PPD are plotted against varying inter-pulse intervals in the top and bottom panels, respectively, for all 12 devices.}
    \label{FigureS14}
\end{figure}

\newpage

\begin{figure}[h]
    \centering
    \includegraphics[width=4.5 in]{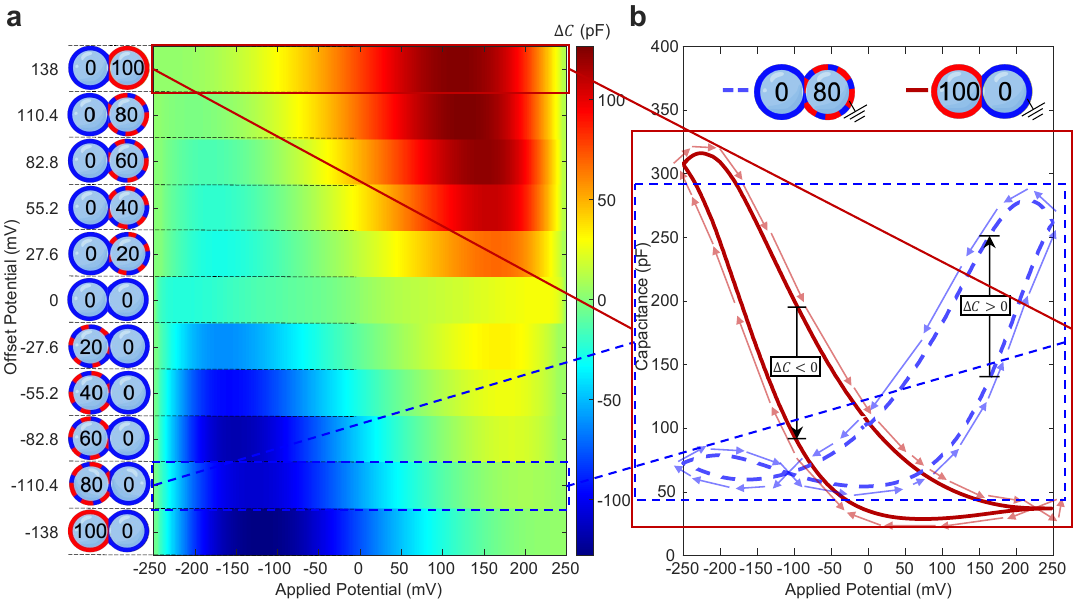}
    \caption{The internal offsets affect the hysteresis lobe area's rotation direction as expressed by the difference in capacitance in response to the increasing voltage sweep (from -250mV to +250mV) and the capacitance in response to the decreasing voltage sweep (from +250mV to -250mV), denoted $\Delta C$. \textbf{a} A 2-D map demonstrating $\Delta C$ as a function of the applied voltage across the eleven memcapacitors. As shown from the highly varying gradient (from $\sim+150$ pF to $\sim-150$ pF), the $\Delta C$ fully switches direction as a function of the internal offset, which depends on the membrane composition. \textbf{b} C-V curves for two different devices (0-80 and 100-0), one exhibiting increased capacitance with increased voltage, i.e., PPF (dashed blue for 0-80), while the other exhibiting decreased capacitance with increased voltage, i.e., PPD (solid red for 100-0). As shown, the positive $\Delta C$ (PPF) is associated with a counterclockwise rotation of the hysteresis lobe. In contrast, the negative $\Delta C$ (PPD) is associated with a clockwise rotation of the hysteresis lobe.}
    \label{FigureS15}
\end{figure}

\newpage

\begin{figure}[h]
    \centering
    \includegraphics[width=4.5 in]{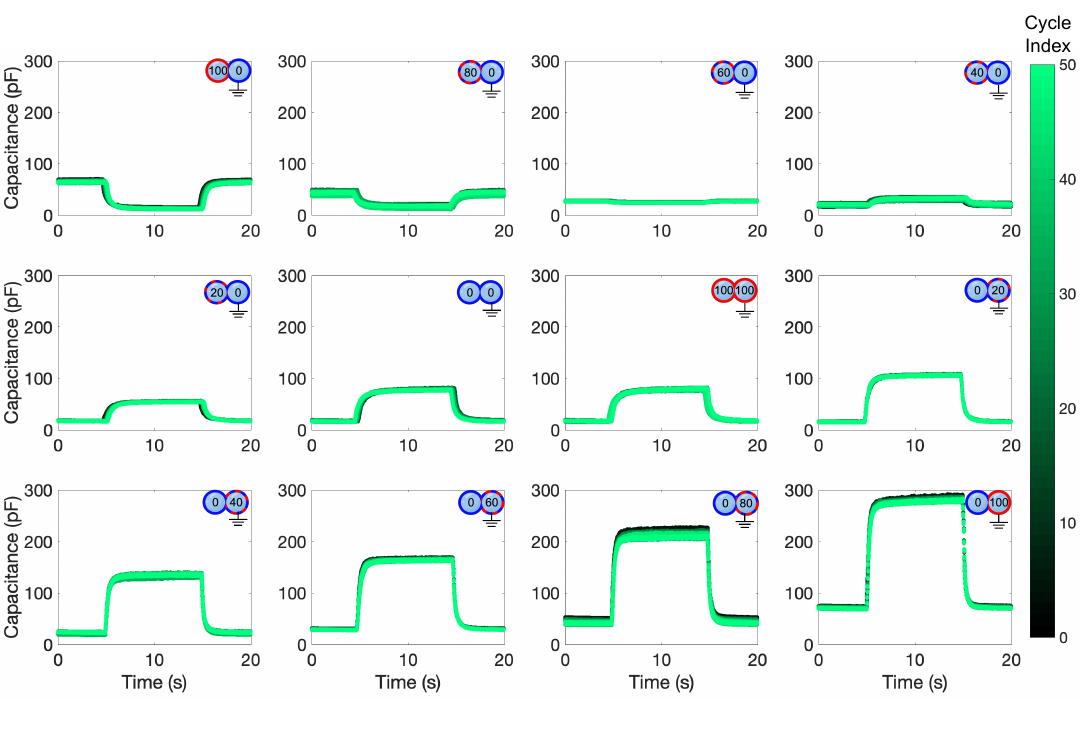}
    \caption{For each of the 12 devices, 50 temporal capacitance profiles are overlaid on each other in response to 50 consecutive cycles of 150-mV square pulses (applied at times between 5 s and 15 s). This plot serves to visualize the minimal cycle-to-cycle variations for all 12 devices. Minimized cycle-to-cycle variations are critical to ensure common-signal-induced synchronization \cite{Inubushi2017ReservoirTrade-off}.}
    \label{FigureS16}
\end{figure}

\newpage

\begin{figure}[h]
    \centering
    \includegraphics[width=4.5 in]{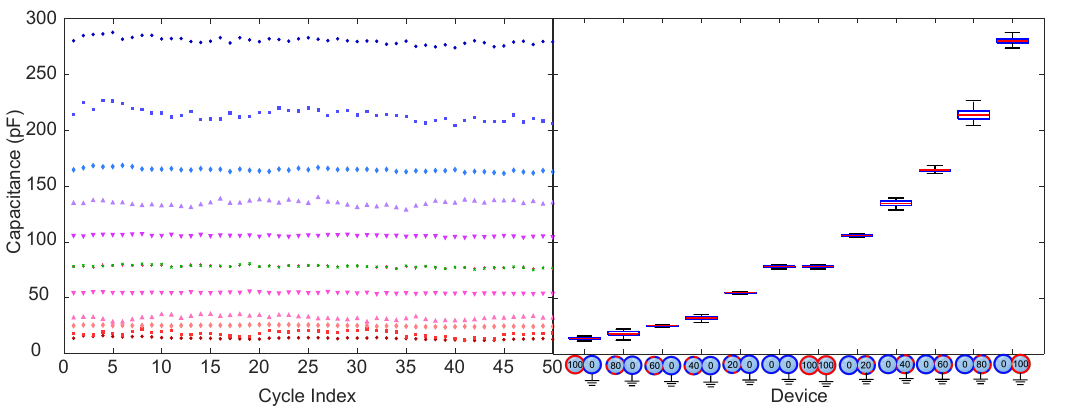}
    \caption{Cycle-to-cycle variations of all 12 devices. \textbf{a} Capacitance responses as a function of cycle index. Each point denotes the capacitance value at the end of the pulse on time (at 15s) from \textbf{Fig. 15}. \textbf{b} Box plot summarising the statistics of cycle-to-cycle variations for all 12 devices.}
    \label{FigureS17}
\end{figure}

\newpage

\begin{figure}[h]
    \centering
    \includegraphics[width=4.5 in]{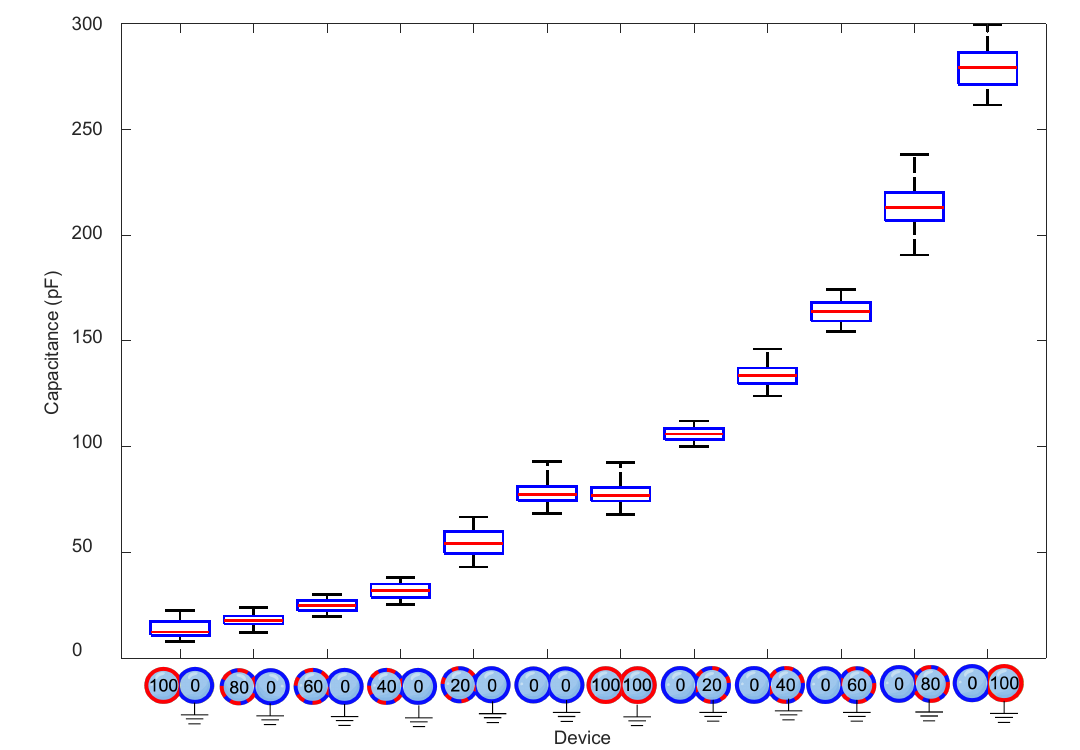}
    \caption{Box plot summarising the statistics of device-to-device variations for all 12 devices. For each of the 12 memcapacitors, the procedure used to obtain the capacitance values is similar to that used for cycle-to-cycle variations (\textbf{Fig. 15} and \textbf{Fig. 16}), yet implemented on five samples of each of the 12 memcapacitor types instead of one. In concept, large device-to-device variations help increase the reservoir's high-dimensional mapping efficacy. However, the device-to-device variations associated with our devices are relatively small compared to other devices \cite{Moon2019TemporalSystem, Du2017ReservoirProcessing} used in PRC to be considered a main contributor towards the high performance observed with our PRC in SONDS and Henon Map tasks.}
    \label{FigureS18}
\end{figure}

\newpage

\begin{figure}[h]
    \centering
    \includegraphics[width=4.5 in]{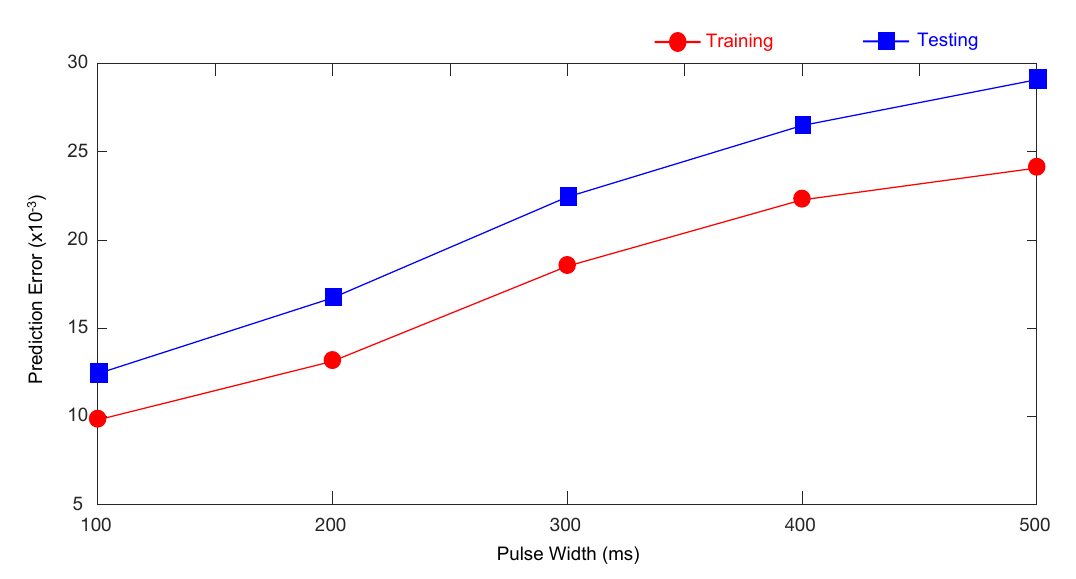}
    \caption{SONDS training and testing prediction errors using a heterogeneous reservoir with externally applied offsets that nullify the internal offsets. For each memcapacitor, the overall offset was approximately zero, and the device's heterogeneity purely stems from differences in device temporal dynamics (main text \textbf{Fig. 3} and \textbf{Supplementary Fig. S22}). The PEs are almost two orders of magnitude larger than those obtained using devices with offsets (main text \textbf{Fig. 4}), highlighting the offsets' role in maximizing the reservoir's dimensionality.}
    \label{FigureS19}
\end{figure}

\newpage

\begin{figure}[h]
    \centering
    \includegraphics[width=4.5 in]{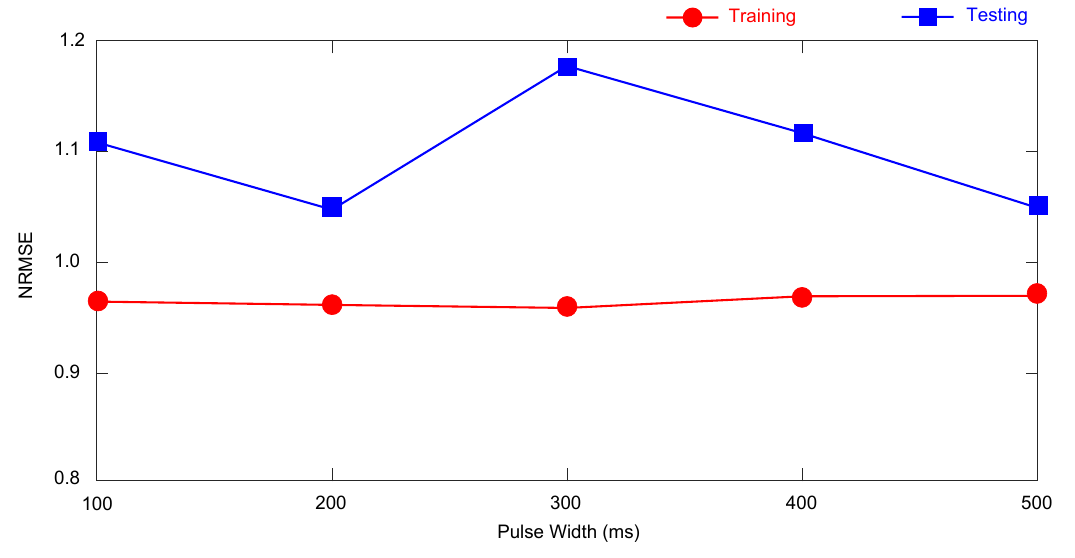}
    \caption{Hénon Map training and testing NRMSEs using a heterogeneous reservoir with externally applied offsets that nullify the internal offsets. For each memcapacitor, the overall offset was approximately zero, and the device's heterogeneity purely stems from differences in device temporal dynamics (main text \textbf{Fig. 3} and \textbf{Supplementary Fig. S22}). As shown ($\mathrm{NRMSE \simeq 1}$ for almost all pulse widths), without the offsets, the devices are unable to predict the Hénon map, underscoring the importance of offsets for realizing non-monotonic input-state correlations, crucial for predicting the Hénon Map.}
    \label{FigureS20}
\end{figure}

\newpage

\section*{Supplementary Methods}\label{SectionS3}
\subsection*{Memcapacitor model parameters and temporal dynamics}\label{SubsectionS3.1}

\begin{figure}[h]
    \centering
    \includegraphics[width=4.5 in]{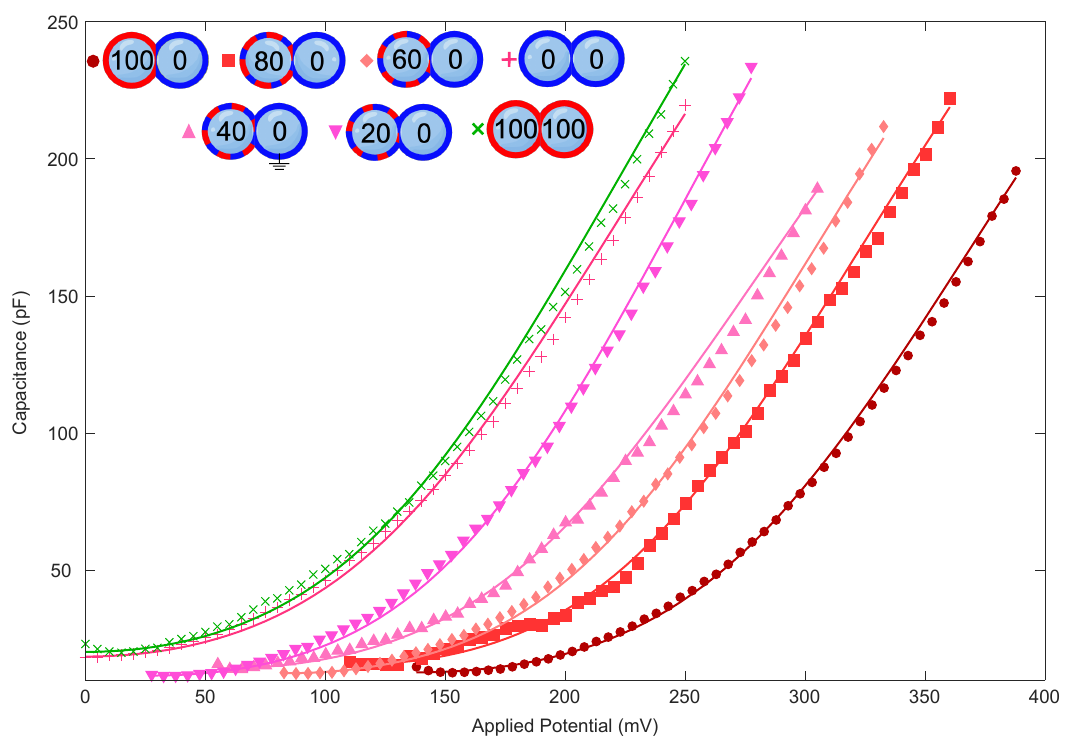}
    \caption{Steady-state capacitance as a function of the applied potential for all seven leaflet types. For each membrane composition, the starting potential was approximately $-v_\Phi$ pertinent to the membrane type. The steady-state solution for \textbf{Eq. \ref{equationR}} and  {Eq. \ref{equationW}} was used to fit the measured memcapacitive responses.}
    \label{FigureS21}
\end{figure}

To extract unique parameters for the memcapacitor model (\textbf{Eq. \ref{equationR}} and  {Eq. \ref{equationW}}), we use a two-step process. First, we extract $k_{ew}$ and $k_{ec}$ by assuming steady-state conditions for \textbf{Eq. \ref{equationR}} and  {Eq. \ref{equationW}} (i.e. $dR/dt=dW/dt=0$). By setting the time derivatives to zero and performing algebraic manipulations, the resulting steady-state equations are:

\begin{equation}
a_{5}W_\infty^5 + a_{4}W_\infty^4 + a_{3}W_\infty^3 + a_{2}W_\infty^2
 + a_{1}W_\infty + a_{0} = 0
 \label{equationWSS}
\end{equation}	
\begin{equation}
R_\infty = \frac{a\varepsilon\varepsilon_0v_m^2}{2k_{ew}} \frac{1}{W_\infty} + R_0
\label{equationRSS}
\end{equation}	

where 
\begin{equation*}
a_{5} = -8k_{ec}k_{ew}^2
\end{equation*}
\begin{equation*} 
a_{4} =8W_{0}k_{ec}k_{ew}^2
\end{equation*}
\begin{equation*} 
a_{3} = 0
\end{equation*}
\begin{equation*} 
a_{2} = 
4R_{0}^2v_m^2a\varepsilon\varepsilon_0k_{ew}^2\pi
\end{equation*}
\begin{equation*} 
a_{1} = 4R_0v_m^4a^2\varepsilon^2\varepsilon_0^2k_{ew}\pi
\end{equation*}
\begin{equation*} 
a_{0} = v_m^6a^3\varepsilon^3\varepsilon_0^3\pi
\end{equation*} 

\begin{figure}[h]
    \centering
    \includegraphics[width=4.5 in]{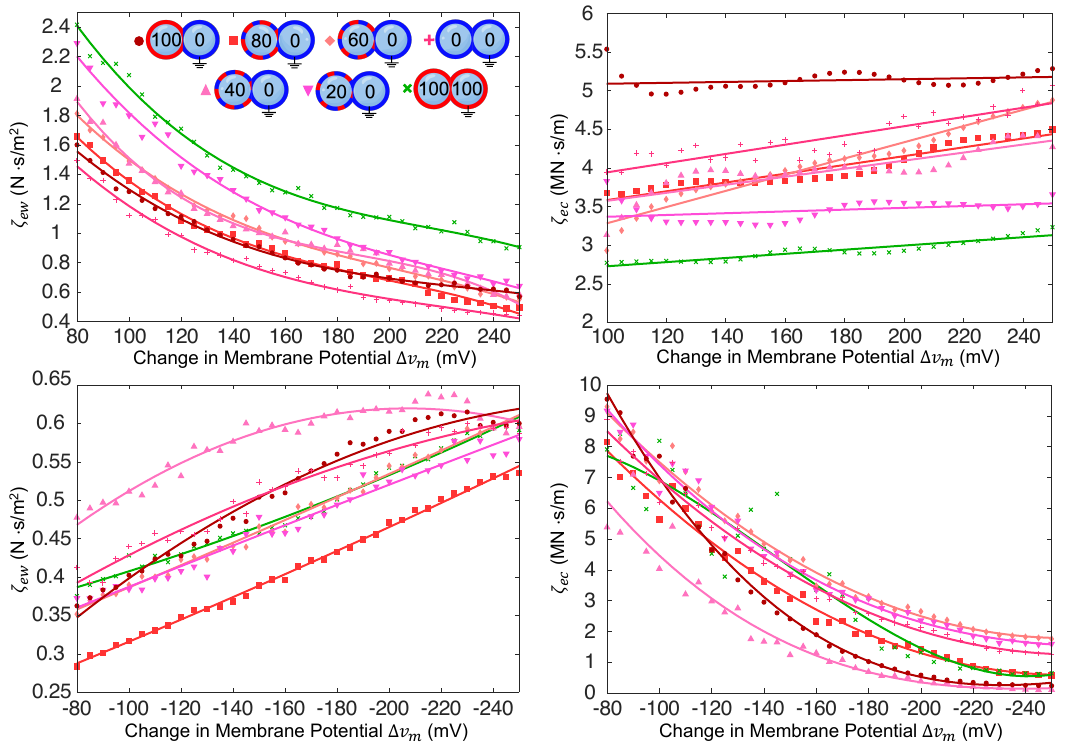}
    \caption{Temporal parameters $k_{ew}$ and $k_{ec}$ as functions of $\Delta v_m$ for different $v_{m0}$ magnitudes. Values of $k_{ew}$ and $k_{ec}$ for $\Delta v_m <80$ mV were discarded as the potential change is too small to cause an accurately measurable $k_{ew}$ and $k_{ec}$ values. \textbf{a} $k_{ew}$ as a function of $\Delta v_m$ for $v_{m0} = 0$ mV. \textbf{b} $k_{ec}$ as a function of $\Delta v_m$ for $v_{m0} = 0$ mV. \textbf{c} $k_{ew}$ as a function of $\Delta v_m$ for $v_{m0} = 250$ mV. \textbf{d} $k_{ec}$ as a function of $\Delta v_m$ for $v_{m0} = 250$ mV.}
    \label{FigureS22}
\end{figure}
\noindent Using the numerical solution from these equations along with the recorded steady-state area and computed thickness (see \textbf{Supplementary Note 2} and main text Methods), we can fit for $k_{ew}$ $k_{ec}$. The values for $a$, $\varepsilon$, and $\varepsilon_0$ for bilayers in decane oil can be found in \cite{Najem2019DynamicalMembranes, Taylor2015DirectBilayer}. The resulting model $R_\infty$ and $W_\infty$ are then used to calculate $C_\infty$ for all seven types of leaflets discussed in this study. The resulting steady-state capacitances are plotted against the measured steady-state capacitances in Supplementary Fig. S21. Furthermore, the resultant values for $k_{ew}$ and $k_{ec}$ are $2.02 \pm 0.2$ and $2.33\times10^5\pm 0.31\times10^5$, respectively, in agreement with previously reported values \cite{Najem2019DynamicalMembranes}. Unlike this work, in \cite{Najem2019DynamicalMembranes}, the steady-state capacitance was also shown to fit well to an empirical quartic function which is not derived from Eq. \ref{equationR} and  {Eq. \ref{equationW}. These two parameters do not vary as Do composition is increased on either droplet.
 Second, after obtaining $k_{ew}$ and $k_{ec}$, we use them to fit for $\zeta_{ew}$ and $\zeta_{ec}$. To fit for these two parameters, we fit \textbf{Eq. \ref{equationR}} and  {Eq. \ref{equationW}} simultaneously to the measured dynamic area and corresponding computed thickness, respectively, for step membrane voltage changes ($\Delta v_m = v_m - v_{m0}$) with amplitudes varying from 0 to 250 mV in increments of 5 mV. For every step, the step voltage starts at 0 mV (i.e. $v_{m0}=0$ mV) and ends at a chosen $v_m$, where values for $\zeta_{ew}$ and $\zeta_{ec}$ were obtained, resulting in a total of 51 values. Due to the small nondeterministic area and thickness values corresponding to small $\Delta v_m$, values where $\Delta v_m < 80 mV$ were discarded, and the resulting $\zeta_{ew}$ and $\zeta_{ec}$ as a function of membrane potential is plotted in \textbf{Supplementary Fig. S22a,b}. As observed, $\zeta_{ew}$ exhibits a cubic decaying relationship with an increasing $\Delta v_m$ as discussed earlier (\textbf{Fig. 3c} in the main text), while $\zeta_{ec}$ has a subtle linear dependence on $\Delta v_m$. Surprisingly, when the applied $\Delta v_m$ was switched to start at 250 mV (i.e. $v_{m0}=250$ mV) and decay to a desired $v_m$, the dependence of $\zeta_{ew}$ and $\zeta_{ec}$ drastically changes as in \textbf{Supplementary Fig. S22c,d}. For a decreasing $\Delta v_m$, $\zeta_{ew}$ exhibits a linear dependence on $\Delta v_m$, while $\zeta_{ec}$ appears to have a cubic dependence. This concludes that $\zeta_{ew}$ and $\zeta_{ec}$ are not only sensitive to $\Delta v_m$, but also to the starting $v_{m0}$. We interpret such complex dynamics as a consequence of the reversible oil expulsion and inclusion into the membrane core resulting from EW and EC \cite{Najem2019DynamicalMembranes, Taylor2015DirectBilayer}. That is, we expect the parameters $\zeta_{ew}$ and $\zeta_{ec}$ to be functions of the trapped oil volume in the membrane rather than simpler functions of the membrane potential. Modeling this phenomenon is beyond the scope of this study and a subject for future investigation.
 
\subsection*{Prediction Error and Normalised mean squared error}
In this work, we have used the metric prediction error (PE) to calculate the error between the actual and predicted signal when solving SONDS, which is defined as:
\begin{equation}
PE=\tfrac{\sum_i\left({z_i(t)- y_i(t)}\right)^2}{\sum_i{y_i}^{2}(t)}
\label{PE_Du}
\end{equation}
where $z(t)$ is the predicted signal and $y(t)$ is the actual signal. As the actual signal normalizes the result, the error is unitless.

So far, we have computed the PE using the previous equation to facilitate equitable comparisons with the work of another group \cite{Du2017ReservoirProcessing}. Moreover, other studies have emphasized the utility of another metric, the normalised mean squared error (NMSE) for SONDS. The NMSE value is defined as:
\begin{equation}
NMSE=\tfrac{\sum_i\left({z_i(t)- y_i(t)}\right)^2}{\sum_i({y_i(t)- \bar{y}(t)})^{2}}
\label{new_eqn}
\end{equation}

The difference between the PE and the NMSE is in the denominator, where the PE includes the sum of the squared difference between $z(t)$ and $y(t)$ by the square of the summation of square of the actual signal $\sum_i{y_i(t)^{2}}$. At the same time, the NMSE uses instead the summation of variance of the actual signal $\sum_i({y_i(t)- \bar{y}(t)})^{2}$. While the NMSE is a more standardized metric of error in signal prediction, we opted to use the PE due to its prevalence in literature when quantifying the signal error in SONDS for fair comparison. The NMSE is equivalent to the PE only if the mean $\bar{y}(t)$ of the true signal is zero, which is not the case for SONDS (see \textbf{Fig. 4} in the main text). 

\subsection*{Normalised root mean squared error}
In this work, we have used the metric normalized root mean squared error (NRMSE) to calculate the error between the actual and predicted signal when predicting the Hénon map, which is defined as:
\begin{equation}
NRMSE=\sqrt{\tfrac{\sum_i\left({z_i(t)- y_i(t)}\right)^2}{\sum_i{y_i}^{2}(t)}}
\label{NRMSE}
\end{equation}
where $z(t)$ is the predicted signal and $y(t)$ is the actual signal. As the actual signal normalizes the result, the error is unitless.

\clearpage
\newpage

\section*{Supplementary Discussion}\label{SectionS4}
\par Given the success of employing external offsets in a symmetric memcapacitor (\textbf{Fig. 5e} and \textbf{Fig. 4d} and its straightforward implementation, we aimed to extend its utility to other dynamic devices utilized in PRC. In this section, we evaluate the application of two memristor models previously employed in PRC: those proposed by Du \textit{et al.} \cite{Du2017ReservoirProcessing} and Armendarez \textit{et al.} \cite{Armendarez2024Brain-InspiredPlasticity}. In simulation, we incorporate external offsets to predict SONDS and the Hénon Map problems. Our selection of these two models is intended to demonstrate the versatility of our approach, even with devices exhibiting diverse temporal dynamics, in which the former memristor model assumes a constant time constant, whereas the latter features a voltage-sensitive time constant. 
\newpage

\subsection*{Predicting SONDS with External offsets using Du \textit{et al.} memristor}\label{SubsectionS4.1}

\begin{figure}[h]
    \centering
    \includegraphics[width=4.5 in]{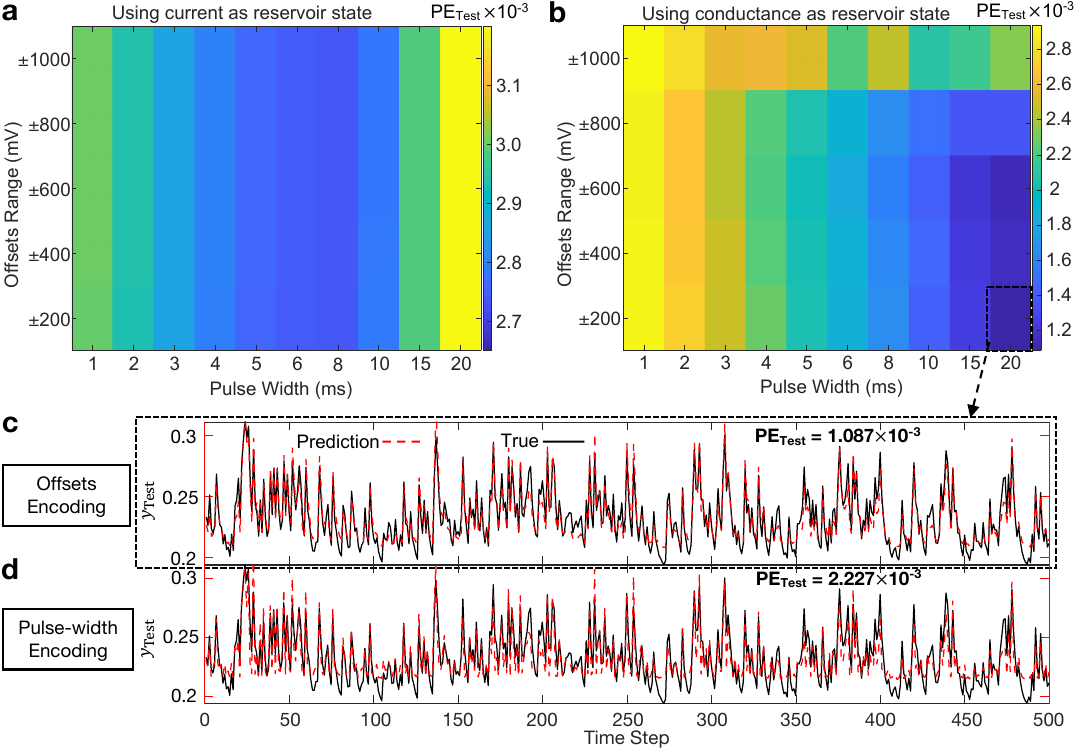}
    \caption{SONDS test data set prediction results using the newly introduced externally applied offsets-encoding method on Du's memristor model \cite{Du2017ReservoirProcessing}. \textbf{a} Testing prediction error as a function of pulse width and offset range using memristor's current as the reservoir state (Eq. \ref{DusIModel}). Only one pulse width and offset range were used for each run. \textbf{b} Testing prediction error as a pulse width and offset range function using memristor's conductance as the reservoir state ($G = I/v$). \textbf{c} Sample 500 points of the reservoir's predicted output vs the true SONDS output ($\mathrm{y_{test}}$) for the lowest PE achieved using the offsets-encoding method with conductance as the reservoir state. The lowest test PE achieved is $1.087\times10^{-3}$ \textbf{d} Sample 500 points of the reservoir's predicted output vs the true SONDS output ($\mathrm{y_{test}}$) for the PE achieved using the pulse-width encoding method with conductance as the reservoir state. The test PE achieved is $2.227\times10^{-3}$.}
    \label{FigureS23}
\end{figure}

\noindent We modeled Du's memristor using these equations as in \cite{Du2017ReservoirProcessing}:
\begin{equation} 
\frac{dw}{dt} = \lambda \sinh(\eta v) - w/\tau
\end{equation}\label{DusModel}
\begin{equation} 
I = (1 - w)\alpha [1 - exp(-\beta v)] + w\gamma \sinh(\delta \mid v\mid)
\end{equation}\label{DusIModel}

\noindent where the values of the parameters $\lambda$, $\eta$, $\tau$, $\alpha$, $\beta$, $\gamma$, and $\delta$ can be found in \cite{Du2017ReservoirProcessing}. Herein, we predicted SONDS using pulse-width encoding with 11 different pulse widths and once using one pulse-width along with 11 different external voltage offsets, denoted as 'offsets-encoding'. Full details on the prediction workflow of SONDS can be found in \textbf{Fig. 4} the main text. For the pulse-width encoding run, the pulse widths used were 1, 2, 3, 4, 5, 6, 8, 10, 15, 20, and 40 ms were used as in \cite{Du2017ReservoirProcessing}. For the offsets-encoding, we encoded the data in 50 different ways, where we mixed 10 different pulse widths (1, 2, 3, 4, 5, 6, 8, 10, 15, or 20 ms) with 5 different offset ranges; however, each run used only one pulse width and 11 offsets linearly spaced from one offset range as depicted by the 2-D map in \textbf{Supplementary Fig. S23}. For example, the rectangular area situated in the lower-rightmost position in \textbf{Supplementary Fig. S23} represents the prediction error (PE) pertaining to an offsets-encoding run with a pulse width of 20 ms and 11 offsets that are linearly spaced between -200 mV and +200 mV (i.e. the 11 offsets used where -200, -160, -120, -80, -40, 0, +40, +80, +120, +160, and +200 mV). In summary, we ran 1 pulse-width encoding run and 50 different offset-encoding runs to show that for most encoding parameters, i.e., pulse width and offset range, the offset-encoding run produced prediction errors that are either comparable to or less than that of the pulse-width encoding method. For both runs, the same 2000-points (1000 for training and 1000 for testing) uniformly random input stream $u\in (0,0.5)$ was used for comparison. For all runs, the input $u$ was mapped into voltage pulses ranging from 0.8 V to 1.8 V with a $100\%$ duty cycle as in \cite{Du2017ReservoirProcessing}, and the capacitance state was captured at the end of each pulse for both pulse-width and offsets encoding. In \textbf{Supplementary Fig. S23}, we show the testing data set's PE as a function of pulse widths and voltage offset ranges, where pulse widths of 1, 2, 3, 4, 5, 6, 8, 10, 15, or 20 ms (horizontal axis) and voltage offset ranges of $\pm 200$, $\pm 400$, $\pm 600$, $\pm 800$, or $\pm 1000$ mV (vertical axis) were employed. Furthermore, we conducted two sets of offset-encoding runs: one using the memristor's output current as the reservoir state (\textbf{Supplementary Fig. S23a}), monotonic input-state correlation, and another using the memristor's conductance (\textbf{Supplementary Fig. S23b}), non-monotonic input-state correlation, to observe the significance of input-state non-monotonicity when deploying the offsets-encoding approach. As observed in \textbf{Supplementary Fig. S23a}, when using a monotonic function (current), varying the offset range has almost no effect on the reservoir's performance as the PE does not vary when varying the offset range (across the vertical axis). Conversely, the PE varies significantly when a non-monotonic function (conductance) is used as the reservoir state. In addition, the testing PEs obtained using current as a reservoir state are, on average, larger than those obtained when using conductance as a reservoir state. We also observe across the entire 2-D map in \textbf{Supplementary Fig. S223b} that the PE achieved using offsets-encoding, in most cases, is lower than that obtained using pulse-widths encoding ($2.227\times10^{-3}$ as in \textbf{Supplementary Fig. S23d}), with the lowest PE being $1.087\times10^{-3}$ corresponding to a pulse width of 20 ms and an offset range of $\pm 200$ mV (\textbf{Supplementary Fig. S23c}). With these results, we highlight the success of implementing the offsets-encoding method. Additionally, by comparing the PEs achieved using current and conductance as reservoir states, we remark on the importance of using reservoir states that have a non-monotonic relationship with the input, as using offsets-encoding along with non-monotonic correlations appears to maximize the reservoir's dimensionality. 

\newpage

\subsection*{Predicting SONDS with External offsets using Armendarez \textit{et al.} memristor}\label{SubsectionS4.2}
\begin{figure}[h]
    \centering
    \includegraphics[width=4.5 in]{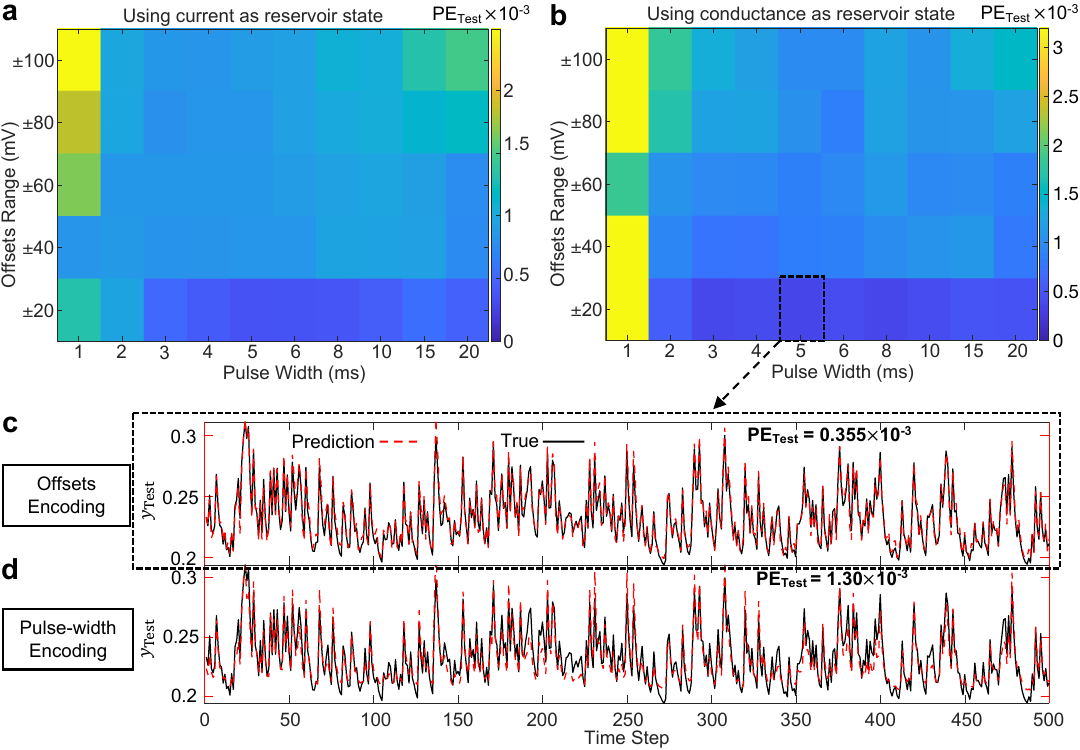}
    \caption{SONDS test data set prediction results using the newly introduced externally applied offsets-encoding method on Armendarez's memristor model \cite{Armendarez2024Brain-InspiredPlasticity}. \textbf{a} Testing prediction error as a function of pulse width and offset range using memristor's current as the reservoir state (Eq. \ref{ArmendarezIModel}). Only one pulse width and offset range were used for each run. \textbf{b} Testing prediction error as a pulse width and offset range function using memristor's conductance as the reservoir state ($G = I/v$). \textbf{c} Sample 500 points of the reservoir's predicted output vs the true SONDS output ($\mathrm{y_{test}}$) for the lowest PE achieved using the offsets-encoding method with conductance as the reservoir state. The lowest test PE achieved is $0.355\times10^{-3}$ \textbf{d} Sample 500 points of the reservoir's predicted output vs the true SONDS output ($\mathrm{y_{test}}$) for the PE achieved using the pulse-width encoding method with conductance as the reservoir state. The test PE achieved is $1.30\times10^{-3}$.}
    \label{FigureS24}
\end{figure}

\noindent We modeled Armendarez's memristor using these equations as in \cite{Armendarez2024Brain-InspiredPlasticity}:
\begin{equation} 
\frac{dN}{dt} = \frac{1}{\tau(v_m)}(N_0e^{\frac{\mid v_m\mid}{v_{e}}} - N)
\end{equation}
\begin{equation} 
I = g_uA_0Nv_m
\end{equation}\label{ArmendarezIModel}

\noindent where the voltage-dependent time constant $\tau(v_m) = \tau_0e^{\frac{\mid v_m\mid}{v_{\tau}}}$ and the values of the parameters $\tau_0$, $v_{\tau}$, $N_0$, $v_{e}$, $g_u$, and $A_0$ can be found in \cite{Armendarez2024Brain-InspiredPlasticity}. Similar to the previous subsection, we predicted SONDS once using pulse-width encoding with 11 different pulse widths (0.5, 1, 2, 5, 8, 10, 15, 20, 25, 40, and 50 ms) and once using one pulse-width along with 11 different external voltage offsets with 5 different ranges as shown in \textbf{Supplementary Fig. S24a-b} ($\pm 20$, $\pm 40$, $\pm 60$, $\pm 80$, and $\pm 100 mV$). In both cases, the random input was encoded to voltage pulses with amplitudes between 100 mV and 160 mV and 90\% duty cycle. Similar to the previous section, we have solved SONDS using current \textbf{Supplementary Fig. S24a} as the reservoir state and once using conductance as the reservoir state \textbf{Supplementary Fig. S24b}. Unlike predicting SONDS with Du's model in which using the current as a reservoir state appeared to be insensitive to the offset-encoding range (\textbf{Supplementary Fig. S23a}), using current as the reservoir state with Armendarez's memristor appears to be a function of the offset-encoding range as shown in \textbf{Supplementary Fig. S24a}. Since both Du's and Armendarez's memristor are modeled using state-linear first-order differential equations (only input-nonlinear), we attribute the offset-encoding-range sensitivity to the time constant's voltage dependence ($\tau = \tau(v_m)$). The memristor's current responses become slower with larger voltages, thus varying the PE with larger voltage offsets, unlike Du's memristor where $\tau$ (Eq. \ref{DusModel}) is a constant. As shown in \textbf{Supplementary Fig. S24a-b}, using either the current or the conductance as the reservoir state, predicting SONDS with offsets-encoding, in most cases, produces PEs that are either comparable to or lower than these obtained from pulse-width encoding (\textbf{Supplementary Fig. S24d}) with the best being $0.355\times10^{-3}$ (\textbf{Supplementary Fig. S24c}). This proves that offsets-encoding achieves marked prediction accuracies for voltage-sensitive (Armendarez's memristor) and voltage-insensitive (Du's memristor) temporal dynamics.

\newpage
\subsection*{Predicting Hénon map with External offsets using Du \textit{et al.} memristor and Armendarez \textit{et al.} memristor models}

\begin{figure}[h]
    \centering
    \includegraphics[width=4.5 in]{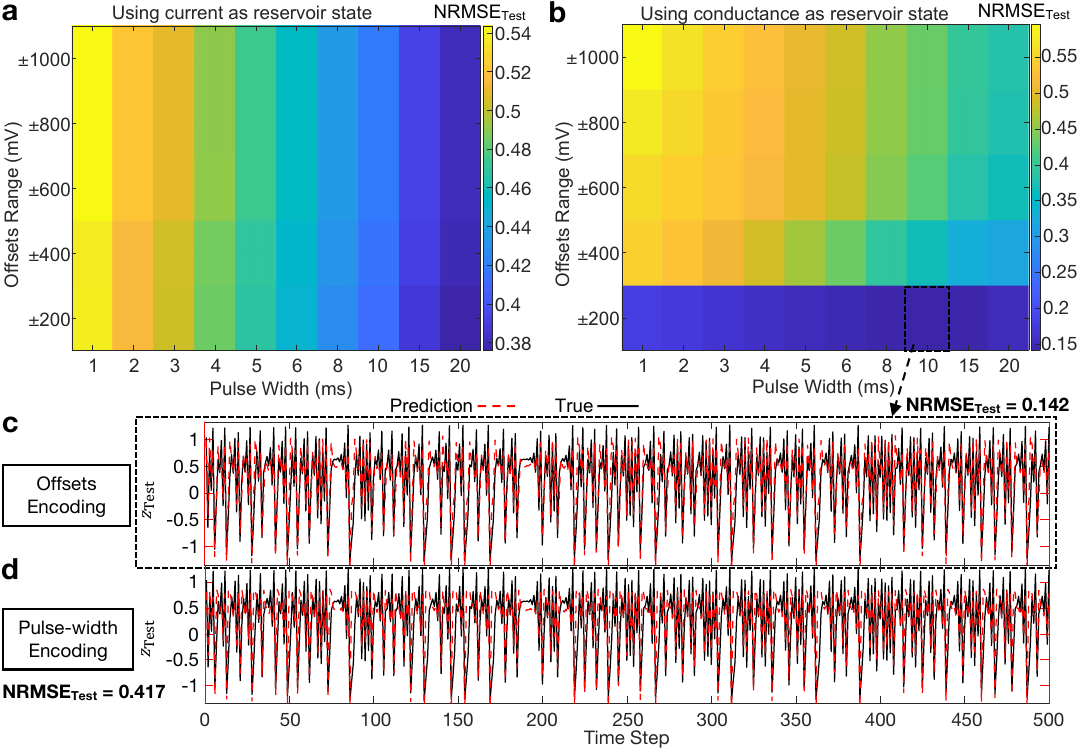}
    \caption{Hénon map test data set prediction results using the newly introduced externally applied offsets-encoding method on Du's memristor model \cite{Du2017ReservoirProcessing}. \textbf{a} Testing NRMSE as a function of pulse width and offset range using memristor's current as the reservoir state (Eq. \ref{DusIModel}). Only one pulse width and offset range were used for each run. \textbf{b} Testing NRMSE as a pulse width and offset range function using memristor's conductance as the reservoir state ($G = I/v$). \textbf{c} Sample 500 points of the reservoir's predicted output vs the true Hénon map output ($\mathrm{y_{test}}$) for the lowest NRMSE achieved using the offsets-encoding method with conductance as the reservoir state. The lowest test NRMSE achieved is $0.142$ \textbf{d} Sample 500 points of the reservoir's predicted output vs the true Hénon map output ($\mathrm{y_{test}}$) for the NRMSE achieved using the pulse-width encoding method with conductance as the reservoir state. The test NRMSE achieved is $0.417$.}
    \label{FigureS25}
\end{figure}
\noindent For predicting the Hénon map, we have followed the same procedure as outlined in the main text but using the models defined in the previous two sections. The same encoding and model parameters used for solving SONDs were replicated for solving the Hénon map. As shown in \textbf{Supplementary Fig. S25c} and \textbf{Supplementary Fig. S26c}, using pulse-width encoding results in a poor prediction of the Hénon map. On the other hand, for most cases, using either model, particularly when using the conductance as the reservoir state, offsets-encoding succeeds in achieving remarkable predictions of the Hénon map. This emphasizes the importance of incorporating non-monotonic input-state correlations, realized using offsets-encoding, for solving rapidly varying functions like the Hénon map. 
\begin{figure}[h]
    \centering
    \includegraphics[width=4.5 in]{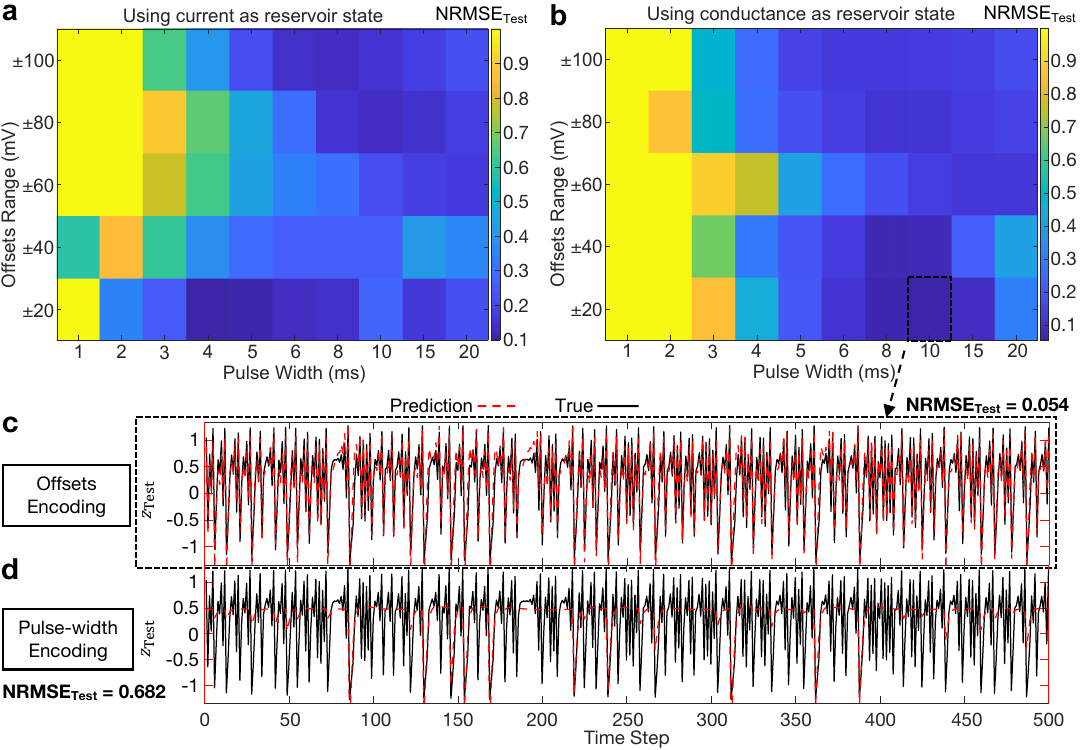}
    \caption{Hénon map test data set prediction results using the newly introduced externally applied offsets-encoding method on Armendarez's memristor model \cite{Armendarez2024Brain-InspiredPlasticity}. \textbf{a} Testing NRMSE as a function of pulse width and offset range using memristor's current as the reservoir state (Eq. \ref{ArmendarezIModel}). Only one pulse width and offset range were used for each run. \textbf{b} Testing NRMSE as a pulse width and offset range function using memristor's conductance as the reservoir state ($G = I/v$). \textbf{c} Sample 500 points of the reservoir's predicted output vs the true Hénon map output ($\mathrm{y_{test}}$) for the lowest NRMSE achieved using the offsets-encoding method with conductance as the reservoir state. The lowest test NRMSE achieved is $0.054$ \textbf{d} Sample 500 points of the reservoir's predicted output vs the true Hénon map output ($\mathrm{y_{test}}$) for the NRMSE achieved using the pulse-width encoding method with conductance as the reservoir state. The test NRMSE achieved is $0.682$.}
    \label{FigureS26}
\end{figure}
We also note the importance of using non-monotonic states, such as conductance, to maximize offsets-encoding benefit, as concluded from \textbf{Supplementary Fig. S25a-b}. With these results, we suggest, with no strict rules, that the offsets-encoding approach is most effective with devices that possess non-monotonicity input-state relationships (similar to that of memcapacitors C-V relationships in this work and conductance for Du's and Armendarez's memristor models). Additionally, we believe that devices possessing unrectified and non-switching, continuous input-state relationships (similar to that of the memcapacitors C-V relationships in this work) would be most effective for practical implementations of offsets encoding. This is because the success of offsets-encoding is majorly dependent on the span of states (i.e., the more independent states, the better the reservoir separability \cite{Appeltant2011InformationSystem}) that a device can realize within a given input range, which any rectifying effects may restrict. Switching behaviors may limit the span of available states since the device is only responsive to a limited input range and pose a practical limitation for measuring device states using one measuring device. For instance, Armendarez's switching memristor has conductance states varying from tens of $pS$ (single-channel recordings) \cite{Eisenberg1973TheMembranes} which can only be measured using patch-clamp amplifiers to tens of $\mu S$ (macroscopic conductance) which require the use of trans-impedance amplifiers to record \cite{Armendarez2024Brain-InspiredPlasticity}. To conclude, besides any physical limitations on the devices' input or output ranges, we believe that input-state 1) non-monotonicity and 2) continuity are the main device requirements for optimum use of the outlined offsets-encoding method. 
\clearpage


\begin{thebibliography}{80}
\ifx \bisbn   \undefined \def \bisbn  #1{ISBN #1}\fi
\ifx \binits  \undefined \def \binits#1{#1}\fi
\ifx \bauthor  \undefined \def \bauthor#1{#1}\fi
\ifx \batitle  \undefined \def \batitle#1{#1}\fi
\ifx \bjtitle  \undefined \def \bjtitle#1{#1}\fi
\ifx \bvolume  \undefined \def \bvolume#1{\textbf{#1}}\fi
\ifx \byear  \undefined \def \byear#1{#1}\fi
\ifx \bissue  \undefined \def \bissue#1{#1}\fi
\ifx \bfpage  \undefined \def \bfpage#1{#1}\fi
\ifx \blpage  \undefined \def \blpage #1{#1}\fi
\ifx \burl  \undefined \def \burl#1{\textsf{#1}}\fi
\ifx \doiurl  \undefined \def \doiurl#1{\url{https://doi.org/#1}}\fi
\ifx \betal  \undefined \def \betal{\textit{et al.}}\fi
\ifx \binstitute  \undefined \def \binstitute#1{#1}\fi
\ifx \binstitutionaled  \undefined \def \binstitutionaled#1{#1}\fi
\ifx \bctitle  \undefined \def \bctitle#1{#1}\fi
\ifx \beditor  \undefined \def \beditor#1{#1}\fi
\ifx \bpublisher  \undefined \def \bpublisher#1{#1}\fi
\ifx \bbtitle  \undefined \def \bbtitle#1{#1}\fi
\ifx \bedition  \undefined \def \bedition#1{#1}\fi
\ifx \bseriesno  \undefined \def \bseriesno#1{#1}\fi
\ifx \blocation  \undefined \def \blocation#1{#1}\fi
\ifx \bsertitle  \undefined \def \bsertitle#1{#1}\fi
\ifx \bsnm \undefined \def \bsnm#1{#1}\fi
\ifx \bsuffix \undefined \def \bsuffix#1{#1}\fi
\ifx \bparticle \undefined \def \bparticle#1{#1}\fi
\ifx \barticle \undefined \def \barticle#1{#1}\fi
\bibcommenthead
\ifx \bconfdate \undefined \def \bconfdate #1{#1}\fi
\ifx \botherref \undefined \def \botherref #1{#1}\fi
\ifx \url \undefined \def \url#1{\textsf{#1}\fi
\ifx \bchapter \undefined \def \bchapter#1{#1}\fi
\ifx \bbook \undefined \def \bbook#1{#1}\fi
\ifx \bcomment \undefined \def \bcomment#1{#1}\fi
\ifx \oauthor \undefined \def \oauthor#1{#1}\fi
\ifx \citeauthoryear \undefined \def \citeauthoryear#1{#1}\fi
\ifx \endbibitem  \undefined \def \endbibitem {}\fi
\ifx \bconflocation  \undefined \def \bconflocation#1{#1}\fi
\ifx \arxivurl  \undefined \def \arxivurl#1{\textsf{#1}}\fi
\csname PreBibitemsHook\endcsname

\bibitem[\protect\citeauthoryear{Verstraeten et~al.}{2006}]{Verstraeten2006Reservoir-basedRecognition}
\begin{botherref}
\oauthor{\bsnm{Verstraeten}, \binits{D.}},
\oauthor{\bsnm{Schrauwen}, \binits{B.}},
\oauthor{\bsnm{Stroobandt}, \binits{D.}}:
{Reservoir-based techniques for speech recognition}.
IEEE International Conference on Neural Networks - Conference Proceedings,
1050--1053
(2006)
\doiurl{10.1109/IJCNN.2006.246804}
\end{botherref}
\endbibitem

\bibitem[\protect\citeauthoryear{Gauthier et~al.}{2021}]{Gauthier2021NextComputing}
\begin{barticle}
\bauthor{\bsnm{Gauthier}, \binits{D.J.}},
\bauthor{\bsnm{Bollt}, \binits{E.}},
\bauthor{\bsnm{Griffith}, \binits{A.}},
\bauthor{\bsnm{Barbosa}, \binits{W.A.S.}}:
\batitle{{Next generation reservoir computing}}.
\bjtitle{Nature Communications 2021 12:1}
\bvolume{12}(\bissue{1}),
\bfpage{1}--\blpage{8}
(\byear{2021})
\doiurl{10.1038/s41467-021-25801-2}
\end{barticle}
\endbibitem

\bibitem[\protect\citeauthoryear{Zhong et~al.}{2021}]{Zhong2021DynamicProcessing}
\begin{barticle}
\bauthor{\bsnm{Zhong}, \binits{Y.}},
\bauthor{\bsnm{Tang}, \binits{J.}},
\bauthor{\bsnm{Li}, \binits{X.}},
\bauthor{\bsnm{Gao}, \binits{B.}},
\bauthor{\bsnm{Qian}, \binits{H.}},
\bauthor{\bsnm{Wu}, \binits{H.}}:
\batitle{{Dynamic memristor-based reservoir computing for high-efficiency temporal signal processing}}.
\bjtitle{Nature Communications 2021 12:1}
\bvolume{12}(\bissue{1}),
\bfpage{1}--\blpage{9}
(\byear{2021})
\doiurl{10.1038/s41467-020-20692-1}
\end{barticle}
\endbibitem

\bibitem[\protect\citeauthoryear{Cucchi et~al.}{2022}]{Cucchi2022Hands-onImplementation}
\begin{barticle}
\bauthor{\bsnm{Cucchi}, \binits{M.}},
\bauthor{\bsnm{Abreu}, \binits{S.}},
\bauthor{\bsnm{Ciccone}, \binits{G.}},
\bauthor{\bsnm{Brunner}, \binits{D.}},
\bauthor{\bsnm{Kleemann}, \binits{H.}}:
\batitle{{Hands-on reservoir computing: a tutorial for practical implementation}}.
\bjtitle{Neuromorphic Computing and Engineering}
\bvolume{2}(\bissue{3}),
\bfpage{032002}
(\byear{2022})
\doiurl{10.1088/2634-4386/AC7DB7}
\end{barticle}
\endbibitem

\bibitem[\protect\citeauthoryear{Nakajima}{2020}]{Nakajima2020PhysicalPerspective}
\begin{barticle}
\bauthor{\bsnm{Nakajima}, \binits{K.}}:
\batitle{{Physical reservoir computing—an introductory perspective}}.
\bjtitle{Japanese Journal of Applied Physics}
\bvolume{59}(\bissue{6}),
\bfpage{060501}
(\byear{2020})
\doiurl{10.35848/1347-4065/AB8D4F}
\end{barticle}
\endbibitem

\bibitem[\protect\citeauthoryear{Tanaka et~al.}{2019}]{Tanaka2019RecentReview}
\begin{barticle}
\bauthor{\bsnm{Tanaka}, \binits{G.}},
\bauthor{\bsnm{Yamane}, \binits{T.}},
\bauthor{\bsnm{H{\'{e}}roux}, \binits{J.B.}},
\bauthor{\bsnm{Nakane}, \binits{R.}},
\bauthor{\bsnm{Kanazawa}, \binits{N.}},
\bauthor{\bsnm{Takeda}, \binits{S.}},
\bauthor{\bsnm{Numata}, \binits{H.}},
\bauthor{\bsnm{Nakano}, \binits{D.}},
\bauthor{\bsnm{Hirose}, \binits{A.}}:
\batitle{{Recent advances in physical reservoir computing: A review}}.
\bjtitle{Neural Networks}
\bvolume{115},
\bfpage{100}--\blpage{123}
(\byear{2019})
\doiurl{10.1016/J.NEUNET.2019.03.005}
\end{barticle}
\endbibitem

\bibitem[\protect\citeauthoryear{Schrauwen et~al.}{2007}]{Schrauwen2007AnImplementations}
\begin{botherref}
\oauthor{\bsnm{Schrauwen}, \binits{B.}},
\oauthor{\bsnm{Verstraeten}, \binits{D.}},
\oauthor{\bsnm{Campenhout}, \binits{J.V.}}:
{An overview of reservoir computing: theory, applications and implementations}.
The European Symposium on Artificial Neural Networks
(2007)
\end{botherref}
\endbibitem

\bibitem[\protect\citeauthoryear{Luko{\v{s}}evi{\v{c}}ius and Jaeger}{2009}]{Lukosevicius2009ReservoirTraining}
\begin{barticle}
\bauthor{\bsnm{Luko{\v{s}}evi{\v{c}}ius}, \binits{M.}},
\bauthor{\bsnm{Jaeger}, \binits{H.}}:
\batitle{{Reservoir computing approaches to recurrent neural network training}}.
\bjtitle{Computer Science Review}
\bvolume{3}(\bissue{3}),
\bfpage{127}--\blpage{149}
(\byear{2009})
\doiurl{10.1016/J.COSREV.2009.03.005}
\end{barticle}
\endbibitem

\bibitem[\protect\citeauthoryear{Appeltant et~al.}{2011}]{Appeltant2011InformationSystem}
\begin{barticle}
\bauthor{\bsnm{Appeltant}, \binits{L.}},
\bauthor{\bsnm{Soriano}, \binits{M.C.}},
\bauthor{\bsnm{Van Der~Sande}, \binits{G.}},
\bauthor{\bsnm{Danckaert}, \binits{J.}},
\bauthor{\bsnm{Massar}, \binits{S.}},
\bauthor{\bsnm{Dambre}, \binits{J.}},
\bauthor{\bsnm{Schrauwen}, \binits{B.}},
\bauthor{\bsnm{Mirasso}, \binits{C.R.}},
\bauthor{\bsnm{Fischer}, \binits{I.}}:
\batitle{{Information processing using a single dynamical node as complex system}}.
\bjtitle{Nature Communications 2011 2:1}
\bvolume{2}(\bissue{1}),
\bfpage{1}--\blpage{6}
(\byear{2011})
\doiurl{10.1038/ncomms1476}
\end{barticle}
\endbibitem

\bibitem[\protect\citeauthoryear{Misra and Saha}{2010}]{Misra2010ArtificialProgress}
\begin{barticle}
\bauthor{\bsnm{Misra}, \binits{J.}},
\bauthor{\bsnm{Saha}, \binits{I.}}:
\batitle{{Artificial neural networks in hardware: A survey of two decades of progress}}.
\bjtitle{Neurocomputing}
\bvolume{74}(\bissue{1-3}),
\bfpage{239}--\blpage{255}
(\byear{2010})
\doiurl{10.1016/J.NEUCOM.2010.03.021}
\end{barticle}
\endbibitem

\bibitem[\protect\citeauthoryear{Torrejon et~al.}{2017}]{Torrejon2017NeuromorphicOscillators}
\begin{barticle}
\bauthor{\bsnm{Torrejon}, \binits{J.}},
\bauthor{\bsnm{Riou}, \binits{M.}},
\bauthor{\bsnm{Araujo}, \binits{F.A.}},
\bauthor{\bsnm{Tsunegi}, \binits{S.}},
\bauthor{\bsnm{Khalsa}, \binits{G.}},
\bauthor{\bsnm{Querlioz}, \binits{D.}},
\bauthor{\bsnm{Bortolotti}, \binits{P.}},
\bauthor{\bsnm{Cros}, \binits{V.}},
\bauthor{\bsnm{Yakushiji}, \binits{K.}},
\bauthor{\bsnm{Fukushima}, \binits{A.}},
\bauthor{\bsnm{Kubota}, \binits{H.}},
\bauthor{\bsnm{Yuasa}, \binits{S.}},
\bauthor{\bsnm{Stiles}, \binits{M.D.}},
\bauthor{\bsnm{Grollier}, \binits{J.}}:
\batitle{{Neuromorphic computing with nanoscale spintronic oscillators}}.
\bjtitle{Nature 2017 547:7664}
\bvolume{547}(\bissue{7664}),
\bfpage{428}--\blpage{431}
(\byear{2017})
\doiurl{10.1038/nature23011}
\end{barticle}
\endbibitem

\bibitem[\protect\citeauthoryear{Sillin et~al.}{2013}]{Sillin2013AComputing}
\begin{barticle}
\bauthor{\bsnm{Sillin}, \binits{H.O.}},
\bauthor{\bsnm{Aguilera}, \binits{R.}},
\bauthor{\bsnm{Shieh}, \binits{H.H.}},
\bauthor{\bsnm{Avizienis}, \binits{A.V.}},
\bauthor{\bsnm{Aono}, \binits{M.}},
\bauthor{\bsnm{Stieg}, \binits{A.Z.}},
\bauthor{\bsnm{Gimzewski}, \binits{J.K.}}:
\batitle{{A theoretical and experimental study of neuromorphic atomic switch networks for reservoir computing}}.
\bjtitle{Nanotechnology}
\bvolume{24}(\bissue{38}),
\bfpage{384004}
(\byear{2013})
\doiurl{10.1088/0957-4484/24/38/384004}
\end{barticle}
\endbibitem

\bibitem[\protect\citeauthoryear{Vandoorne et~al.}{2014}]{Vandoorne2014ExperimentalChip}
\begin{barticle}
\bauthor{\bsnm{Vandoorne}, \binits{K.}},
\bauthor{\bsnm{Mechet}, \binits{P.}},
\bauthor{\bsnm{Van~Vaerenbergh}, \binits{T.}},
\bauthor{\bsnm{Fiers}, \binits{M.}},
\bauthor{\bsnm{Morthier}, \binits{G.}},
\bauthor{\bsnm{Verstraeten}, \binits{D.}},
\bauthor{\bsnm{Schrauwen}, \binits{B.}},
\bauthor{\bsnm{Dambre}, \binits{J.}},
\bauthor{\bsnm{Bienstman}, \binits{P.}}:
\batitle{{Experimental demonstration of reservoir computing on a silicon photonics chip}}.
\bjtitle{Nature Communications 2014 5:1}
\bvolume{5}(\bissue{1}),
\bfpage{1}--\blpage{6}
(\byear{2014})
\doiurl{10.1038/ncomms4541}
\end{barticle}
\endbibitem

\bibitem[\protect\citeauthoryear{Van Der~Sande et~al.}{2017}]{VanDerSande2017AdvancesComputing}
\begin{barticle}
\bauthor{\bsnm{Van Der~Sande}, \binits{G.}},
\bauthor{\bsnm{Brunner}, \binits{D.}},
\bauthor{\bsnm{Soriano}, \binits{M.C.}}:
\batitle{{Advances in photonic reservoir computing}}.
\bjtitle{Nanophotonics}
\bvolume{6}(\bissue{3}),
\bfpage{561}--\blpage{576}
(\byear{2017})
\doiurl{10.1515/NANOPH-2016-0132/ASSET/GRAPHIC/J{\_}NANOPH-2016-0132{\_}FIG{\}}}
\end{barticle}
\endbibitem

\bibitem[\protect\citeauthoryear{Toprasertpong et~al.}{2022}]{Toprasertpong2022ReservoirTransistor}
\begin{barticle}
\bauthor{\bsnm{Toprasertpong}, \binits{K.}},
\bauthor{\bsnm{Nako}, \binits{E.}},
\bauthor{\bsnm{Wang}, \binits{Z.}},
\bauthor{\bsnm{Nakane}, \binits{R.}},
\bauthor{\bsnm{Takenaka}, \binits{M.}},
\bauthor{\bsnm{Takagi}, \binits{S.}}:
\batitle{{Reservoir computing on a silicon platform with a ferroelectric field-effect transistor}}.
\bjtitle{Communications Engineering 2022 1:1}
\bvolume{1}(\bissue{1}),
\bfpage{1}--\blpage{9}
(\byear{2022})
\doiurl{10.1038/s44172-022-00021-8}
\end{barticle}
\endbibitem

\bibitem[\protect\citeauthoryear{Duong et~al.}{2023}]{Duong2023DynamicProcessing}
\begin{botherref}
\oauthor{\bsnm{Duong}, \binits{N.T.}},
\oauthor{\bsnm{Chien}, \binits{Y.-C.}},
\oauthor{\bsnm{Xiang}, \binits{H.}},
\oauthor{\bsnm{Li}, \binits{S.}},
\oauthor{\bsnm{Zheng}, \binits{H.}},
\oauthor{\bsnm{Shi}, \binits{Y.}},
\oauthor{\bsnm{Ang}, \binits{K.-W.}}:
{Dynamic Ferroelectric Transistor‐Based Reservoir Computing for Spatiotemporal Information Processing}.
Advanced Intelligent Systems,
2300009
(2023)
\doiurl{10.1002/AISY.202300009}
\end{botherref}
\endbibitem

\bibitem[\protect\citeauthoryear{Du et~al.}{2017}]{Du2017ReservoirProcessing}
\begin{barticle}
\bauthor{\bsnm{Du}, \binits{C.}},
\bauthor{\bsnm{Cai}, \binits{F.}},
\bauthor{\bsnm{Zidan}, \binits{M.A.}},
\bauthor{\bsnm{Ma}, \binits{W.}},
\bauthor{\bsnm{Lee}, \binits{S.H.}},
\bauthor{\bsnm{Lu}, \binits{W.D.}}:
\batitle{{Reservoir computing using dynamic memristors for temporal information processing}}.
\bjtitle{Nature Communications 2017 8:1}
\bvolume{8}(\bissue{1}),
\bfpage{1}--\blpage{10}
(\byear{2017})
\doiurl{10.1038/s41467-017-02337-y}
\end{barticle}
\endbibitem

\bibitem[\protect\citeauthoryear{Zhu et~al.}{2020}]{Zhu2020MemristorAnalysis}
\begin{botherref}
\oauthor{\bsnm{Zhu}, \binits{X.}},
\oauthor{\bsnm{Wang}, \binits{Q.}},
\oauthor{\bsnm{Lu}, \binits{W.D.}}:
{Memristor networks for real-time neural activity analysis}.
Nature Communications
\textbf{11}(1)
(2020)
\doiurl{10.1038/S41467-020-16261-1}
\end{botherref}
\endbibitem

\bibitem[\protect\citeauthoryear{Midya et~al.}{2019}]{Midya2019ReservoirMemristors}
\begin{barticle}
\bauthor{\bsnm{Midya}, \binits{R.}},
\bauthor{\bsnm{Wang}, \binits{Z.}},
\bauthor{\bsnm{Asapu}, \binits{S.}},
\bauthor{\bsnm{Zhang}, \binits{X.}},
\bauthor{\bsnm{Rao}, \binits{M.}},
\bauthor{\bsnm{Song}, \binits{W.}},
\bauthor{\bsnm{Zhuo}, \binits{Y.}},
\bauthor{\bsnm{Upadhyay}, \binits{N.}},
\bauthor{\bsnm{Xia}, \binits{Q.}},
\bauthor{\bsnm{Joshua~Yang}, \binits{J.}},
\bauthor{\bsnm{Midya}, \binits{R.}},
\bauthor{\bsnm{Wang}, \binits{Z.}},
\bauthor{\bsnm{Asapu}, \binits{S.}},
\bauthor{\bsnm{Zhang}, \binits{X.}},
\bauthor{\bsnm{Rao}, \binits{M.}},
\bauthor{\bsnm{Song}, \binits{W.}},
\bauthor{\bsnm{Zhuo}, \binits{Y.}},
\bauthor{\bsnm{Upadhyay}, \binits{N.}},
\bauthor{\bsnm{Xia}, \binits{Q.}},
\bauthor{\bsnm{Yang}, \binits{J.J.}}:
\batitle{{Reservoir Computing Using Diffusive Memristors}}.
\bjtitle{Advanced Intelligent Systems}
\bvolume{1}(\bissue{7}),
\bfpage{1900084}
(\byear{2019})
\doiurl{10.1002/AISY.201900084}
\end{barticle}
\endbibitem

\bibitem[\protect\citeauthoryear{Hossain et~al.}{2023}]{Hossain2023Biomembrane-BasedProcessing}
\begin{barticle}
\bauthor{\bsnm{Hossain}, \binits{M.R.}},
\bauthor{\bsnm{Mohamed}, \binits{A.S.}},
\bauthor{\bsnm{Armendarez}, \binits{N.X.}},
\bauthor{\bsnm{Najem}, \binits{J.S.}},
\bauthor{\bsnm{Hasan}, \binits{M.S.}}:
\batitle{{Biomembrane-Based Memcapacitive Reservoir Computing System for Energy-Efficient Temporal Data Processing}}.
\bjtitle{Advanced Intelligent Systems}
\bvolume{5}(\bissue{12}),
\bfpage{2300346}
(\byear{2023})
\doiurl{10.1002/AISY.202300346}
\end{barticle}
\endbibitem

\bibitem[\protect\citeauthoryear{Di~Ventra et~al.}{2009}]{DiVentra2009CircuitMeminductors}
\begin{barticle}
\bauthor{\bsnm{Di~Ventra}, \binits{M.}},
\bauthor{\bsnm{Pershin}, \binits{Y.V.}},
\bauthor{\bsnm{Chua}, \binits{L.O.}}:
\batitle{{Circuit elements with memory: memristors, memcapacitors and meminductors}}.
\bjtitle{Proceedings of the IEEE}
\bvolume{97}(\bissue{10}),
\bfpage{1717}--\blpage{1724}
(\byear{2009})
\doiurl{10.1109/JPROC.2009.2021077}
\end{barticle}
\endbibitem

\bibitem[\protect\citeauthoryear{Liu et~al.}{2021}]{Liu2021TheEmulator}
\begin{barticle}
\bauthor{\bsnm{Liu}, \binits{Y.}},
\bauthor{\bsnm{Iu}, \binits{H.H.C.}},
\bauthor{\bsnm{Guo}, \binits{Z.}},
\bauthor{\bsnm{Si}, \binits{G.}}:
\batitle{{The Simple Charge-Controlled Grounded/Floating Mem-Element Emulator}}.
\bjtitle{IEEE Transactions on Circuits and Systems II: Express Briefs}
\bvolume{68}(\bissue{6}),
\bfpage{2177}--\blpage{2181}
(\byear{2021})
\doiurl{10.1109/TCSII.2020.3041862}
\end{barticle}
\endbibitem

\bibitem[\protect\citeauthoryear{Chua}{1971}]{Chua1971MemristorTheElement}
\begin{barticle}
\bauthor{\bsnm{Chua}, \binits{L.O.}}:
\batitle{{Memristor—The Missing Circuit Element}}.
\bjtitle{IEEE Transactions on Circuit Theory}
\bvolume{18}(\bissue{5}),
\bfpage{507}--\blpage{519}
(\byear{1971})
\doiurl{10.1109/TCT.1971.1083337}
\end{barticle}
\endbibitem

\bibitem[\protect\citeauthoryear{Strukov et~al.}{2008}]{Strukov2008TheFound}
\begin{barticle}
\bauthor{\bsnm{Strukov}, \binits{D.B.}},
\bauthor{\bsnm{Snider}, \binits{G.S.}},
\bauthor{\bsnm{Stewart}, \binits{D.R.}},
\bauthor{\bsnm{Williams}, \binits{R.S.}}:
\batitle{{The missing memristor found}}.
\bjtitle{Nature 2008 453:7191}
\bvolume{453}(\bissue{7191}),
\bfpage{80}--\blpage{83}
(\byear{2008})
\doiurl{10.1038/NATURE06932}
\end{barticle}
\endbibitem

\bibitem[\protect\citeauthoryear{Najem et~al.}{2019}]{Najem2019DynamicalMembranes}
\begin{barticle}
\bauthor{\bsnm{Najem}, \binits{J.S.}},
\bauthor{\bsnm{Hasan}, \binits{M.S.}},
\bauthor{\bsnm{Williams}, \binits{R.S.}},
\bauthor{\bsnm{Weiss}, \binits{R.J.}},
\bauthor{\bsnm{Rose}, \binits{G.S.}},
\bauthor{\bsnm{Taylor}, \binits{G.J.}},
\bauthor{\bsnm{Sarles}, \binits{S.A.}},
\bauthor{\bsnm{Collier}, \binits{C.P.}}:
\batitle{{Dynamical nonlinear memory capacitance in biomimetic membranes}}.
\bjtitle{Nature Communications 2019 10:1}
\bvolume{10}(\bissue{1}),
\bfpage{1}--\blpage{11}
(\byear{2019})
\doiurl{10.1038/s41467-019-11223-8}
\end{barticle}
\endbibitem

\bibitem[\protect\citeauthoryear{Dinavahi et~al.}{2023}]{Dinavahi2023PhysicalElement}
\begin{barticle}
\bauthor{\bsnm{Dinavahi}, \binits{A.}},
\bauthor{\bsnm{Yamamoto}, \binits{A.}},
\bauthor{\bsnm{Harris}, \binits{H.R.}}:
\batitle{{Physical evidence of meminductance in a passive, two-terminal circuit element}}.
\bjtitle{Scientific Reports 2023 13:1}
\bvolume{13}(\bissue{1}),
\bfpage{1}--\blpage{10}
(\byear{2023})
\doiurl{10.1038/S41598-022-24914-Y}
\end{barticle}
\endbibitem

\bibitem[\protect\citeauthoryear{Biolek et~al.}{2016}]{Biolek2016EveryTheorem}
\begin{barticle}
\bauthor{\bsnm{Biolek}, \binits{D.}},
\bauthor{\bsnm{Biolek}, \binits{Z.}},
\bauthor{\bsnm{Biolkova}, \binits{V.}}:
\batitle{{Every nonlinear element from Chua's table can generate pinched hysteresis loops: generalised homothety theorem}}.
\bjtitle{Electronics Letters}
\bvolume{52}(\bissue{21}),
\bfpage{1744}--\blpage{1746}
(\byear{2016})
\doiurl{10.1049/EL.2016.2961}
\end{barticle}
\endbibitem

\bibitem[\protect\citeauthoryear{Abdelouahab et~al.}{2014}]{Abdelouahab2014Memfractance:Memory}
\begin{botherref}
\oauthor{\bsnm{Abdelouahab}, \binits{M.S.}},
\oauthor{\bsnm{Lozi}, \binits{R.}},
\oauthor{\bsnm{Chua}, \binits{L.}}:
{Memfractance: A Mathematical Paradigm for Circuit Elements with Memory}.
https://doi.org/10.1142/S0218127414300237
\textbf{24}(9)
(2014)
\doiurl{10.1142/S0218127414300237}
\end{botherref}
\endbibitem

\bibitem[\protect\citeauthoryear{Dambre et~al.}{2012}]{Dambre2012InformationSystems}
\begin{barticle}
\bauthor{\bsnm{Dambre}, \binits{J.}},
\bauthor{\bsnm{Verstraeten}, \binits{D.}},
\bauthor{\bsnm{Schrauwen}, \binits{B.}},
\bauthor{\bsnm{Massar}, \binits{S.}}:
\batitle{{Information Processing Capacity of Dynamical Systems}}.
\bjtitle{Scientific Reports 2012 2:1}
\bvolume{2}(\bissue{1}),
\bfpage{1}--\blpage{7}
(\byear{2012})
\doiurl{10.1038/SREP00514}
\end{barticle}
\endbibitem

\bibitem[\protect\citeauthoryear{Nishioka et~al.}{2022}]{Nishioka2022Edge-of-chaosReservoir}
\begin{botherref}
\oauthor{\bsnm{Nishioka}, \binits{D.}},
\oauthor{\bsnm{Tsuchiya}, \binits{T.}},
\oauthor{\bsnm{Namiki}, \binits{W.}},
\oauthor{\bsnm{Takayanagi}, \binits{M.}},
\oauthor{\bsnm{Imura}, \binits{M.}},
\oauthor{\bsnm{Koide}, \binits{Y.}},
\oauthor{\bsnm{Higuchi}, \binits{T.}},
\oauthor{\bsnm{Terabe}, \binits{K.}}:
{Edge-of-chaos learning achieved by ion-electron- coupled dynamics in an ion-gating reservoir}.
Science Advances
\textbf{8}(50)
(2022)
\doiurl{10.1126/SCIADV.ADE1156/SUPPL{\_}FILE/SCIADV.ADE1156{\_}MOVIE{}}
\end{botherref}
\endbibitem

\bibitem[\protect\citeauthoryear{Ghenzi et~al.}{2024}]{Ghenzi2024HeterogeneousMemristors}
\begin{barticle}
\bauthor{\bsnm{Ghenzi}, \binits{N.}},
\bauthor{\bsnm{Park}, \binits{T.W.}},
\bauthor{\bsnm{Kim}, \binits{S.S.}},
\bauthor{\bsnm{Kim}, \binits{H.J.}},
\bauthor{\bsnm{Jang}, \binits{Y.H.}},
\bauthor{\bsnm{Woo}, \binits{K.S.}},
\bauthor{\bsnm{Hwang}, \binits{C.S.}}:
\batitle{{Heterogeneous reservoir computing in second-order Ta 2 O 5 /HfO 2 memristors}}.
\bjtitle{Nanoscale Horizons}
(\byear{2024})
\doiurl{10.1039/D3NH00493G}
\end{barticle}
\endbibitem

\bibitem[\protect\citeauthoryear{Armendarez et~al.}{2024}]{Armendarez2024Brain-InspiredPlasticity}
\begin{barticle}
\bauthor{\bsnm{Armendarez}, \binits{N.X.}},
\bauthor{\bsnm{Mohamed}, \binits{A.S.}},
\bauthor{\bsnm{Dhungel}, \binits{A.}},
\bauthor{\bsnm{Hossain}, \binits{M.R.}},
\bauthor{\bsnm{Hasan}, \binits{M.S.}},
\bauthor{\bsnm{Najem}, \binits{J.S.}}:
\batitle{{Brain-Inspired Reservoir Computing Using Memristors with Tunable Dynamics and Short-Term Plasticity}}.
\bjtitle{ACS Applied Materials {\&} Interfaces}
(\byear{2024})
\doiurl{10.1021/ACSAMI.3C16003}
\end{barticle}
\endbibitem

\bibitem[\protect\citeauthoryear{Moon et~al.}{2019}]{Moon2019TemporalSystem}
\begin{barticle}
\bauthor{\bsnm{Moon}, \binits{J.}},
\bauthor{\bsnm{Ma}, \binits{W.}},
\bauthor{\bsnm{Shin}, \binits{J.H.}},
\bauthor{\bsnm{Cai}, \binits{F.}},
\bauthor{\bsnm{Du}, \binits{C.}},
\bauthor{\bsnm{Lee}, \binits{S.H.}},
\bauthor{\bsnm{Lu}, \binits{W.D.}}:
\batitle{{Temporal data classification and forecasting using a memristor-based reservoir computing system}}.
\bjtitle{Nature Electronics 2019 2:10}
\bvolume{2}(\bissue{10}),
\bfpage{480}--\blpage{487}
(\byear{2019})
\doiurl{10.1038/s41928-019-0313-3}
\end{barticle}
\endbibitem

\bibitem[\protect\citeauthoryear{Rueda-Orozco et~al.}{}]{Rueda-OrozcoBriefNeurons}
\begin{botherref}
\oauthor{\bsnm{Rueda-Orozco}, \binits{P.E.}},
\oauthor{\bsnm{Mendoza}, \binits{E.}},
\oauthor{\bsnm{Hernandez}, \binits{R.}},
\oauthor{\bsnm{Aceves}, \binits{J.J.}},
\oauthor{\bsnm{Ibanez-Sandoval}, \binits{O.}},
\oauthor{\bsnm{Galarraga}, \binits{E.}},
\oauthor{\bsnm{Bargas}, \binits{J.}}:
{Brief Communication Diversity in long-term synaptic plasticity at inhibitory synapses of striatal spiny neurons}
\doiurl{10.1101/lm.1439909}
\end{botherref}
\endbibitem

\bibitem[\protect\citeauthoryear{Debanne et~al.}{1996}]{Debanne1996Paired-pulseRelease}
\begin{barticle}
\bauthor{\bsnm{Debanne}, \binits{D.}},
\bauthor{\bsnm{Gu{\'{e}}rineau}, \binits{N.C.}},
\bauthor{\bsnm{G{\"{a}}hwiler}, \binits{B.H.}},
\bauthor{\bsnm{Thompson}, \binits{S.M.}}:
\batitle{{Paired-pulse facilitation and depression at unitary synapses in rat hippocampus: quantal fluctuation affects subsequent release}}.
\bjtitle{The Journal of physiology}
\bvolume{491 ( Pt 1)}(\bissue{Pt 1}),
\bfpage{163}--\blpage{176}
(\byear{1996})
\doiurl{10.1113/JPHYSIOL.1996.SP021204}
\end{barticle}
\endbibitem

\bibitem[\protect\citeauthoryear{Wu et~al.}{2024}]{Wu2024ACuInP2S6}
\begin{barticle}
\bauthor{\bsnm{Wu}, \binits{Y.}},
\bauthor{\bsnm{Duong}, \binits{N.T.}},
\bauthor{\bsnm{Chien}, \binits{Y.C.}},
\bauthor{\bsnm{Liu}, \binits{S.}},
\bauthor{\bsnm{Ang}, \binits{K.W.}}:
\batitle{{A Dynamic Memory for Reservoir Computing Utilizing Ion Migration in CuInP2S6}}.
\bjtitle{Advanced Electronic Materials}
\bvolume{10}(\bissue{1}),
\bfpage{2300481}
(\byear{2024})
\doiurl{10.1002/AELM.202300481}
\end{barticle}
\endbibitem

\bibitem[\protect\citeauthoryear{Pei et~al.}{2023}]{Pei2023Power-EfficientArrays}
\begin{barticle}
\bauthor{\bsnm{Pei}, \binits{M.}},
\bauthor{\bsnm{Zhu}, \binits{Y.}},
\bauthor{\bsnm{Liu}, \binits{S.}},
\bauthor{\bsnm{Cui}, \binits{H.}},
\bauthor{\bsnm{Li}, \binits{Y.}},
\bauthor{\bsnm{Yan}, \binits{Y.}},
\bauthor{\bsnm{Li}, \binits{Y.}},
\bauthor{\bsnm{Wan}, \binits{C.}},
\bauthor{\bsnm{Wan}, \binits{Q.}}:
\batitle{{Power-Efficient Multisensory Reservoir Computing Based on Zr-Doped HfO2 Memcapacitive Synapse Arrays}}.
\bjtitle{Advanced Materials}
\bvolume{35}(\bissue{41}),
\bfpage{2305609}
(\byear{2023})
\doiurl{10.1002/ADMA.202305609}
\end{barticle}
\endbibitem

\bibitem[\protect\citeauthoryear{Chen et~al.}{2023}]{Chen2023All-ferroelectricComputing}
\begin{barticle}
\bauthor{\bsnm{Chen}, \binits{Z.}},
\bauthor{\bsnm{Li}, \binits{W.}},
\bauthor{\bsnm{Fan}, \binits{Z.}},
\bauthor{\bsnm{Dong}, \binits{S.}},
\bauthor{\bsnm{Chen}, \binits{Y.}},
\bauthor{\bsnm{Qin}, \binits{M.}},
\bauthor{\bsnm{Zeng}, \binits{M.}},
\bauthor{\bsnm{Lu}, \binits{X.}},
\bauthor{\bsnm{Zhou}, \binits{G.}},
\bauthor{\bsnm{Gao}, \binits{X.}},
\bauthor{\bsnm{Liu}, \binits{J.M.}}:
\batitle{{All-ferroelectric implementation of reservoir computing}}.
\bjtitle{Nature Communications 2023 14:1}
\bvolume{14}(\bissue{1}),
\bfpage{1}--\blpage{12}
(\byear{2023})
\doiurl{10.1038/S41467-023-39371-Y}
\end{barticle}
\endbibitem

\bibitem[\protect\citeauthoryear{Fang et~al.}{2024}]{Fang2024Oxide-BasedComputing}
\begin{botherref}
\oauthor{\bsnm{Fang}, \binits{R.}},
\oauthor{\bsnm{Wang}, \binits{S.}},
\oauthor{\bsnm{Zhang}, \binits{W.}},
\oauthor{\bsnm{Ren}, \binits{K.}},
\oauthor{\bsnm{Sun}, \binits{W.}},
\oauthor{\bsnm{Wang}, \binits{F.}},
\oauthor{\bsnm{Lai}, \binits{J.}},
\oauthor{\bsnm{Zhang}, \binits{P.}},
\oauthor{\bsnm{Xu}, \binits{X.}},
\oauthor{\bsnm{Luo}, \binits{Q.}},
\oauthor{\bsnm{Li}, \binits{L.}},
\oauthor{\bsnm{Wang}, \binits{Z.}},
\oauthor{\bsnm{Shang}, \binits{D.}}:
{Oxide-Based Electrolyte-Gated Transistors with Stable and Tunable Relaxation Responses for Deep Time-Delayed Reservoir Computing}.
Advanced Electronic Materials,
2300652
(2024)
\doiurl{10.1002/AELM.202300652}
\end{botherref}
\endbibitem

\bibitem[\protect\citeauthoryear{Feng et~al.}{2023}]{Feng2023FullyStates}
\begin{barticle}
\bauthor{\bsnm{Feng}, \binits{Y.}},
\bauthor{\bsnm{Tang}, \binits{M.}},
\bauthor{\bsnm{Sun}, \binits{Z.}},
\bauthor{\bsnm{Qi}, \binits{Y.}},
\bauthor{\bsnm{Zhan}, \binits{X.}},
\bauthor{\bsnm{Liu}, \binits{J.}},
\bauthor{\bsnm{Zhang}, \binits{J.}},
\bauthor{\bsnm{Wu}, \binits{J.}},
\bauthor{\bsnm{Chen}, \binits{J.}}:
\batitle{{Fully Flash-Based Reservoir Computing Network With Low Power and Rich States}}.
\bjtitle{IEEE Transactions on Electron Devices}
\bvolume{70}(\bissue{9}),
\bfpage{4972}--\blpage{4975}
(\byear{2023})
\doiurl{10.1109/TED.2023.3295791}
\end{barticle}
\endbibitem

\bibitem[\protect\citeauthoryear{Taylor et~al.}{2015}]{Taylor2015DirectBilayer}
\begin{barticle}
\bauthor{\bsnm{Taylor}, \binits{G.J.}},
\bauthor{\bsnm{Venkatesan}, \binits{G.A.}},
\bauthor{\bsnm{Collier}, \binits{C.P.}},
\bauthor{\bsnm{Sarles}, \binits{S.A.}}:
\batitle{{Direct in situ measurement of specific capacitance, monolayer tension, and bilayer tension in a droplet interface bilayer}}.
\bjtitle{Soft Matter}
\bvolume{11}(\bissue{38}),
\bfpage{7592}--\blpage{7605}
(\byear{2015})
\doiurl{10.1039/C5SM01005E}
\end{barticle}
\endbibitem

\bibitem[\protect\citeauthoryear{Requena and Haydon}{1975}]{Requena1975TheFilms}
\begin{barticle}
\bauthor{\bsnm{Requena}, \binits{J.}},
\bauthor{\bsnm{Haydon}, \binits{D.A.}}:
\batitle{{The lippmann equation and the characterization of black lipid films}}.
\bjtitle{Journal of Colloid and Interface Science}
\bvolume{51}(\bissue{2}),
\bfpage{315}--\blpage{327}
(\byear{1975})
\doiurl{10.1016/0021-9797(75)90119-8}
\end{barticle}
\endbibitem

\bibitem[\protect\citeauthoryear{Evans and Simon}{1975}]{Evans1975MechanicsMembranes}
\begin{barticle}
\bauthor{\bsnm{Evans}, \binits{E.A.}},
\bauthor{\bsnm{Simon}, \binits{S.}}:
\batitle{{Mechanics of electrocompression of lipid bilayer membranes}}.
\bjtitle{Biophysical Journal}
\bvolume{15}(\bissue{8}),
\bfpage{850}--\blpage{852}
(\byear{1975})
\doiurl{10.1016/S0006-3495(75)85860-7}
\end{barticle}
\endbibitem

\bibitem[\protect\citeauthoryear{Taylor et~al.}{2019}]{Taylor2019ElectrophysiologicalFlip-flop}
\begin{barticle}
\bauthor{\bsnm{Taylor}, \binits{G.}},
\bauthor{\bsnm{Nguyen}, \binits{M.A.}},
\bauthor{\bsnm{Koner}, \binits{S.}},
\bauthor{\bsnm{Freeman}, \binits{E.}},
\bauthor{\bsnm{Collier}, \binits{C.P.}},
\bauthor{\bsnm{Sarles}, \binits{S.A.}}:
\batitle{{Electrophysiological interrogation of asymmetric droplet interface bilayers reveals surface-bound alamethicin induces lipid flip-flop}}.
\bjtitle{Biochimica et Biophysica Acta (BBA) - Biomembranes}
\bvolume{1861}(\bissue{1}),
\bfpage{335}--\blpage{343}
(\byear{2019})
\doiurl{10.1016/J.BBAMEM.2018.07.001}
\end{barticle}
\endbibitem

\bibitem[\protect\citeauthoryear{Romero et~al.}{2021}]{Romero2021MemcapacitorReview}
\begin{barticle}
\bauthor{\bsnm{Romero}, \binits{F.J.}},
\bauthor{\bsnm{Ohata}, \binits{A.}},
\bauthor{\bsnm{Toral-Lopez}, \binits{A.}},
\bauthor{\bsnm{Godoy}, \binits{A.}},
\bauthor{\bsnm{Morales}, \binits{D.P.}},
\bauthor{\bsnm{Rodriguez}, \binits{N.}}:
\batitle{{Memcapacitor and Meminductor Circuit Emulators: A Review}}.
\bjtitle{Electronics 2021, Vol. 10, Page 1225}
\bvolume{10}(\bissue{11}),
\bfpage{1225}
(\byear{2021})
\doiurl{10.3390/ELECTRONICS10111225}
\end{barticle}
\endbibitem

\bibitem[\protect\citeauthoryear{{Herbert Jaeger}}{2002}]{HerbertJaeger2002Short-termNetworks}
\begin{botherref}
\oauthor{\bsnm{{Herbert Jaeger}}}:
{Short-term Memory in Echo State Networks}.
GMD Report
(2002)
\end{botherref}
\endbibitem

\bibitem[\protect\citeauthoryear{Richtmyer}{1978}]{Richtmyer1978NonlinearDynamics}
\begin{botherref}
\oauthor{\bsnm{Richtmyer}, \binits{R.D.}}:
{Nonlinear Problems: Fluid Dynamics}.
Principles of Advanced Mathematical Physics,
364--408
(1978)
\doiurl{10.1007/978-3-642-46378-5{\_}}
\end{botherref}
\endbibitem

\bibitem[\protect\citeauthoryear{Isidori}{1995}]{Isidori1995NonlinearSystems}
\begin{botherref}
\oauthor{\bsnm{Isidori}, \binits{A.}}:
{Nonlinear Control Systems}
(1995)
\doiurl{10.1007/978-1-84628-615-5}
\end{botherref}
\endbibitem

\bibitem[\protect\citeauthoryear{Adachi et~al.}{2022}]{Adachi2022UniversalDynamics}
\begin{barticle}
\bauthor{\bsnm{Adachi}, \binits{K.}},
\bauthor{\bsnm{Iritani}, \binits{R.}},
\bauthor{\bsnm{Hamazaki}, \binits{R.}}:
\batitle{{Universal constraint on nonlinear population dynamics}}.
\bjtitle{Communications Physics 2022 5:1}
\bvolume{5}(\bissue{1}),
\bfpage{1}--\blpage{7}
(\byear{2022})
\doiurl{10.1038/S42005-022-00912-4}
\end{barticle}
\endbibitem

\bibitem[\protect\citeauthoryear{Gross and Feudel}{2006}]{Gross2006GeneralizedSystems}
\begin{barticle}
\bauthor{\bsnm{Gross}, \binits{T.}},
\bauthor{\bsnm{Feudel}, \binits{U.}}:
\batitle{{Generalized models as a universal approach to the analysis of nonlinear dynamical systems}}.
\bjtitle{Physical Review E - Statistical, Nonlinear, and Soft Matter Physics}
\bvolume{73}(\bissue{1}),
\bfpage{016205}
(\byear{2006})
\doiurl{10.1103/PHYSREVE.73.016205/FIGURES/2/MEDIUM}
\end{barticle}
\endbibitem

\bibitem[\protect\citeauthoryear{Cover}{1965}]{Cover1965GeometricalRecognition}
\begin{barticle}
\bauthor{\bsnm{Cover}, \binits{T.M.}}:
\batitle{{Geometrical and Statistical Properties of Systems of Linear Inequalities with Applications in Pattern Recognition}}.
\bjtitle{IEEE Transactions on Electronic Computers}
\bvolume{EC-14}(\bissue{3}),
\bfpage{326}--\blpage{334}
(\byear{1965})
\doiurl{10.1109/PGEC.1965.264137}
\end{barticle}
\endbibitem

\bibitem[\protect\citeauthoryear{Guo et~al.}{2023}]{Guo2023GenerativeVariability}
\begin{barticle}
\bauthor{\bsnm{Guo}, \binits{Y.}},
\bauthor{\bsnm{Duan}, \binits{W.}},
\bauthor{\bsnm{Liu}, \binits{X.}},
\bauthor{\bsnm{Wang}, \binits{X.}},
\bauthor{\bsnm{Wang}, \binits{L.}},
\bauthor{\bsnm{Duan}, \binits{S.}},
\bauthor{\bsnm{Ma}, \binits{C.}},
\bauthor{\bsnm{Li}, \binits{H.}}:
\batitle{{Generative complex networks within a dynamic memristor with intrinsic variability}}.
\bjtitle{Nature Communications 2023 14:1}
\bvolume{14}(\bissue{1}),
\bfpage{1}--\blpage{10}
(\byear{2023})
\doiurl{10.1038/S41467-023-41921-3}
\end{barticle}
\endbibitem

\bibitem[\protect\citeauthoryear{Kaplan et~al.}{1996}]{Kaplan1996SubthresholdAxons}
\begin{barticle}
\bauthor{\bsnm{Kaplan}, \binits{D.T.}},
\bauthor{\bsnm{Clay}, \binits{J.R.}},
\bauthor{\bsnm{Manning}, \binits{T.}},
\bauthor{\bsnm{Glass}, \binits{L.}},
\bauthor{\bsnm{Guevara}, \binits{M.R.}},
\bauthor{\bsnm{Shrier}, \binits{A.}}:
\batitle{{Subthreshold Dynamics in Periodically Stimulated Squid Giant Axons}}.
\bjtitle{Physical Review Letters}
\bvolume{76}(\bissue{21}),
\bfpage{4074}
(\byear{1996})
\doiurl{10.1103/PhysRevLett.76.4074}
\end{barticle}
\endbibitem

\bibitem[\protect\citeauthoryear{Linde}{1986}]{Linde1986EternallyUniverse}
\begin{barticle}
\bauthor{\bsnm{Linde}, \binits{A.D.}}:
\batitle{{Eternally existing self-reproducing chaotic inflanationary universe}}.
\bjtitle{Physics Letters B}
\bvolume{175}(\bissue{4}),
\bfpage{395}--\blpage{400}
(\byear{1986})
\doiurl{10.1016/0370-2693(86)90611-8}
\end{barticle}
\endbibitem

\bibitem[\protect\citeauthoryear{Hudson and Mankin}{1981}]{Hudson1981ChaosReaction}
\begin{barticle}
\bauthor{\bsnm{Hudson}, \binits{J.L.}},
\bauthor{\bsnm{Mankin}, \binits{J.C.}}:
\batitle{{Chaos in the Belousov–Zhabotinskii reaction}}.
\bjtitle{The Journal of Chemical Physics}
\bvolume{74}(\bissue{11}),
\bfpage{6171}--\blpage{6177}
(\byear{1981})
\doiurl{10.1063/1.441007}
\end{barticle}
\endbibitem

\bibitem[\protect\citeauthoryear{Toker et~al.}{2020}]{Toker2020ANature}
\begin{barticle}
\bauthor{\bsnm{Toker}, \binits{D.}},
\bauthor{\bsnm{Sommer}, \binits{F.T.}},
\bauthor{\bsnm{D’Esposito}, \binits{M.}}:
\batitle{{A simple method for detecting chaos in nature}}.
\bjtitle{Communications Biology 2020 3:1}
\bvolume{3}(\bissue{1}),
\bfpage{1}--\blpage{13}
(\byear{2020})
\doiurl{10.1038/S42003-019-0715-9}
\end{barticle}
\endbibitem

\bibitem[\protect\citeauthoryear{Wen}{2014}]{Wen2014AInterpretations}
\begin{botherref}
\oauthor{\bsnm{Wen}, \binits{H.}}:
{A review of the H{\'{e}}non map and its physical interpretations}
(2014)
\end{botherref}
\endbibitem

\bibitem[\protect\citeauthoryear{Zhong et~al.}{2022}]{Zhong2022AProcessing}
\begin{barticle}
\bauthor{\bsnm{Zhong}, \binits{Y.}},
\bauthor{\bsnm{Tang}, \binits{J.}},
\bauthor{\bsnm{Li}, \binits{X.}},
\bauthor{\bsnm{Liang}, \binits{X.}},
\bauthor{\bsnm{Liu}, \binits{Z.}},
\bauthor{\bsnm{Li}, \binits{Y.}},
\bauthor{\bsnm{Xi}, \binits{Y.}},
\bauthor{\bsnm{Yao}, \binits{P.}},
\bauthor{\bsnm{Hao}, \binits{Z.}},
\bauthor{\bsnm{Gao}, \binits{B.}},
\bauthor{\bsnm{Qian}, \binits{H.}},
\bauthor{\bsnm{Wu}, \binits{H.}}:
\batitle{{A memristor-based analogue reservoir computing system for real-time and power-efficient signal processing}}.
\bjtitle{Nature Electronics 2022 5:10}
\bvolume{5}(\bissue{10}),
\bfpage{672}--\blpage{681}
(\byear{2022})
\doiurl{10.1038/s41928-022-00838-3}
\end{barticle}
\endbibitem

\bibitem[\protect\citeauthoryear{Voinigescu}{2013}]{Voinigescu2013High-FrequencyCircuits}
\begin{bbook}
\bauthor{\bsnm{Voinigescu}, \binits{S.}}:
\bbtitle{{High-Frequency Integrated Circuits}}.
\bpublisher{Cambridge University Press}, \blocation{???}
(\byear{2013}).
\doiurl{10.1017/CBO9781139021128} .
\burl{https://www.cambridge.org/core/books/highfrequency-integrated-circuits/4BF00791FE2320D7FE4368E8F2424E4D}
\end{bbook}
\endbibitem

\bibitem[\protect\citeauthoryear{Hou et~al.}{2023}]{Hou2023LearningNanofluidics}
\begin{barticle}
\bauthor{\bsnm{Hou}, \binits{Y.}},
\bauthor{\bsnm{Ling}, \binits{Y.}},
\bauthor{\bsnm{Wang}, \binits{Y.}},
\bauthor{\bsnm{Wang}, \binits{M.}},
\bauthor{\bsnm{Chen}, \binits{Y.}},
\bauthor{\bsnm{Li}, \binits{X.}},
\bauthor{\bsnm{Hou}, \binits{X.}}:
\batitle{{Learning from the Brain: Bioinspired Nanofluidics}}.
\bjtitle{Journal of Physical Chemistry Letters}
\bvolume{14}(\bissue{11}),
\bfpage{2891}--\blpage{2900}
(\byear{2023})
\doiurl{10.1021/ACS.JPCLETT.2C03930/ASSET/IMAGES/LARGE/JZ2C03930{\_}0005.J}
\end{barticle}
\endbibitem

\bibitem[\protect\citeauthoryear{Schimel et~al.}{2021}]{Schimel2021Pressure-drivenNetworks}
\begin{barticle}
\bauthor{\bsnm{Schimel}, \binits{T.M.}},
\bauthor{\bsnm{Nguyen}, \binits{M.A.}},
\bauthor{\bsnm{Sarles}, \binits{S.A.}},
\bauthor{\bsnm{Lenaghan}, \binits{S.C.}}:
\batitle{{Pressure-driven generation of complex microfluidic droplet networks}}.
\bjtitle{Microfluidics and Nanofluidics}
\bvolume{25}(\bissue{9}),
\bfpage{1}--\blpage{12}
(\byear{2021})
\doiurl{10.1007/S10404-021-02477-0/FIGURES/8}
\end{barticle}
\endbibitem

\bibitem[\protect\citeauthoryear{Nguyen and Sarles}{2016}]{Nguyen2016MicrofluidicBilayers}
\begin{botherref}
\oauthor{\bsnm{Nguyen}, \binits{M.A.}},
\oauthor{\bsnm{Sarles}, \binits{S.A.}}:
{Microfluidic Generation, Encapsulation and Characterization of Asymmetric Droplet Interface Bilayers}.
ASME 2016 Conference on Smart Materials, Adaptive Structures and Intelligent Systems, SMASIS 2016
\textbf{2}
(2016)
\doiurl{10.1115/SMASIS2016-9034}
\end{botherref}
\endbibitem

\bibitem[\protect\citeauthoryear{Nguyen et~al.}{2016}]{Nguyen2016HydrodynamicArrays}
\begin{barticle}
\bauthor{\bsnm{Nguyen}, \binits{M.A.}},
\bauthor{\bsnm{Srijanto}, \binits{B.}},
\bauthor{\bsnm{Collier}, \binits{C.P.}},
\bauthor{\bsnm{Retterer}, \binits{S.T.}},
\bauthor{\bsnm{Sarles}, \binits{S.A.}}:
\batitle{{Hydrodynamic trapping for rapid assembly and in situ electrical characterization of droplet interface bilayer arrays}}.
\bjtitle{Lab on a Chip}
\bvolume{16}(\bissue{18}),
\bfpage{3576}--\blpage{3588}
(\byear{2016})
\doiurl{10.1039/C6LC00810K}
\end{barticle}
\endbibitem

\bibitem[\protect\citeauthoryear{Nguyen et~al.}{2017}]{Nguyen2017AMembranes}
\begin{botherref}
\oauthor{\bsnm{Nguyen}, \binits{M.A.}},
\oauthor{\bsnm{Taylor}, \binits{G.}},
\oauthor{\bsnm{Sarles}, \binits{S.A.}}:
{A Microfluidic Assembly and Simultaneous Interrogation of Networks of Asymmetric Biomimetic Membranes}.
ASME 2017 Conference on Smart Materials, Adaptive Structures and Intelligent Systems, SMASIS 2017
\textbf{1}
(2017)
\doiurl{10.1115/SMASIS2017-3878}
\end{botherref}
\endbibitem

\bibitem[\protect\citeauthoryear{Alcinesio et~al.}{2022}]{Alcinesio2022FunctionalNetworks}
\begin{barticle}
\bauthor{\bsnm{Alcinesio}, \binits{A.}},
\bauthor{\bsnm{Krishna~Kumar}, \binits{R.}},
\bauthor{\bsnm{Bayley}, \binits{H.}}:
\batitle{{Functional Multivesicular Structures with Controlled Architecture from 3D-Printed Droplet Networks}}.
\bjtitle{ChemSystemsChem}
\bvolume{4}(\bissue{1}),
\bfpage{202100036}
(\byear{2022})
\doiurl{10.1002/SYST.202100036}
\end{barticle}
\endbibitem

\bibitem[\protect\citeauthoryear{Sarles and Leo}{2010}]{Sarles2010PhysicalNetworks}
\begin{barticle}
\bauthor{\bsnm{Sarles}, \binits{S.A.}},
\bauthor{\bsnm{Leo}, \binits{D.J.}}:
\batitle{{Physical encapsulation of droplet interface bilayers for durable, portable biomolecular networks}}.
\bjtitle{Lab on a Chip}
\bvolume{10}(\bissue{6}),
\bfpage{710}--\blpage{717}
(\byear{2010})
\doiurl{10.1039/B916736F}
\end{barticle}
\endbibitem

\bibitem[\protect\citeauthoryear{Challita et~al.}{2018}]{Challita2018EncapsulatingOrganogel}
\begin{barticle}
\bauthor{\bsnm{Challita}, \binits{E.J.}},
\bauthor{\bsnm{Najem}, \binits{J.S.}},
\bauthor{\bsnm{Monroe}, \binits{R.}},
\bauthor{\bsnm{Leo}, \binits{D.J.}},
\bauthor{\bsnm{Freeman}, \binits{E.C.}}:
\batitle{{Encapsulating Networks of Droplet Interface Bilayers in a Thermoreversible Organogel}}.
\bjtitle{Scientific Reports 2018 8:1}
\bvolume{8}(\bissue{1}),
\bfpage{1}--\blpage{11}
(\byear{2018})
\doiurl{10.1038/S41598-018-24720-5}
\end{barticle}
\endbibitem

\bibitem[\protect\citeauthoryear{Villar et~al.}{2013}]{Villar2013AMaterial}
\begin{barticle}
\bauthor{\bsnm{Villar}, \binits{G.}},
\bauthor{\bsnm{Graham}, \binits{A.D.}},
\bauthor{\bsnm{Bayley}, \binits{H.}}:
\batitle{{A tissue-like printed material}}.
\bjtitle{Science}
\bvolume{340}(\bissue{6128}),
\bfpage{48}--\blpage{52}
(\byear{2013})
\doiurl{10.1126/SCIENCE.1229495/SUPPL{\_}FILE/VILLAR.SM.}
\end{barticle}
\endbibitem

\bibitem[\protect\citeauthoryear{Alcinesio et~al.}{2020}]{Alcinesio2020ControlledTissues}
\begin{barticle}
\bauthor{\bsnm{Alcinesio}, \binits{A.}},
\bauthor{\bsnm{Meacock}, \binits{O.J.}},
\bauthor{\bsnm{Allan}, \binits{R.G.}},
\bauthor{\bsnm{Monico}, \binits{C.}},
\bauthor{\bsnm{Restrepo~Schild}, \binits{V.}},
\bauthor{\bsnm{Cazimoglu}, \binits{I.}},
\bauthor{\bsnm{Cornall}, \binits{M.T.}},
\bauthor{\bsnm{Krishna~Kumar}, \binits{R.}},
\bauthor{\bsnm{Bayley}, \binits{H.}}:
\batitle{{Controlled packing and single-droplet resolution of 3D-printed functional synthetic tissues}}.
\bjtitle{Nature Communications 2020 11:1}
\bvolume{11}(\bissue{1}),
\bfpage{1}--\blpage{13}
(\byear{2020})
\doiurl{10.1038/S41467-020-15953-Y}
\end{barticle}
\endbibitem

\bibitem[\protect\citeauthoryear{Graham et~al.}{2017}]{Graham2017High-ResolutionPrinting}
\begin{barticle}
\bauthor{\bsnm{Graham}, \binits{A.D.}},
\bauthor{\bsnm{Olof}, \binits{S.N.}},
\bauthor{\bsnm{Burke}, \binits{M.J.}},
\bauthor{\bsnm{Armstrong}, \binits{J.P.K.}},
\bauthor{\bsnm{Mikhailova}, \binits{E.A.}},
\bauthor{\bsnm{Nicholson}, \binits{J.G.}},
\bauthor{\bsnm{Box}, \binits{S.J.}},
\bauthor{\bsnm{Szele}, \binits{F.G.}},
\bauthor{\bsnm{Perriman}, \binits{A.W.}},
\bauthor{\bsnm{Bayley}, \binits{H.}}:
\batitle{{High-Resolution Patterned Cellular Constructs by Droplet-Based 3D Printing}}.
\bjtitle{Scientific Reports 2017 7:1}
\bvolume{7}(\bissue{1}),
\bfpage{1}--\blpage{11}
(\byear{2017})
\doiurl{10.1038/S41598-017-06358-X}
\end{barticle}
\endbibitem

\bibitem[\protect\citeauthoryear{Mugele and Baret}{2005}]{Mugele2005Electrowetting:Applications}
\begin{barticle}
\bauthor{\bsnm{Mugele}, \binits{F.}},
\bauthor{\bsnm{Baret}, \binits{J.C.}}:
\batitle{{Electrowetting: from basics to applications}}.
\bjtitle{Journal of Physics: Condensed Matter}
\bvolume{17}(\bissue{28}),
\bfpage{705}
(\byear{2005})
\doiurl{10.1088/0953-8984/17/28/R01}
\end{barticle}
\endbibitem

\bibitem[\protect\citeauthoryear{Li and Kim}{2020}]{Li2020CurrentMicrofluidics}
\begin{barticle}
\bauthor{\bsnm{Li}, \binits{J.}},
\bauthor{\bsnm{Kim}, \binits{C.J.}}:
\batitle{{Current commercialization status of electrowetting-on-dielectric (EWOD) digital microfluidics}}.
\bjtitle{Lab on a Chip}
\bvolume{20}(\bissue{10}),
\bfpage{1705}--\blpage{1712}
(\byear{2020})
\doiurl{10.1039/D0LC00144A}
\end{barticle}
\endbibitem

\bibitem[\protect\citeauthoryear{Najem et~al.}{2019}]{Najem2019AssemblyMembranes}
\begin{barticle}
\bauthor{\bsnm{Najem}, \binits{J.S.}},
\bauthor{\bsnm{Taylor}, \binits{G.J.}},
\bauthor{\bsnm{Armendarez}, \binits{N.}},
\bauthor{\bsnm{Weiss}, \binits{R.J.}},
\bauthor{\bsnm{Hasan}, \binits{M.S.}},
\bauthor{\bsnm{Rose}, \binits{G.S.}},
\bauthor{\bsnm{Schuman}, \binits{C.D.}},
\bauthor{\bsnm{Belianinov}, \binits{A.}},
\bauthor{\bsnm{Sarles}, \binits{S.A.}},
\bauthor{\bsnm{Collier}, \binits{C.P.}}:
\batitle{{Assembly and Characterization of Biomolecular Memristors Consisting of Ion Channel-doped Lipid Membranes}}.
\bjtitle{JoVE (Journal of Visualized Experiments)}
\bvolume{2019}(\bissue{145}),
\bfpage{58998}
(\byear{2019})
\doiurl{10.3791/58998}
\end{barticle}
\endbibitem

\bibitem[\protect\citeauthoryear{Jaeger}{2001}]{Jaeger2001TheNetworks}
\begin{barticle}
\bauthor{\bsnm{Jaeger}, \binits{H.}}:
\batitle{{The “echo state” approach to analysing and training recurrent neural networkroach to analysing and training recurrent neural networks}}.
\bjtitle{Bonn, Germany: German National Research Center for Information Technology GMD Technical Report}
\bvolume{148}(\bissue{34}),
\bfpage{13}
(\byear{2001})
\end{barticle}
\endbibitem

\bibitem[\protect\citeauthoryear{Liang et~al.}{2022}]{Liang2022RotatingComputing}
\begin{botherref}
\oauthor{\bsnm{Liang}, \binits{X.}},
\oauthor{\bsnm{Zhong}, \binits{Y.}},
\oauthor{\bsnm{Tang}, \binits{J.}},
\oauthor{\bsnm{Liu}, \binits{Z.}},
\oauthor{\bsnm{Yao}, \binits{P.}},
\oauthor{\bsnm{Sun}, \binits{K.}},
\oauthor{\bsnm{Zhang}, \binits{Q.}},
\oauthor{\bsnm{Gao}, \binits{B.}},
\oauthor{\bsnm{Heidari}, \binits{H.}},
\oauthor{\bsnm{Qian}, \binits{H.}},
\oauthor{\bsnm{Wu}, \binits{H.}}:
{Rotating neurons for all-analog implementation of cyclic reservoir computing}.
Nature communications
\textbf{13}(1)
(2022)
\doiurl{10.1038/S41467-022-29260-1}
\end{botherref}
\endbibitem

\bibitem[\protect\citeauthoryear{Yang et~al.}{2013}]{Yang2013MemristiveComputing}
\begin{barticle}
\bauthor{\bsnm{Yang}, \binits{J.J.}},
\bauthor{\bsnm{Strukov}, \binits{D.B.}},
\bauthor{\bsnm{Stewart}, \binits{D.R.}}:
\batitle{{Memristive devices for computing}}.
\bjtitle{Nature nanotechnology}
\bvolume{8}(\bissue{1}),
\bfpage{13}--\blpage{24}
(\byear{2013})
\doiurl{10.1038/NNANO.2012.240}
\end{barticle}
\endbibitem

\bibitem[\protect\citeauthoryear{Sarles and Leo}{2010}]{Sarles2010RegulatedSubstrates}
\begin{barticle}
\bauthor{\bsnm{Sarles}, \binits{S.A.}},
\bauthor{\bsnm{Leo}, \binits{D.J.}}:
\batitle{{Regulated attachment method for reconstituting lipid bilayers of prescribed size within flexible substrates}}.
\bjtitle{Analytical Chemistry}
\bvolume{82}(\bissue{3}),
\bfpage{959}--\blpage{966}
(\byear{2010})
\doiurl{10.1021/AC902555Z/AC902555Z{\_}WEO{\_}0}
\end{barticle}
\endbibitem

\bibitem[\protect\citeauthoryear{Najem et~al.}{2018}]{Najem2018MemristiveMimics}
\begin{barticle}
\bauthor{\bsnm{Najem}, \binits{J.S.}},
\bauthor{\bsnm{Taylor}, \binits{G.J.}},
\bauthor{\bsnm{Weiss}, \binits{R.J.}},
\bauthor{\bsnm{Hasan}, \binits{M.S.}},
\bauthor{\bsnm{Rose}, \binits{G.}},
\bauthor{\bsnm{Schuman}, \binits{C.D.}},
\bauthor{\bsnm{Belianinov}, \binits{A.}},
\bauthor{\bsnm{Collier}, \binits{C.P.}},
\bauthor{\bsnm{Sarles}, \binits{S.A.}}:
\batitle{{Memristive Ion Channel-Doped Biomembranes as Synaptic Mimics}}.
\bjtitle{ACS Nano}
\bvolume{12}(\bissue{5}),
\bfpage{4702}--\blpage{4711}
(\byear{2018})
\doiurl{10.1021/ACSNANO.8B01282/ASSET/IMAGES/LARGE/NN-2018-01282F{\_}0005.J}
\end{barticle}
\endbibitem

\bibitem[\protect\citeauthoryear{Inubushi and Yoshimura}{2017}]{Inubushi2017ReservoirTrade-off}
\begin{barticle}
\bauthor{\bsnm{Inubushi}, \binits{M.}},
\bauthor{\bsnm{Yoshimura}, \binits{K.}}:
\batitle{{Reservoir Computing Beyond Memory-Nonlinearity Trade-off}}.
\bjtitle{Scientific Reports 2017 7:1}
\bvolume{7}(\bissue{1}),
\bfpage{1}--\blpage{10}
(\byear{2017})
\doiurl{10.1038/S41598-017-10257-6}
\end{barticle}
\endbibitem

\bibitem[\protect\citeauthoryear{Eisenberg et~al.}{1973}]{Eisenberg1973TheMembranes}
\begin{barticle}
\bauthor{\bsnm{Eisenberg}, \binits{M.}},
\bauthor{\bsnm{Hall}, \binits{J.E.}},
\bauthor{\bsnm{Mead}, \binits{C.A.}}:
\batitle{{The nature of the voltage-dependent conductance induced by alamethicin in black lipid membranes}}.
\bjtitle{The Journal of Membrane Biology}
\bvolume{14}(\bissue{1}),
\bfpage{143}--\blpage{176}
(\byear{1973})
\doiurl{10.1007/BF01868075/METRICS}
\end{barticle}
\endbibitem

\end{thebibliography}
\end{document}